\documentclass[preprint,letterpaper,preprintnumbers,superscriptaddress,aps,nofootinbib,tightenlines,floatfix]{revtex4}
\usepackage{amsmath,amssymb}
\usepackage{graphicx}
\usepackage{epstopdf}
\usepackage{bm}
\usepackage{comment}
\newcommand{\beq}{\begin{equation}}
\newcommand{\eeq}{\end{equation}}
\newcommand{\bea}{\begin{eqnarray}}
\newcommand{\eea}{\end{eqnarray}}
\newcommand{\Choose}[2]{{^{#1}C_{#2}}}
\def\abar{\overline{a}}

\newcommand{\gsim}{\raisebox{-0.7ex}{$\stackrel{\textstyle >}{\sim}$ }}

\def\bx{\bf x}
\def\bxp{{\bf x}^{\prime}}
\def\by{\bf y}
\def\byp{{\bf y}^{\prime}}
\def\bp{\bf p}
\def\bpp{{\bf p}^{\prime}}

\def\OMIT#1{{}}

\newcommand{\lsim}{\raisebox{-0.7ex}{$\stackrel{\textstyle <}{\sim}$ }}
\newcount\hour \newcount\hourminute \newcount\minute
\hour=\time \divide \hour by 60
\hourminute=\hour \multiply \hourminute by 60
\minute=\time \advance \minute by -\hourminute
\newcommand{\mydate}{\ \today \ - \number\hour :\number\minute}
\begin{document}
 \preprint{\vbox{ \hbox{JLAB-THY-12-1548} }}
\title{Lattice QCD at non-zero isospin chemical potential} 
\author{William Detmold} \author{Kostas Orginos} \author{Zhifeng Shi }
\affiliation{Department of Physics, The College of William \& Mary, Williamsburg, VA 23187, USA}
\affiliation{Jefferson Lab, Newport News, VA 23606, USA }
\date{\mydate}
\begin{abstract}
  \noindent
  Quantum chromodynamics (QCD) at non-zero isospin chemical potential
  is studied in a canonical approach by analyzing systems of fixed
  isospin number density. To construct these systems, we develop a
  range of new algorithms for performing the factorially large numbers
  of Wick contractions required in multi-hadron systems.  We then use
  these methods to study systems with the quantum numbers of up to 72
  $\pi^+$'s on three ensembles of gauge configurations with spatial
  extents $L\sim$ 2.0, 2.5 and 3.0 fm, and light quark masses
  corresponding to a pion mass of \mbox{$390$ MeV}. The ground state
  energies of these systems are extracted and the volume dependence of
  these energies is utilized to determine the two- and three- body
  interactions amongst $\pi^+$'s.  The systems studied correspond to
  isospin densities of up to $\rho_I\sim 9\ {\rm fm}^{-3}$ and probe
  isospin chemical potentials, $\mu_I$, in the range $m_\pi\ \lsim \mu_I\ 
  \lsim 4.5\ m_\pi$, allowing us to investigate aspects of the QCD
  phase diagram at low temperature and for varying isospin chemical
  potential. By studying the energy density of the system, we provide
  numerical evidence for the conjectured transition of the system
  to a Bose-Einstein condensed phase at $\mu_I\ \gsim m_\pi$.

\end{abstract}
\maketitle
\section{Introduction}
An important goal of nuclear physics is to study the interactions and
properties of systems comprised of large number of hadrons.  Nuclear
physics is an emergent phenomenon of the Standard Model and as this
goal requires an understanding of the strong interaction dynamics of
multi-hadron systems, it necessitates lattice QCD calculations.  In
recent years, preliminary studies of three- and four- baryon systems
have been undertaken \cite{Beane:2009gs,Yamazaki:2009ua} and more
investigations are underway. In addition, systems involving up to
twelve $\pi^{\pm}$'s~\cite{Beane:2007es, Detmold:2008fn} or twelve
$K^{\pm}$'s~\cite{Detmold:2008yn} and systems comprised of more than
one species~\cite{Detmold_Brian:2011} have been studied, allowing
the various two- and three-body interaction parameters
of these systems to be determined from the energy shift of $N$-meson
system at finite density.

The study of systems comprised of large numbers of hadrons can provide
vital insight into the structure of high density mater, which may
exist in the interiors of neutron stars~\cite{Kaplan:1986}, and it is
also interesting from a purely theoretical point of view to study the
rich phase structure of QCD.  For systems of high isopin density,
$\rho_I$, and non-zero isospin chemical potentials, $\mu_I$, as will
be studied here, a complex phase structure has been
conjectured~\cite{Son:Stephanov}.  When $\mu_I$ reaches the mass of
one pion, pions can be produced out of the vacuum and a Bose-Einstein
condensate (BEC) is expected to form. At asymptotically large $\mu_I$,
the system is known to be a colour superconducting BCS-like state and
at an intermediate isospin chemical potential, a BEC/BCS transition is
expected to occur. However, the exact locations and natures of these
BEC and the BEC/BCS transitions are unknown and can only be determined
by non-perturbative QCD calculations.  QCD studies of systems with
finite baryon density are hampered by the sign problem resulting from
the non-positivity of the determinant of the Dirac operator.  However
for systems with non-zero isospin chemical potential, there is no sign
problem. By introducing an isospin chemical potential into the QCD
action, non-zero isospin density systems have been studied
in Ref.~\cite{Kogut:2004zg,Sinclair:2006zm,deForcrand:Wenger2007}, showing
hints of some aspects of the expected phase structure.  Non-zero
isospin density system can also be studied by the direct computation
of correlation function of increasing numbers of
pions~\cite{Beane:2007es,Detmold:2008fn,Detmold:2008yn} and the
extension of these methods is the subject of the current work.

Calculating correlation functions involving many-meson systems (here
we will focus on many $\pi^+$ systems) involves computing all possible
contractions between quark field operators, the number of which naively
grows as $N!N!$ for mesonic systems. Even considering symmetries
between up and down quarks and identifying vanishing and redundant
contribution, the number of remaining contractions grows exponentially
with $N$. In order to overcome this problem, much progress has been
made in studying many meson systems in Ref.~\cite{Detmold:Savage} by
constructing a recursion relation for correlation functions of systems
having different number of mesons, taking advantage of the fact that
many contractions in the correlation function of an \mbox{$N$-meson}
system have been partly computed for an \mbox{$(N-1)$-meson}
system. Comparing with direct contractions, the recursion
relation~\cite{Detmold:Savage} saves tremendous amount of time, and
systems having up to $24\ \pi^+$'s have been studied
in~\cite{Z_W_2011_proceeding} using it.  Since the
Pauli principle excludes putting more than $N_c N_s = 12$ quarks
(where $N_c$ and $N_s$ are the number of color and spin components of
the quark fields, respectively) in the same source location,
additional sources are required for $N>12$-meson\footnote{ For
  simplicity we will refer to ``$N$-meson'' systems.  More correctly,
  we deal with systems in which a conserved quantum number (such as
  isospin) is fixed to $N$.}  systems.  In additional to requiring
more quark propagators, this complicates the recursion relation and
increases the computational cost of contractions such that the cost
for $N=24$ $\pi^+$'s is $\sim 100$ times that of $12$ $\pi^+$'s. Studying
systems of $36\ \pi^+$'s becomes extremely time consuming even with
the recursion relation.

In the current work, we construct new methods to compute correlation
function of systems containing large numbers of mesons of one species
and also for muti-species systems by utilizing the fact that the
ground state energy is independent of how the $\pi^+$'s are
distributed among different source
locations~\cite{Z_W_2011_proceeding}.  The new methods that are
presented herein significantly speed up the contractions, and enable
us to study even higher density systems, that are impractical with
other methods.  Using one of our new approaches,
 systems comprised of up to $72$
$\pi^+$'s are studied on four ensembles of anisotropic clover
lattices~\cite{heuy:david2008} with dimensions $16^3 \times 128$,
$20^3 \times 128$, $24^3 \times 128$ and $20^3 \times 256$. This
allows us to investigate multi-pion interactions and study the phase
structure of QCD at non-zero $\mu_I$.  In this work, we are able to
probe the QCD phase diagram from $\mu_I = m_{\pi}$ up to $\mu_I
\approx 4.5\ m_{\pi}$.  We provide strong evidence for the
Bose-Einstein condensation of the system and attempt to investigate
the BEC/BCS transition at larger $\mu_I$.

The layout of this paper is as follows. In
Section.~\ref{sec:problem_to_solve}, we describe new methods to
compute the correlation functions corresponding to many pion systems,
and compare the performance and scalability of each method.  Details
of the lattice ensembles and our computation of the relevant
correlation functions in momentum space are discussed in
Section.~\ref{sec:lattice_detail}.  By applying the most efficient
contraction method, correlation function of systems comprised of up to
$72\ \pi^+$'s are computed, and the ground state energies are
extracted in Section.~\ref{sec:results}.  In
Section.~\ref{scattering_parameters}, the two-body and three-body
interaction parameters are studied.  In
Section.~\ref{sec:qcd_phase_diagram}, the QCD phase diagram at
non-zero isospin chemical potential is investigated, and the
transition to a BEC state is identified.

\section{Methodology of multi-meson contractions}
\label{sec:problem_to_solve}
Non-zero isospin density meson systems can be studied by evaluating
correlation functions of many $\pi^+$'s at finite volume (as we work
in the context of a relativistic field theory, the pion number is
ill-defined, however the net isospin of the system is specified in the
correlation functions below).  A correlation function for a system of
$\overline{n}=\sum_{i=1}^Nn_i$ $\pi^+$'s with $n_i\ \pi^+$'s from
source locations (${\bf y}_i,0$) is defined as:
\begin{eqnarray}
  C_{n_1, ... ,n_N}(t) & = &
  \left <\
    \left(\ \sum_{\bf x}\ \pi^+({\bf x},t)\ \right)^{\overline{n}}
    \left( \phantom{\sum_{\bf x}}\hskip -0.2in
      \pi^-({\bf y_1},0)\ \right)^{n_1} \ldots
    \left( \phantom{\sum_{\bf x}}\hskip -0.2in
      \pi^-({\bf y_N},0)\ \right)^{n_N}\
  \right >
  \ ,
  \label{eq:mpi}
\end{eqnarray}
where the interpolating operator ${\pi^+({\bx},t)} = {\overline
  d({\bx},t)} \gamma_5 u({\bx},t)$ and ${\pi^-({\bx},t)} = {\overline
  u({\bx},t)} \gamma_5 d({\bx},t)$.  The correlator $C_{n_1,
  ... ,n_N}$ can be identified as the term with prefactor
$\prod_{i=1}^N \lambda_i^{n_i}$ from the expansion of \mbox{$\det
  [1+\lambda_1 P_1+\lambda_2 P_2+\ldots+\lambda_N P_N]$}, where $N$ is
the number of sources, and the $12N \times 12N$ matrices $P_k$ are
given by:
\begin{eqnarray}
  P_k & = &
  \left(\begin{array}{c|c|c|c}
      0 & 0 & 0 & 0 \\
      \hline
      \vdots & \ldots & \ldots & \ldots \\
      \hline
      P_{k,1}&P_{k,2} & \ldots & P_{k,N}\\
      \hline
      \vdots & \ldots & \ldots & \ldots \\
      \hline
      0 & 0 & 0 & 0
    \end{array}
  \right),
  \label{eq:Pdef_descending}
\end{eqnarray}
with $12\times12$ sub-blocks
\begin{eqnarray}
  P_{k,i}(t) = \sum_{\bf x}S({\bf x},t; {\bf y}_i,0)S^{\dagger}({\bf x},t;{\bf y}_k,0),
\end{eqnarray}
where $S({\bf x},t; {\bf y},0)$ is a quark propagator between
two points.  Each $P_{k,i}$ is an uncontracted correlator describing a
quark propagating from source $i$ to source $k$ through the sink at
$\bx$ with the quantum number of a $\pi^+$.

As shown in Ref.~\cite{Detmold:Savage}, a recursion relation for the
$C_{n_1,..,n_N}(t)$ can be derived by studying the properties of the
expansion of the above determinant.  The $C_{n_1,...,n_N}(t)$'s have
the same energy spectrum for all combinations of $n_i$'s as long as
${\overline n}$ is fixed, so separately computing correlation
functions of all possible combinations of $n_i$'s is redundant.  We
can thus identify a combined correlator $C_{{\overline n}\pi}(t)$ as
the term having prefactor $\lambda ^n$ from the expansion of
\mbox{$\det[1+\lambda A]$}, with
\begin{eqnarray}
  A = P_1 + P_2+\ldots+P_N = 
  \left(\begin{array}{c|c|c|c}
      P_{1,1}&P_{1,2}& \ldots & P_{1,N}\\
      \hline
      \vdots & \ldots & \ldots & \ldots \\
      \hline
      P_{k,1}&P_{k,2}& \ldots & P_{k,N}\\
      \hline
      \vdots & \ldots & \ldots & \ldots \\
      \hline
      P_{N,1}&P_{N,2} & \ldots & P_{N,N}\\
    \end{array}
  \right).
  \label{eq:Matrix_A}
\end{eqnarray}

$C_{{\overline n}\pi}(t)$ is a complicated summation of all possible
$C_{n_1,n_2,...,n_N}(t)$ with fixed ${\overline n}$, in which we do not
identify which pions originate at which source. For multiple source
contractions, even terms representing more than $12$ $\pi^+$'s located
in a single source are included, however such terms vanish identically
and so do not produce additional noise in numerical calculations.  As
fewer correlation functions are needed, computing $C_{{\overline
    n}\pi}(t)$ is a computationally simpler task than recursively
computing all $C_{n_1,n_2,...,n_N}(t)$.  In the following subsections,
we will construct four algorithms to further speed up the calculation
of $C_{\bar{n}\pi}(t)$ and compare each algorithm in terms of
precision requirement and numerical cost.

\subsection{Vandermonde Matrix method (VMm)}

As described above, a correlation function of an $n$-$\pi^+$ system
($C_{n\pi}$) can be identified as the coefficient of $\lambda^n$ from
the power series expansion of \mbox{$\det[1+\lambda A]$}
\begin{eqnarray}
  \det[1+\lambda A] = 1 + \lambda C_{1\pi} + \lambda^2 C_{2\pi} + \ldots +
  \lambda^{12N} C_{12N\pi},
  \label{equ:det_expansion}
\end{eqnarray}
where $A$ is a $12N \times 12N$ matrix constructed from uncontracted
correlators following Eq.~(\ref{eq:Matrix_A}).  A simple way to get
$C_{n\pi}$ is by computing Eq.~(\ref{equ:det_expansion}) for $12N$
different choices of $\lambda$ ($\lambda_1,\ldots,\lambda_{12N}$).
The resulting system of equations can be written in the following
matrix form
\begin{eqnarray}
  \left (
    \begin{array}{c}
      \frac{\det[1+\lambda_1A]-1}{\lambda_1} \\
      \frac{\det[1+\lambda_2A]-1}{\lambda_2} \\
      \vdots\\
      \frac{\det[1+\lambda_{12N}A]-1}{\lambda_{12N}}
    \end{array}
  \right )
  =
  \left (
    \begin{array}{c c c c c}
      1 & \lambda_1 & \lambda_1^2 & \ldots & \lambda_1^{12N-1} \\
      1 & \lambda_2 & \lambda_2^2 & \ldots & \lambda_2^{12N-1} \\
      \vdots \\
      1 & \lambda_n & \lambda_n^2 & \ldots & \lambda_n^{12N-1}
    \end{array}
  \right )
  \cdot
  \left (
    \begin{array}{c}
      C_{1\pi} \\
      C_{2\pi} \\
      \vdots \\
      C_{12N\pi}
    \end{array}
  \right ).
  \label{eq:VMm_1}
\end{eqnarray}
The matrix on the RHS of Eq.~(\ref{eq:VMm_1}) is a $12N \times 12N$
Vandermonde matrix, for which there exist analytical forms for the
determinant and inverse (see for example
Ref.~\cite{V_M_inverse:cite}).  The inverse matrix then allows us to
determine the $C_{n\pi}$'s from the numerical calculation of the
determinant vector. However, when the number of sources becomes large,
elements of this matrix can become very small or very large because of
the factors of $\lambda^{1,2,\ldots,12N-1}_i$, making the computation
of the inverse very demanding in precision and eventually resulting in
significant numerical errors.

\subsection{FFT method (FFTm)}
By choosing $\lambda = \exp(i2\pi f_0 \cdot \tau)$ in
Eq.~(\ref{equ:det_expansion}), the expansion becomes
\begin{eqnarray}
  \det [1+\lambda A] = 1 + e^{2i\pi f_0 \cdot \tau}C_{1\pi}+
  e^{4i\pi f_0 \cdot \tau}C_{2\pi}+\ldots
  +e^{24i\pi Nf_0 \cdot \tau}C_{12N\pi}, 
\end{eqnarray}
which contains contributions from signals of frequencies $k f_0$, $k =
1,2,\ldots 12N$.  Because of this feature, the magnitude of each
frequency component can easily be extracted using a Fast Fourier
Transform (FFT).  The magnitude corresponding to frequency $k f_0$ is
equivalent to $C_{k\pi}$ times a normalization constant.  In order to
get better signals, data from multiple $\tau$'s are beneficial, which
results in the need to calculate many determinants, in general making
this method expensive.  On the other hand, specific choices of $f_0$
and $\tau$ can minimize the number of required determinants.  We set
$\tau_n = n\ dt$, for $n = 1,2,\ldots, T$ where $dt$ is the minimal
time step and $T$ is the closest prime number larger than $12N$, and
$f_0 = \frac{1}{dt\cdot T}$ and then compute $\det [1+\lambda_n A]$
with $\lambda_n = \exp(i2\pi f_0 \cdot \tau_n)$.  After applying the
FFT to this series, the amplitude of the frequency $kf_0$ is $T
C_{k\pi}$.  With such choices of $f_0$, $\tau_n$ and $T$, the number
of determinants needed to compute is the same as the Improved
Combination method (ICm) discussed below.

\subsection{Combination method (Cm)}

The FFTm discussed above is constructed from a certain choice of
$\lambda$'s so that the expansion of the determinant can be recognized
as contributions from different frequencies. Similarly, by studying
the properties of Eq.~(\ref{equ:det_expansion}), another choice of
$\lambda$'s can be utilized to eventually separate $\det [1+\lambda
A]$ into groups of functions individually depending only on $3$
correlation functions. This method requires us to determine the
inverse of a $3\times 3$ matrix, rather than a $12N\times 12N$
Vandermande matrix, to solve for the individual correlators.  This
method is applied by the following steps:

Step 1: Choose $f_1=1$ and compute
\begin{eqnarray}
  D_1^{(1)}(f_1\lambda) = \det[1+f_1\lambda A]-1.
\end{eqnarray}
Notice that $D_1^{(1)}(f_1\lambda)$ depends on all correlators
$C_{1\pi}, C_{2\pi}, \ldots, C_{12N\pi}$.

Step 2: Choose $f_2 = \exp (i\pi)$, and construct the following
contractions of the functions $D_1^{(1)}(f_n\lambda)$ to generate the
following two new quantities:
\begin{eqnarray}
  D^{(2)}_1(\lambda) &=& D_1^{(1)}(f_1\lambda) + f_1 D_1^{(1)}(f_2\lambda),  \nonumber  \\
  D^{(2)}_2(\lambda) &=& D_1^{(1)}(f_1\lambda) + f_2 D_1^{(1)}(f_2\lambda).
\end{eqnarray}
By inserting the values of $f_1=1$, $f_2 = -1$, it is clear that the
$D^{(2)}_i(\lambda)$ only depend on
$C_{(1+i)\pi},C_{(3+i)\pi},\ldots$, and so the correlation functions
have been separated into two groups.

Step 3: Choose $f_3 = \exp (i\frac{\pi}{2})$, and construct the
following combinations of the functions $D^{(2)}_1(f_n\lambda)$ and
$D^{(2)}_2(f_n\lambda)$:

\begin{eqnarray}
  D^{(3)}_1(\lambda) &=& D^{(2)}_1(\lambda) + f_1 D^{(2)}_1(f_3\lambda), \\ \nonumber
  D^{(3)}_2(\lambda) &=& D^{(2)}_1(\lambda) + f_2 D^{(2)}_1(f_3\lambda), \\ \nonumber
  D^{(3)}_3(\lambda) &=& D^{(2)}_2(\lambda) + f_1 f_3 D^{(2)}_2(f_3\lambda),  \\ \nonumber
  D^{(3)}_4(\lambda) &=& D^{(2)}_2(\lambda) + f_2 f_3 D^{(2)}_2(f_3\lambda),
\end{eqnarray}
and we see that the $D^{(3)}_i(\lambda)$ for $i=1,2$ depends on
$C_{(0+2i)\pi}, C_{(4+2i)\pi}, \ldots$, and $D^{(3)}_i(\lambda)$ for
$i=3,4$ depends on $C_{(9-2i)\pi}, C_{(13-2i)\pi}, \ldots$.  In each
step, one function depending on a block of $C_{k\pi}$'s is separated
into two functions each depending only on half of the $C_{k\pi}$'s
from the previous function.  We iterate this procedure until 
blocks of only $3$ $C_{k\pi}$'s are reached.

To summarize this method, in ``step $n$'', $f_n = \exp
(i\frac{\pi}{2^{n-2}})$ is chosen, and after this step
$D^{(n-1)}_i(\lambda)$, $i = 1, \ldots, 2^{n-2}$, will be separated
into $2^{n-1}$ functions, $D_i^{(n)}(\lambda)$, each depending on
${12N}/{2^{n-1}}$ $C_{k\pi}$'s.  Assume $D^{(n-1)}_m(\lambda)$ is a
function depending on a block of $C_{k\pi}$'s.  Two functions,
$D^{(n)}_{2m-1} $ and $D^{(n)}_{2m}$, each depending on a half of the
original block of $C_{k\pi}$'s are constructed from
\mbox{$D^{(n-1)}_m(\lambda)+q_{2m-1}\cdot
  D^{(n-1)}_m(f_n\cdot\lambda)$} and
\mbox{$D^{(n-1)}_m(\lambda)+q_{2m}\cdot
  D^{(n-1)}_m(f_n\cdot\lambda)$}, where the $q_{k}$'s,
$k=1,2,\ldots,2^{n-1}$, are prefactors used to construct new functions
depending only on half of the $C_{k\pi}$'s, which
$D^{(n-1)}_m(\lambda)$ depends on.  The prefactor $q_k$ in step $n$ is
constructed in the following way.
\begin{eqnarray}
  \text{Group 1: \hspace{1cm}} q_1 &=& f_1, \nonumber \\
  \text{Group 2: \hspace{1cm}} q_2 &=& f_2 \cdot q_1, \nonumber \\
  \text{Group 3: \hspace{1cm}} q_{k} &=& f_3 \cdot q_{k-2}, k = 3,4, \nonumber \\
  \vdots \nonumber \\
  \text{Group n: \hspace{1cm}} q_{k} &=& f_n \cdot q_{k-2^{n-2}}, k =
  2^{n-2}+1,2^{n-2}+2,...,2^{n-1},
\end{eqnarray}
where ``Group m'' contains $2^{m-2}$ functions for $m=2,3,\ldots, n$.
This process is repeated until functions, $D^{(\tilde n)}_k(\lambda)$, each
depending only on $3$ $C_{i\pi}$'s are reached.  Eventually
\mbox{$\det[1+\lambda A]$} is separated into functions,
$D^{(\tilde n)}_k(\lambda)$, depending on following blocks ($B_k$):
\begin{eqnarray}
  &\text{Group 1: \hspace{1cm}}& B_1 = [C_{4N\pi},C_{8N\pi},C_{12N\pi}] \nonumber \\
  &\text{Group 2: \hspace{1cm}}& B_2 = [C_{2N\pi},C_{6N\pi},C_{10N\pi}] \equiv C_{{\rm Sub}(B_1)-2N} \nonumber \\
  &\text{Group 3: \hspace{1cm}}&  \left\{ \begin{array}{rcl}
      B_3 &=& [C_{3N\pi},C_{7N\pi},C_{11N\pi}] \equiv C_{{\rm Sub}(B_1)-N} \\
      B_4 &=& [C_{N\pi},C_{5N\pi},C_{9N\pi}] \equiv C_{{\rm Sub}(B_2)-N} \end{array} \right. \nonumber \\
  &\vdots& \nonumber \\
  &\text{Group n: \hspace{1cm}}& B_k = C_{{\rm Sub}(B_{k-2^{n-2}})-\frac{4N}{2^{n-2}}}, 
  k = 2^{n-2}+1,2^{n-2}+2,\ldots,2^{n-1} 
\end{eqnarray}
where ${\rm Sub}(B_k)$ are the sub indexes of the $C$'s in $B_k$, for
example ${\rm Sub}(B_1) = \{4N, 8N, 12N\}$ and $C_{{\rm Sub}(B_1)-2N}
= \{ C_{2N\pi}, C_{6N\pi}, C_{10N\pi}\}$.  The dependence of $B_{k}$
on the corresponding $C$'s can be determined from the above recursion
relation.

In order to get the individual $C_{i\pi}$'s,
$D^{(\tilde n)}_{k}(\lambda_j)$ is required for three
different $\lambda_j$'s.  Different choices of $\lambda_j$'s have no
effect on $C_{i\pi}$'s (we have confirmed this numerically). From the
$D^{(\tilde n)}_{k}(\lambda_j)$'s, the $C_{k\pi}$'s are extracted by solving
the following equation, taking the block
$[C_{4N\pi},C_{8N\pi},C_{12N\pi}]$ for example,
\begin{eqnarray}
  \left (
    \begin{array}{c}
      D^{(\tilde n)}_1(\lambda_1) \\
      D^{(\tilde n)}_1(\lambda_2) \\
      D^{(\tilde n)}_1(\lambda_3) \\
    \end{array}
  \right )
  =
  \left (
    \begin{array}{c c c}
      \lambda_1^{4N} & \lambda_1^{8N} & \lambda_1^{12N} \\
      \lambda_2^{4N} & \lambda_2^{8N} & \lambda_2^{12N} \\
      \lambda_3^{4N} & \lambda_3^{8N} & \lambda_3^{12N} \\
    \end{array}
  \right )
  \cdot
  \left (
    \begin{array}{c}
      C_{4N\pi} \\
      C_{8N\pi} \\
      C_{12N\pi}\\
    \end{array}
  \right ).
\end{eqnarray}

\begin{figure}
  \includegraphics[width=5.4cm]{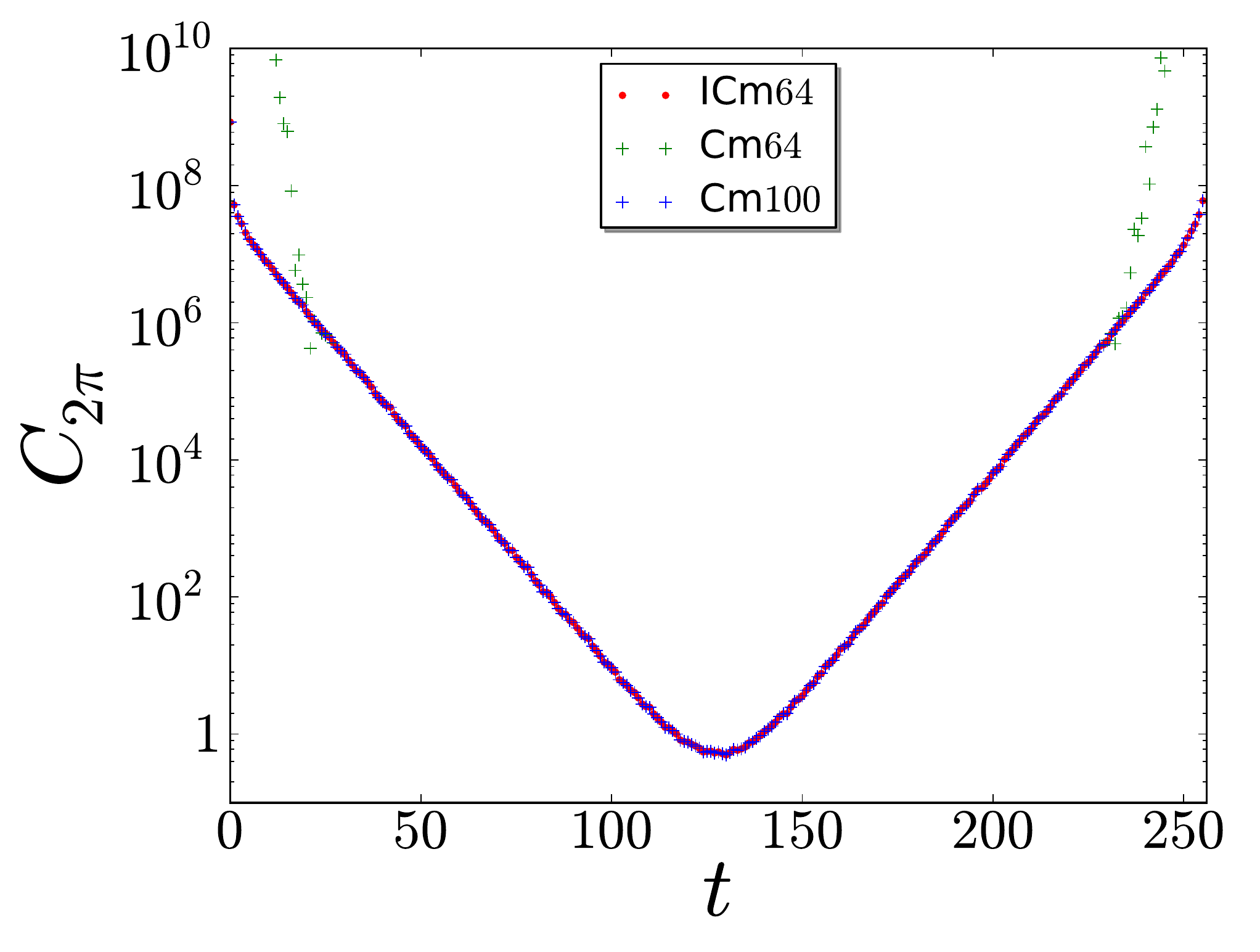}
  \includegraphics[width=5.4cm]{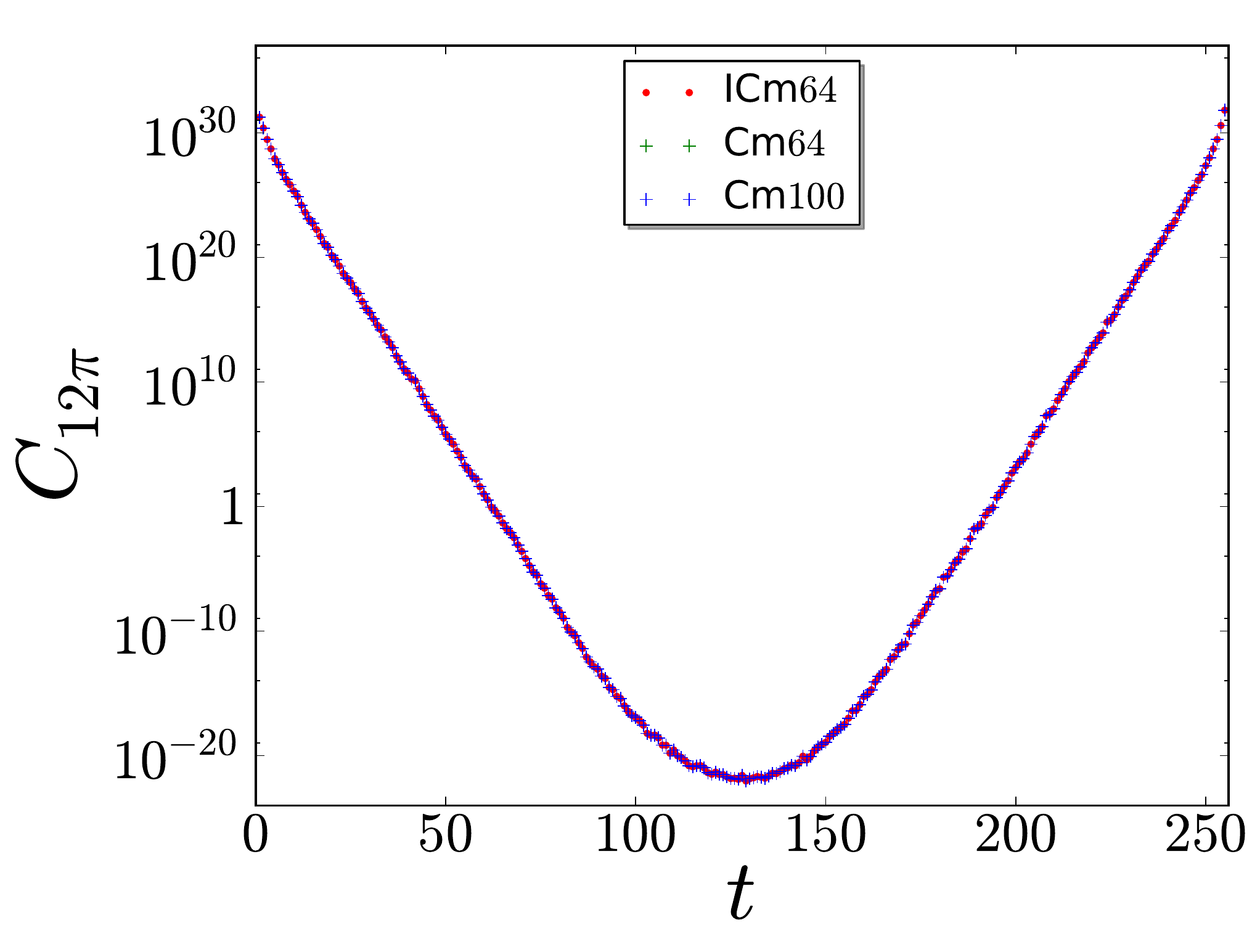}
  \includegraphics[width=5.4cm]{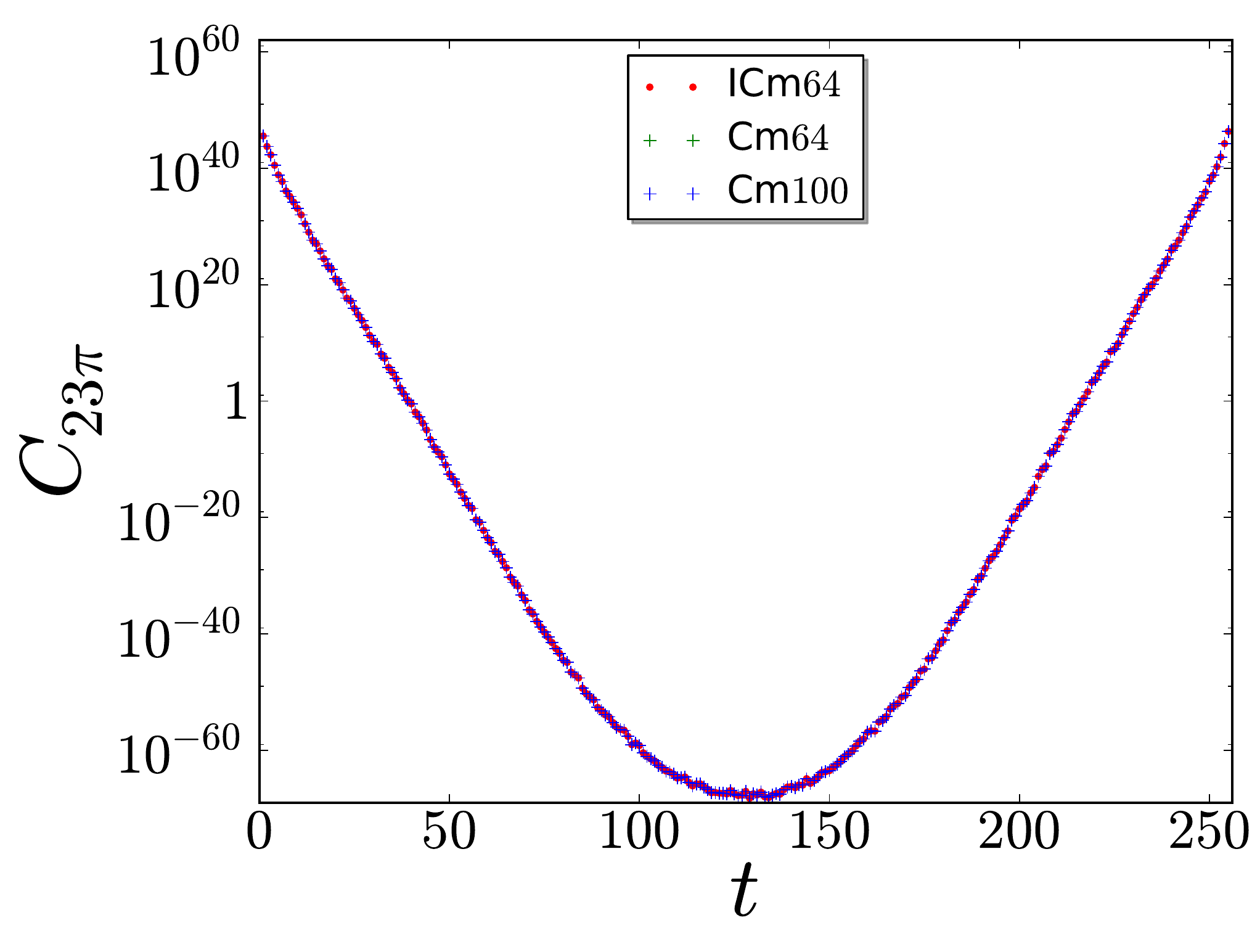}
  \caption{$C_{2\pi}(t)$, $C_{12\pi}(t)$ and $C_{23\pi}(t)$ calculated
    from $2$ sources by ICm with $64$-decimal digital precision,
    denoted as ${\rm ICm}64$, and Cm with $64$($100$)-decimal digital
    precision, denoted as ${\rm Cm}64({\rm Cm}100)$, on a single
    configuration are compared.  Correlation functions from ${\rm
      Cm}100$ agree with those from ${\rm ICm}64$, however for the
    same precision, the ICm gives more accurate result than Cm.
      For $C_{2 \pi}(t)$, Cm64 fails because of the numerical inaccuracy.}
  \label{fig:method_3_3o}
\end{figure}

Inverting this matrix does not suffer from the numerical instabilities
seen in the VMm, however as $12N$ becomes large, even computing the
inverse of these $3 \times 3$ matrices requires high precision.
Fig.~\ref{fig:method_3_3o} shows a comparison of the correlation
functions computed from $2$ sources by applying the Combination method
and Improved Combination method to be discussed below. For the
Combination method at $64$ digit precision, $C_{1\pi}(t)$,
$C_{2\pi}(t)$ and $C_{3\pi}(t)$ show signs of numerical break down at
earlier time slices, which goes away at higher precision ($100$
digit), indicating that even calculating the inverse of the $3 \times
3$ matrix needs high precision to get correct results.

As constructed, this method is only applicable to a $2^n$ source
problem.  In order to solve problems having arbitrary number of
sources, we extended this to an Improved Combination method in the
next section.

\subsection{Improved Combination method (ICm)}

\begin{figure}
  \includegraphics[width=14cm]{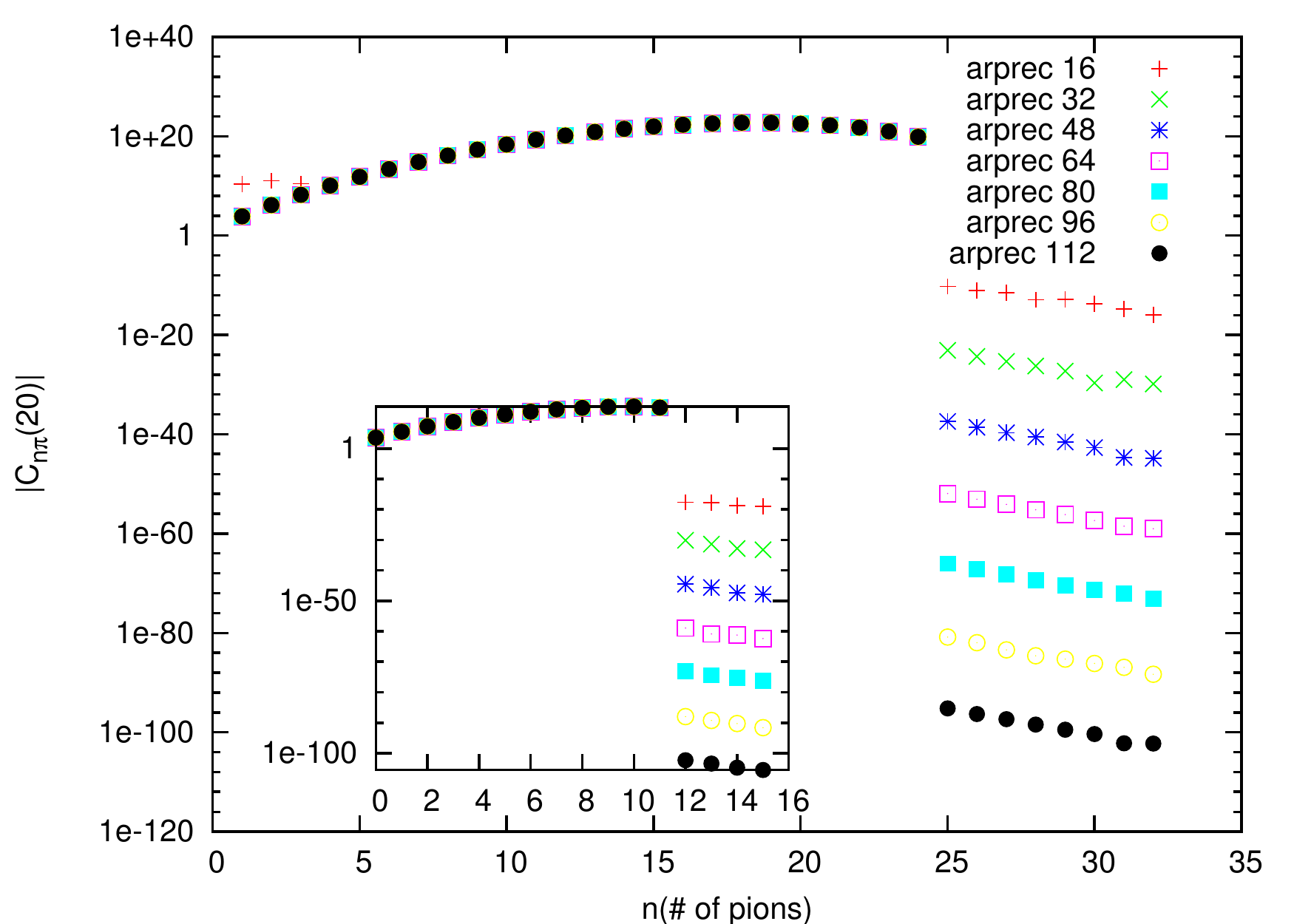}
  \caption{Correlation functions on a single configuration at $t=20$
    from $2$ sources computed with the Improved Combination method
    using the arprec library~\cite{arprec_cite} at various precisions:
    `arprec $X$' denotes that the calculation is done with $X$-decimal
    digit precision.  The $C_{n\pi}(20)$ for $n=1,2,\ldots 24$ all
    agree for the different precision calculations just as they
    should, except for the calculation from $16$-digit precision.
    However $C_{n\pi}(20)$ for $n = 25,26,\ldots, 32$ are all machine
    zero at each precision.  The disagreement of $16$-digit precision
    indicates higher precision is needed. A similar comparison is
    shown for the single source correlation functions in the insert.
  }
  \label{fig:method_3_precision}
\end{figure}

As there are $12N$ terms in the expansion of $\det[1+\lambda A]$, the
Combination method does not allow us to determine functions depending
on less than $3$ $C_{k\pi}$'s.  A similar problem appears in the
application of the FFT.  In order to use FFT, $2^n$ data points are
required. If the number of points in a series is not equal to $2^n$,
points with value zero must be appended to the original series to
produce a series of length $2^n$.  Similarly, we can append additional
$C_{k\pi}$'s to the expansion of $\det[1+\lambda A]$, as:
\begin{eqnarray}
  \det[1+\lambda A] = 1 + \lambda C_{1\pi} + \lambda^2 C_{2\pi} + \ldots +
  \lambda^{12N} C_{12N\pi}
  + \lambda^{12N+1} C_{(12N+1)\pi}+...+\lambda^{2^m} C_{2^m \pi}
\end{eqnarray}
where $C_{p \pi}=0$ for all $p>12N$.  The power $m$ is chosen such
that $2^{m-1} < 12N <2^m$.  With this new arrangement, exactly the
same prescription discussed for the Combination Method can be applied,
but in the last step the $D^{(\tilde n)}_k(\lambda)$ individually depends only
on a single correlation function.

A significant advantage of this method compared with the Cm is that no
matrix inversion is required, so it is consequently less demanding in
numerical precision, see Fig.~\ref{fig:method_3_3o}, and in addition,
problems with arbitrary numbers of sources can be solved with this
method. Correlation functions appended to the series are solved for
simultaneously with the other $C_{k\pi}$'s, providing a numerical
check of the validity of this method. In
Fig.~\ref{fig:method_3_precision}, correlation functions calculated
from $1$-source and $2$-sources on a single configuration are shown
for different precision (we use the ``arprec"
library~\cite{arprec_cite} to perform arbitrary precision
calculations). As expected, all $C_{p\pi}$'s for $p>12N$ are indeed
numerically equivalent to zero, decreasing exponentially as the the
numerical precision is increased.  Since this method is more
numerically stable than the Combination method, and can also solve
problems of arbitrary number of sources, it is used in our further
studies.

\subsection{Generalization to $2$ species from $N$ sources}

The methods discussed above can easily be generalized to two species
by studying properties of the expansion of $\det[1+\lambda_1
A+\lambda_2 B]$, where $A$ and $B$ are uncontracted correlation
functions of two distinct species, for example $\pi^+$ and $\rho^+$.
We can write
\begin{eqnarray}
  \det[1+\lambda_1 A + \lambda_2 B] = 1 + \lambda_2^0 T_0 + \lambda_2^1 T_1+\ldots+\lambda_2^k T_k+\ldots,
\end{eqnarray}
where
\begin{eqnarray}
  T_k(\lambda_1) =\lambda_1^0 C_{0A,kB}+ {k+1 \choose k}\lambda_1 C_{1A,kB} 
  + \ldots +{M \choose k}\lambda_1^{M-k} C_{(M-k)A,kB},
\end{eqnarray}
where $M=12N$ is the dimension of the matrices $A$ and $B$, and the
correlation functions, $C_{mA,nB}$, are complicated combinations of
correlation functions of a system having $m$-$A$'s and $n$-$B$'s
distributed among different sources in all possible ways.

The $T_j(\lambda_1)$, for $j = 0,1,\ldots, M$ for one $\lambda_1$ can
be separated out by applying the methods discussed above with
different choices of $\lambda_2$'s, and then by applying the method
again for different choices $\lambda_1$'s for all $T_j(\lambda_1)$'s,
the $C_{mA,nB}$'s can be separated out.  This can be further
generalized to correlators of arbitrary number of species as
necessary.

\subsection{Performance of different methods}

\begin{figure}
  \includegraphics[width=8cm]{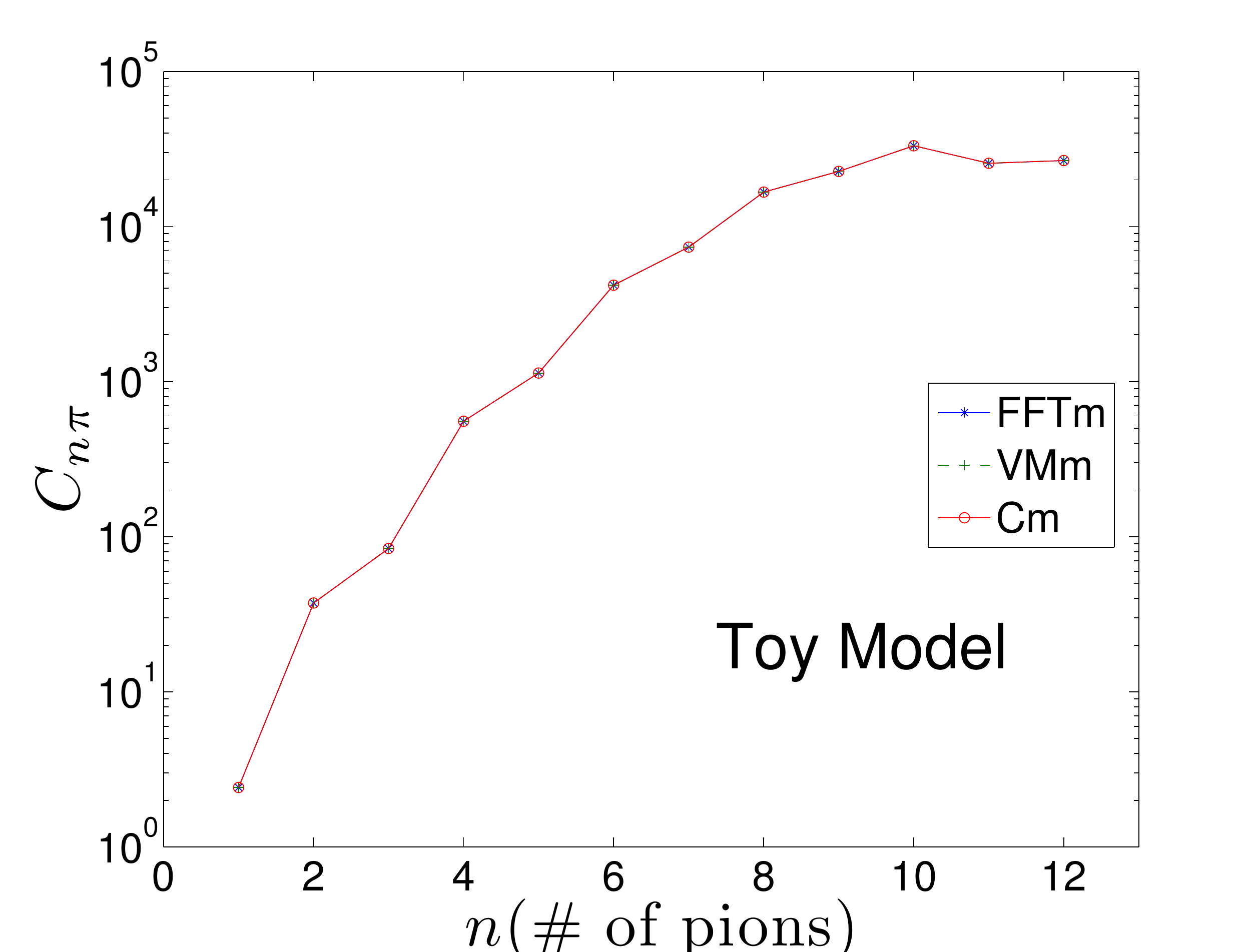}
  \includegraphics[width=8cm]{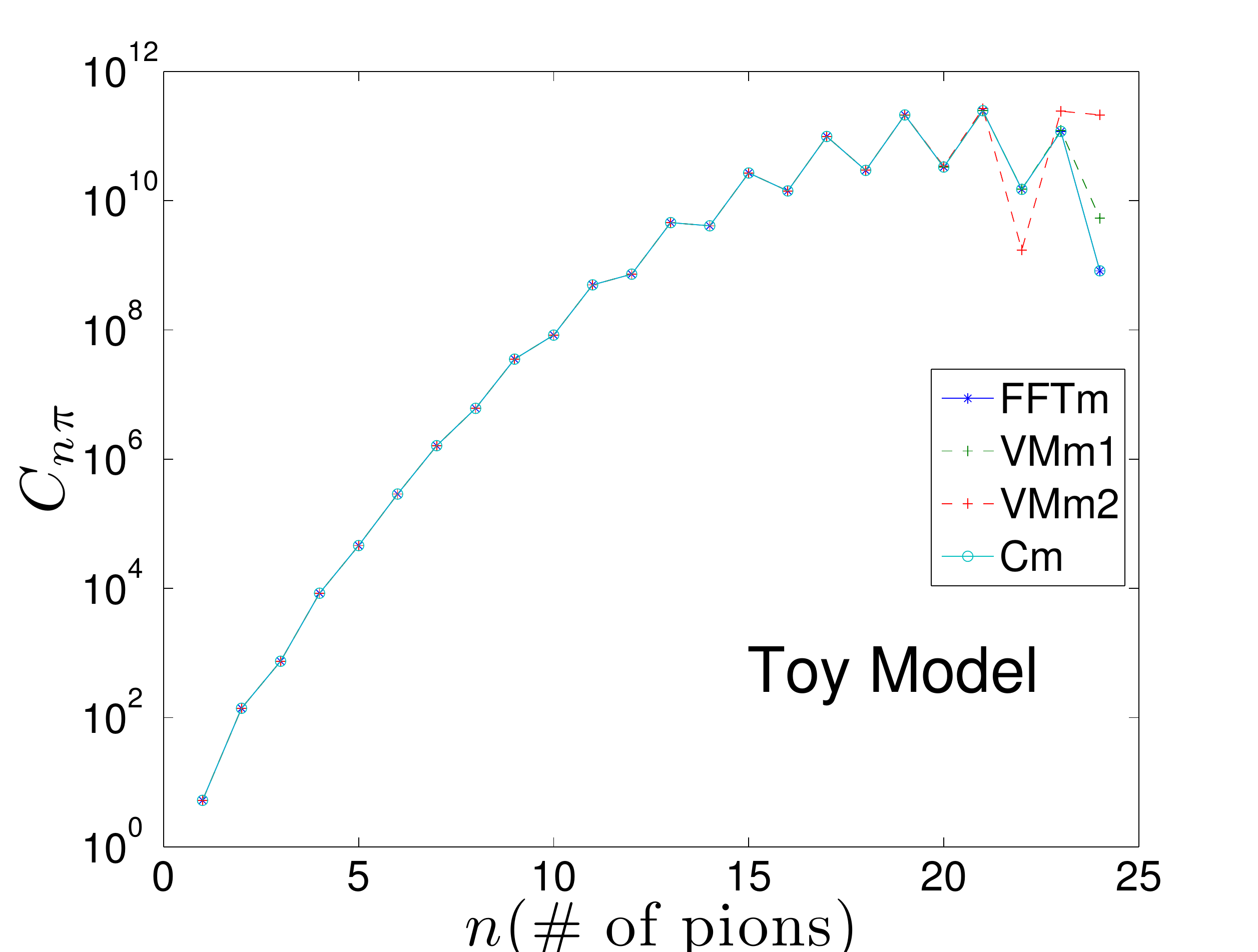} \\
  \includegraphics[width=8cm]{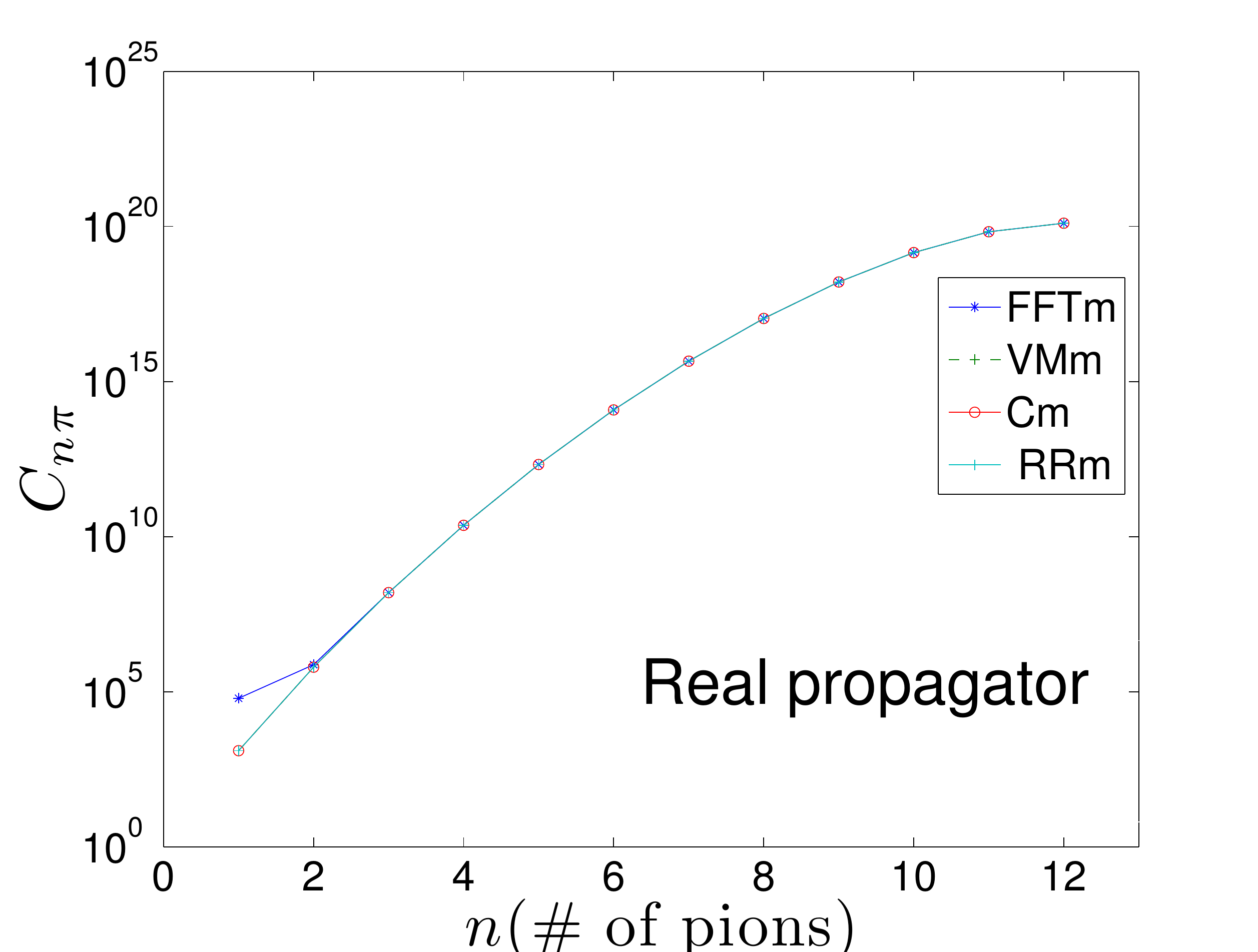}
  \includegraphics[width=8cm]{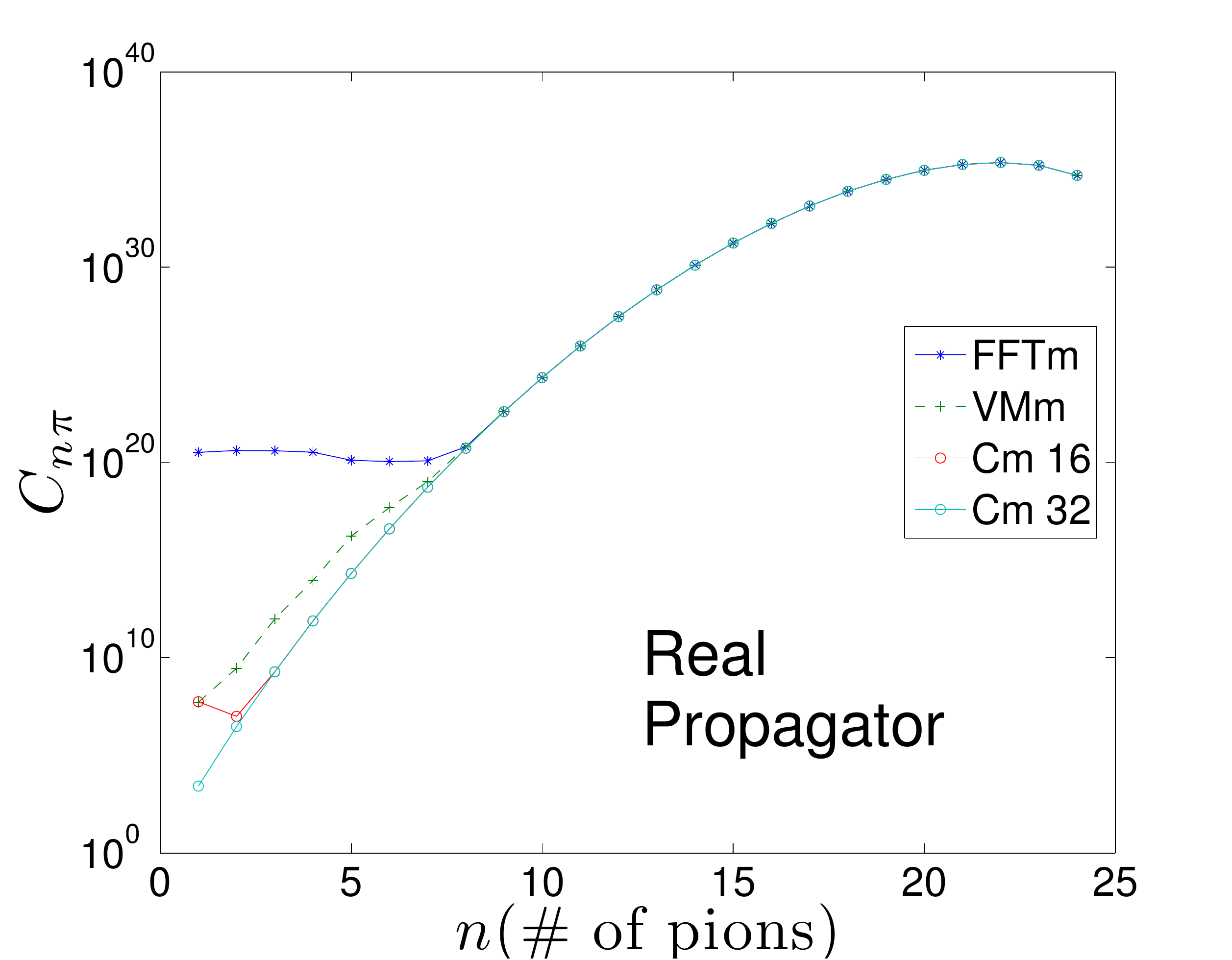}
  \caption{ The left panels compare $1$-source calculations from
    the VMm, FFTm and Cm, and the right panels compare $C_{n\pi}$
    calculated from $2$ sources by the three methods.  
    The real propagator is taken from one time slice, $t=20$.
    The Recursion
    Relation method (RRm) \protect{\cite{Detmold:Savage}} is also
    compared with other methods in the lower left plot as a check on
    the validity of the Cm.  For the $2$-source calculation in the toy
    model (top right) with VMm, two different sets of $\lambda_n$s
    have been used, denoted as VMm1 and VMm2.  For VMm applied to the
    real propagator calculations, only one choice of $\lambda$'s is
    shown. ``Cm 16 (32)'' denotes that calculation is done using Cm
    with 16(32) decimal digit precision.  }
  \label{fig:vmm_fftm}
\end{figure}

In order to test the accuracy of different methods at a fixed
precision, we compared correlation functions calculated from the VMm
(implemented in MATLAB), the FFTm (implemented in MATLAB), and the Cm
(implemented in C++ using the ``arprec'' high precision
library~\cite{arprec_cite}).  We first considered a toy model with
matrix elements \mbox{$A_{n,m}=\sin((m-1)(n-1)+2)+i\cos(2(n-1))$} for
$1$ and $2$ sources in the top half of Fig.~\ref{fig:vmm_fftm}.  For
this test, the $\lambda$'s used in the VMm and Cm are randomly
chosen between $-0.25$ and $0.25$, however $C_{n\pi}(t)$ should be
independent of these choices.  Results from VMm on $1$-source agree
with those from the FFTm and Cm for any set of $\lambda$.  However
for $2$ sources, the FFTm and the Cm give the same results, but the
VMm gives inconsistent results and changes with different choices of
$\lambda$'s, signaling a breakdown of the VMm and the requirement of
higher precision. Similar tests have also been performed with the
matrix elements $A_{n,m}$ extracted from real quark propagators and
the results are shown in the lower half of Fig.~\ref{fig:vmm_fftm}.
In this test, the Recursion Relation method (RRm) has also been used
to compute the $C_{n\pi}$'s in order to validate the new methods. For
the $N > 1$-source calculation no direct comparison with the RRm can be
made, since the $C_{{\overline n}\pi}$ computed from the new methods
are complicated combinations of all $C_{n_1,\ldots,n_N}$'s with
$\sum_{i=1}^N n_i = {\overline n}$.  We verified however that the
energies extracted for these correlators with either method, RRm and
Cm, are in agreement.  In contrast to the toy model, for the real
$A_{n,m}$, the VMm gives more accurate results than the FFTm. However
both tests show that the Cm gives the most accurate results for a
fixed precision.  Tests with real propagators on 2-source shows a
break down of Cm on $C_{1\pi}$ and $C_{2\pi}$, however this breakdown
can easily be corrected by working at higher precision.

\begin{figure}
  \includegraphics[width =  15cm]{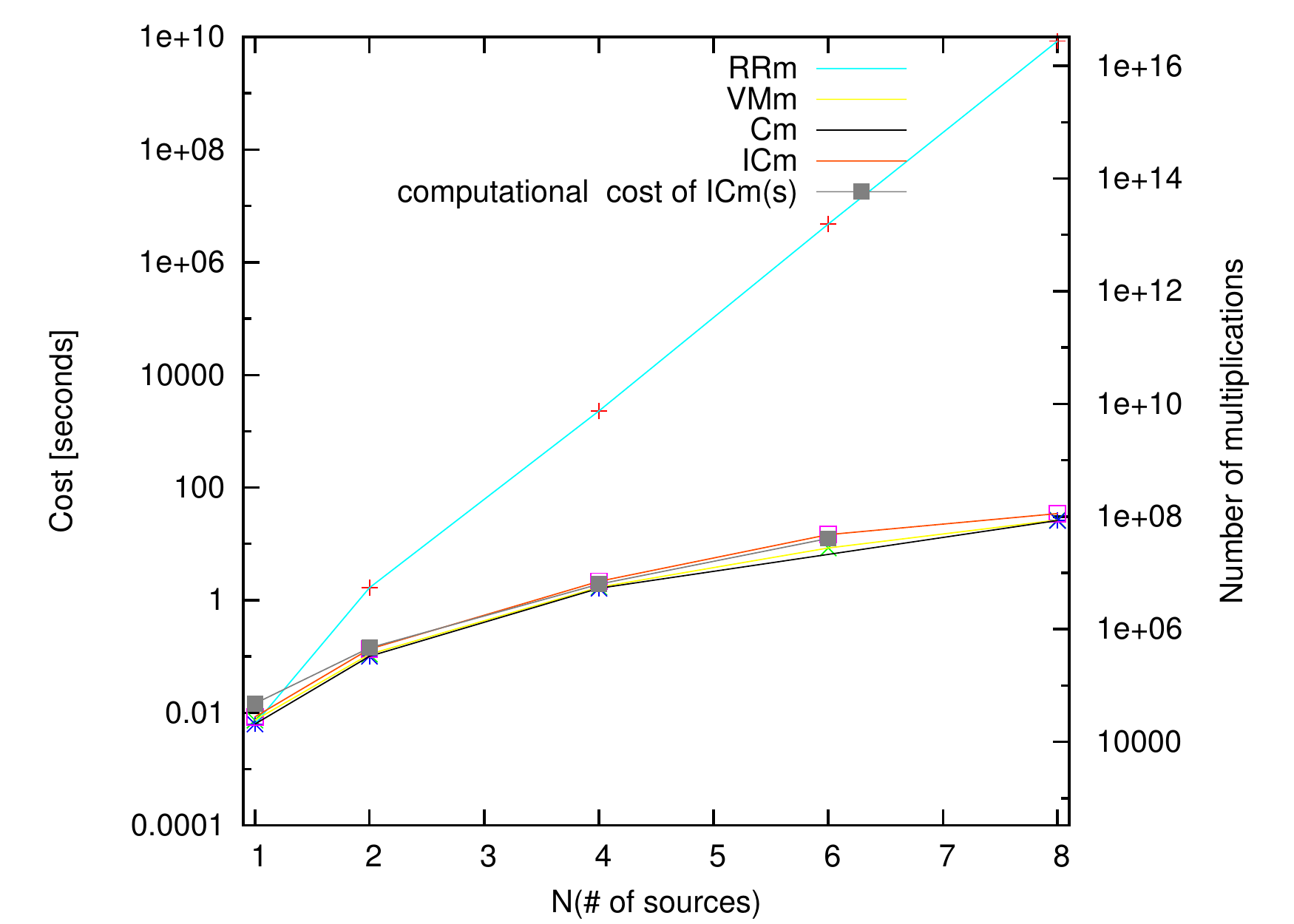}
  \caption{ Comparison of the number of multiplications required for
    each method (RH axis), and the corresponding computation time of
    $C_{n\pi}(t)$ for $n = 1,2,\ldots 12N$ on a single time slice,
    corresponding to one application of the specified contraction
    method in seconds using a single $2.4\ {\rm GHz}$ Xeon core (LH
    axis).  The computational cost of the ICm is taken from the actual
    running time, and it is used to normalize the time scale so that the 
    projected running time of other methods can be read out from the LH axis. }
  \label{fig:computational_time_compare}
\end{figure}

\begin{table}[ht]
  \caption{Scaling of different methods in terms of number of multiplications
    for an $N$ source calculation. }
  \centering
  \begin{tabular}{c|c}
    \hline\hline
    & scaling \\
    \hline
    RRm &  $12^4 N^4 \exp(2.8(N-1))$ \\
    \hline
    VMm &  $(12N+2)(12N)^3$ \\
    \hline
    Cm &  $3\cdot 2^{\log2(4N)}(12N)^3$ \\
    \hline
    ICm & $2^{{\rm floor}(\log_2(12N))+2}(12N)^3$ \\
    \hline
  \end{tabular}
  \label{table:numerical_cost}
\end{table}

The main purpose of constructing these new methods is to expedite the
contractions required in computing correlation functions for systems
comprised large number of mesons.  The numerical scaling of the
Recursion Relation method, Vandermonde Matrix method, Combination
method and Improved Combination method (the FFT method costs the same
amount of time as the ICm if $f_0$, $\tau$ and $T$ are chosen
appropriately) are compared in Table~\ref{table:numerical_cost}.  For
each method, we determine how many multiplications are required. From
Ref.~\cite{Detmold:Savage}, the computational cost of the recursion
relation method is proportional to \mbox{$12^4 N^4 \exp(2.8(N-1))$},
where N is the number of sources.  The VMm requires a calculation of
$12N$ determinants, one inversion of $12N \times 12N$ matrix and the
multiplication of a $12N \times 12N$ matrix and $12N\times 1$ vector.
The dominant contribution to the computational cost of the other two
methods comes from calculating a large number of determinants. For the
Improved Combination method, a step-$n$ calculation requires the
computation of $2^n$ determinants, while the Combination method
requires $3\cdot 2^n$ for a step-$n$ calculation. To solve an
$N$-source problem, the Combination method requires $\log2(4N)$ steps
for $N=2^m$, where m is an integer, and the Improved Combination
method requires ${\rm floor}(\log_2(12N))+2$ steps.  Taking account of
all the determinant calculations that are needed, and the
computational cost of each determinant ($\sim (12N)^3$ using LU
decomposition), the
numerical cost of each method is tabulated in
Table.~\ref{table:numerical_cost}, and compared in
Fig.~\ref{fig:computational_time_compare}.  Although the recursion
relation method significantly reduces the cost of contractions over
the original $(12N!)^2$ scaling, the computational cost of the
recursion relation method is much larger than other methods, all of
which scale similarly.  Using the ICm, we now turn to our numerical
investigations of systems of large number of mesons.

\section{Lattice details}
\label{sec:lattice_detail}

The calculations in this paper are performed on ensembles of
anisotropic gauge field configurations with clover-improved
fermions~\cite{shei_wohl_83} that have been generated by the Hadron
Spectrum Collaboration and the Nuclear Physics with Lattice QCD
collaboration.  The gauge action is a tree-level tadpole-improved
Symanzik-improved action,
and the fermion action~\cite{Okamoto:2001jb,Chen:2000ej} is a
$n_f=2+1$ anisotropic clover action~\cite{Sheik_Woh:85} with two
levels of stout smearing~\cite{colin:mike2004} with weight $\rho =
0.14$ only in spatial directions (see~\cite{heuy:david2008} for more details).
 In order to preserve the
ultra-locality of the action in the temporal direction, no smearing is
performed in that direction.  Furthermore, the tree-level
tadpole-improved Symanzik gauge action without a $1\times2$ rectangle
in the time direction is used.

Four ensembles of gauge fields are used in this paper with volumes
$L^3 \times T$ of $16^3\times 128$, $20^3 \times 128$, $24^3 \times
128$ and $20^3 \times 256$, and with a renormalized anisotropy
$\xi=a_s/a_t=3.5$, where $a_s$ ($a_t$) is the spatial (temporal)
lattice spacing.  The lattice spacing is the same for each ensemble
and is $a_s = 0.1227\pm 0.0008\ {\rm fm}$~\cite{heuy:david2008}, which
gives spatial extents $L \sim 2.0, 2.5, 3.0\ {\rm fm}$ for $L = 16,
20, 24$ respectively.  The same input light quark mass $a_t m_l =
-0.0840$ and input strange quark mass $a_t m_s = -0.0743$ are used in
generating each ensemble, giving a pion mass of $m_{\pi} \sim 390\
{\rm MeV}$ and a kaon mass of $m_{K} \sim 540\ {\rm MeV}$.  The
quantities $m_{\pi}L$ and $m_{\pi}T$, which determine the impact of
the finite volume and temporal extent, are $m_{\pi}L \sim
3.86,4.82,5.79$ for $L = 16,20,24$ lattices and $m_{\pi}T \sim 8.82,
17.64$ for $T=128,256$, respectively. Details of the four ensembles
are summarized in Table~\ref{tab:gauge_ensemble_sum}.

\begin{table}
  \caption{Details of the four gauge ensembles with the same lattice 
    space $a = 0.1227 \pm 0.0008\ {\rm fm}$ used in this paper. 
    $N_{\rm cfg}$ denotes the number of configurations used in the 
    current calculation.
    In the last two columns, $N_{\rm src}$
    is the number of source times used on each configuration and
    $N_{\rm mom}$ is the number of momentum sources used for each
    source time.
  }
  \begin{tabular}{c|c|c|c|c|c|c|c}
    \hline\hline
    $\ $& $L^3 \times T\ (a^{-1})$& $L\ ({\rm fm})$&$m_{\pi}
    L$&$m_{\pi} T$& $N_{\rm cfg}$& $N_{\rm src}$ & $N_{\rm mom}$\\
    \hline
    B1 & $16^3 \times 128$ & $2.0$ & $3.9$ & $8.8$ & $180$ & $8$ &
    33\\
    \hline
    B2 & $20^3 \times 128$ & $2.5$ & $4.8$ & $8.8$ & $51$ & $8$ & 19\\
    \hline
    B3 & $24^3 \times 128$ & $3.0$ & $5.8$ & $8.8$ & $98$ & $8$ & 19\\
    \hline
    B4 & $20^3 \times 256$ & $2.5$ & $4.8$ & $17.6$ & $147$ & $16$
    & 7\\
    \hline
  \end{tabular}
  \label{tab:gauge_ensemble_sum}
\end{table}

In our work, a momentum space representation of the contractions is
used and Coulomb gauge fixed propagators in time-momentum space, which
we refer to as ``colorwave propagators'',
$S_{u/d}({\bp},\tau;{\bpp},0)$, are calculated on each
configuration\footnote{As we compute gauge invariant quantities, our
  results are independent of the gauge fixing procedure.}.  The
colorwave propagator is defined as
\begin{eqnarray}
  S_{u/d}({\bp},t;{\bpp},0) = \sum_{\bf y}e^{-i{\bf p}{\bf x}}S_{u/d}({\bf x}, t; {\bf p}^{\prime},0),
\end{eqnarray}
where
\begin{eqnarray}
  S_{u/d}({\bf x},t; {\bf p}^{\prime},0) = \sum_{\bf y}e^{i{\bf p}^{\prime}{\bf y}}S_{u/d}({\bf x}, t; {\bf y}, 0),  \nonumber
\end{eqnarray}
is a solution of the Dirac equation:
\begin{eqnarray}
  \sum_{{\bf x},t}D({\bf y}, {\tilde t}; {\bf x},t)S_{u/d}({\bf x},t; {\bf p}^{\prime},0)
 = e^{i{\bf p}^{\prime}{\bf y}}\delta_{\tilde t, 0}. \nonumber
\end{eqnarray}
The colorwave propagator, $S_{u/d}({\bp},\tau;{\bpp},0)$ describes a
quark propagating from the source $({\bpp}, 0)$ to the sink $({\bp},
\tau)$ in time-momentum space.  The colorwave propagator and the
position space propagator are related by a $3$ dimensional Fourier
transformation as follows:
\begin{eqnarray}
  \label{def:colorwave_p}
  S_{u/d}({\bp},t;{\bpp},0)&=& \sum_{\bx \by}\ \ e^{-i\bp\bx}\ \ e^{i\bpp \by}\ \ S_{u/d}({\bx},t;{\by},0),
\end{eqnarray}
and the conjugate of a propagator is 
$S^\dagger(-p, t; -p',0)=\gamma_5 S(p', 0;p,t) \gamma_5$.
The $\gamma_5$ hermiticity of the colorwave propagator follows from the
$\gamma_5$ hermiticity of its counterpart in position space.

A correlation function of one pion with momentum ${\bp}_f$ can be
constructed by projecting both sink and source to the same momentum
${\bp}_f$ as:
\begin{eqnarray}
   C_{1\pi}({\bp}_f,t)
  &=& \left \langle \sum_{\bx,\bxp}  e^{-i\left({\bp}_1\bx-{\bp}_2\bxp\right)}
    {\overline d}({\bxp},t)\gamma_5\ u({\bx},t)\ \sum_{\by \byp}\ e^{i\bp\by}
    e^{-i({\bp} - {\bp}_f            
      ) \byp}\ {\overline u}({\by},0)\gamma_5 d({\byp},0) \right \rangle\nonumber \\
  &=& \sum_{\bx,\bxp,\by,\byp}\ \left \langle  
    e^{-i{\bp}_1\bx} e^{i\bp\by}\ \gamma_5
    S_u({\bx},t;{\by},0)\ \gamma_5 e^{i{\bp}_2\bxp}\ \ e^{-i
      ({\bp}-{\bp}_f)\byp}\ 
    (\gamma_5 S_d^{\dagger}({\bxp},t;{\byp},0) \gamma_5 )
  \right \rangle\nonumber\\
  &=& \left \langle \sum_{\bx,\by}\ \gamma_5 (e^{-i{\bp}_1\bx}e^{i\bp\by} 
    S_u({\bx},t;{\by},0))
    \sum_{\bxp , \byp} e^{-i({\bp}-{\bp}_f)\byp} e^{i{\bp}_2\bxp} \gamma_5
    (\gamma_5  S_d^{\dagger}({\bxp},t;{\byp},0) \gamma_5) \right \rangle\nonumber\nonumber\\  
  &=& \left \langle \gamma_5 S_u({\bp}_1,t;{\bp},0)\cdot \gamma_5
    (\gamma_5 S_d^{\dagger}(-{\bp}_2,t;{\bp}_f-{\bp},0) \gamma_5)\right \rangle,       
\end{eqnarray}
where ${\bp}_1 - {\bp}_2 = {\bp}_f$.  Each choice of $\{ {\bp}_1,
{\bp}_2\}$ and $\bp$ satisfying momentum conservation is 
a separate correlation function with distinct 
creation and annihilation interpolating fields,
and we have suppressed the dependence of $C_{1\pi}$ on ${\bp}_1$,
${\bp}_2$ and $\bp$.  
During the calculation, we held ${\bp}_1$, ${\bp}_2$ and ${\bp}_f$ fixed and
summed over all ${\bp}$'s for which we have computed colorwave propagators
(see Table.~\ref{tab:gauge_ensemble_sum})
in order to get more statistics.
In the second step, the definition of propagator
$S_{u/d}({\bxp},t; {\by},0)$ and the $\gamma_5$ hermiticity of the
propagator is used.  The definition of the colorwave propagator,
Eq.~(\ref{def:colorwave_p}), is applied in the last step.
 
The correlation functions of a system having $n\ \pi^+$'s in a single
source, with total momentum ${\bf P}_f = n\ {\bp}_f$ can be
constructed similarly:
\begin{eqnarray}
  C_{n\pi}(t, {\bf P}_f) &=& \left \langle \left( \sum_{\bx,\bxp} e^{-i({\bp}_1\bx-{\bp}_2\bxp)}
      {\overline d}({\bxp},t)\gamma_5 u({\bx},t)\right)^n \right. \nonumber \\
    & & \left. \cdot \left(\sum_{\by, {\by}^{\prime}} e^{i{\bp}\by}e^{-i({\bp}-{\bp}_f){\by}^{\prime}}
      {\overline u}({\by},0)\gamma_5 d({\by}^{\prime},0)\right)^n \right \rangle,
  \label{eq:C_momentum_space}
\end{eqnarray}
where the dependence of $C_{n\pi}$ on ${\bp}_1$, ${\bp}_2$ and $\bp$ has also
been suppressed in Eq.~({\ref{eq:C_momentum_space}}).

Because of the Pauli exclusion principle, systems constructed from a
single source in momentum space can only reach a maximum of $12\
\pi^+$'s. In order to put more pions into a system, additional sources
are required.  Correlation functions of a $N$-source system having
${\overline n}=\sum_{i=1}^N n_i \pi^+$'s with total momentum ${\bf
  P}_f=\sum_{i=1}^{\overline n}{\bp}_{f_i}$ are given by
\begin{eqnarray}
    C_{n_1\pi,\ldots, n_N\pi}(t,{\bf P}_f)
  &=&  \left \langle \prod_{i=1}^N\left( \sum_{{\bx}_i,{\bxp}_i} 
      e^{-i({\bp}^i_1 {\bx}_i-{\bp}^i_2 {\bxp}_i)}
      {\overline d}({\bxp}_i,t)\gamma_5 u({\bx}_i,t)\right)^{n_i} \right. \nonumber \\
  &\times&\left. \prod_{j=1}^{\overline n}\left(\sum_{{\by}_j,{\by}_j^{\prime}}
     e^{i{\bp}_{j} {\by}_j}e^{-i({\bp}_j-{\bp}_{f_j}){\by}_{j}^{\prime}}{\overline u}
      ({\by}_j,0)\gamma_5 d({\by}_j^{\prime},0)\right) \right \rangle,
  \label{eq:C_m_sources_mom}
\end{eqnarray}
where $n_i$ is the number of pions in the $i^{th}$ source, and
momentum conservation \mbox{$\sum_{i=1}^N({\bp}^i_1 - {\bp}^i_2) =
  \sum_{j=1}^{\overline n}{\bp}_{f_j}$}, must be satisfied in order
for the correlation functions to be non-vanishing.  The contraction
methods discussed in the last section apply equally well in momentum
space and are used in our work. The elements of the counterpart of
uncontracted correlation functions defined in
Eq.~(\ref{eq:Pdef_descending}) are:
\begin{eqnarray}
  {\tilde A_{k,i}}\left(t\right) &=& \sum_{{\bp}}^{}
  S\left({\bp}^k_1,{\bp}\right)S^{\dagger}\left(-{\bp}^i_2,{\bp}_{f_i}-{\bp}\right), 
\end{eqnarray}
where $k,i$ label the source and sink, and the dependence on
${\bp}_1^k$, ${\bp}_2^i$, $\bp$ and ${\bf p}_{f_i}$ is suppressed.

For the $T=128$ ($256$) ensembles, $8$ ($16$) colorwave propagators
are generated on each configuration located $16$ time slices apart to
minimize correlations between propagators.
For ensembles $\{B1, B2, B3, B4\}$, $\{180, 51, 147, 98\}$
configurations and $\{33, 19, 19, 7\}$ momenta are used respectively.
In order to reduce contamination from thermal states, a temporal
extent of $T=256$ is desirable for systems of large numbers of
pions. On the B1 and B3 ensembles, the $A\pm P$ (antiperiodic
$\pm$ periodic propagator) method~\cite{rbc:2001, rbc:2006,
  chris:jack2007} is applied to effectively double the temporal
extent. The validity of this method is investigated by comparing
results from ensemble B4 ($20^3 \times 256$) and with those from
ensemble B2 ($20^3 \times 128$) with the $A\pm P$ method and it is
found to be sound at the precision we achieve for the systems under
consideration as discussed below.

\section{Ground state energies}
\label{sec:results}

\begin{figure}
  \mbox{\centerline{
      \includegraphics[width=5.4cm]{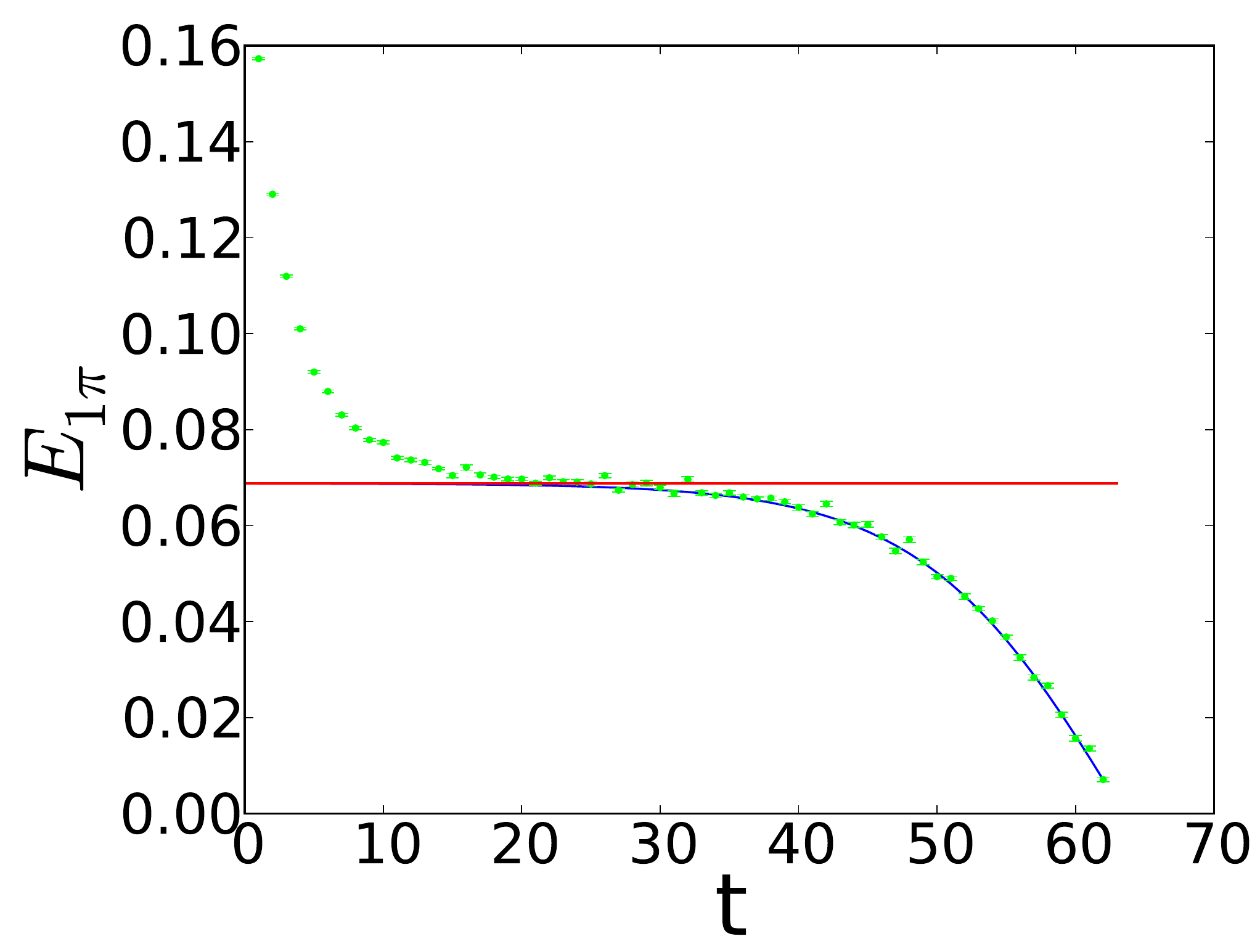}
      \includegraphics[width=5.4cm]{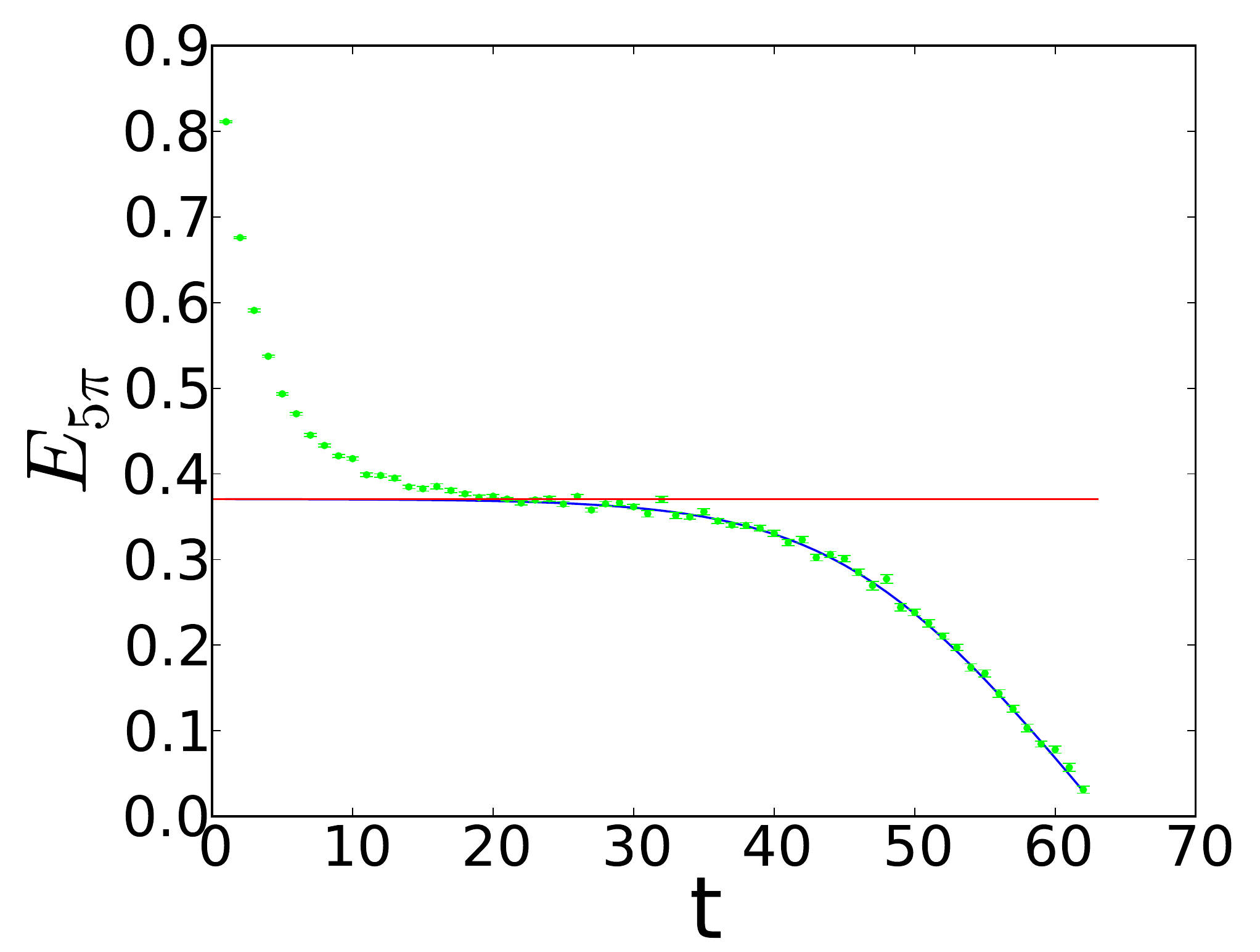}
      \includegraphics[width=5.4cm]{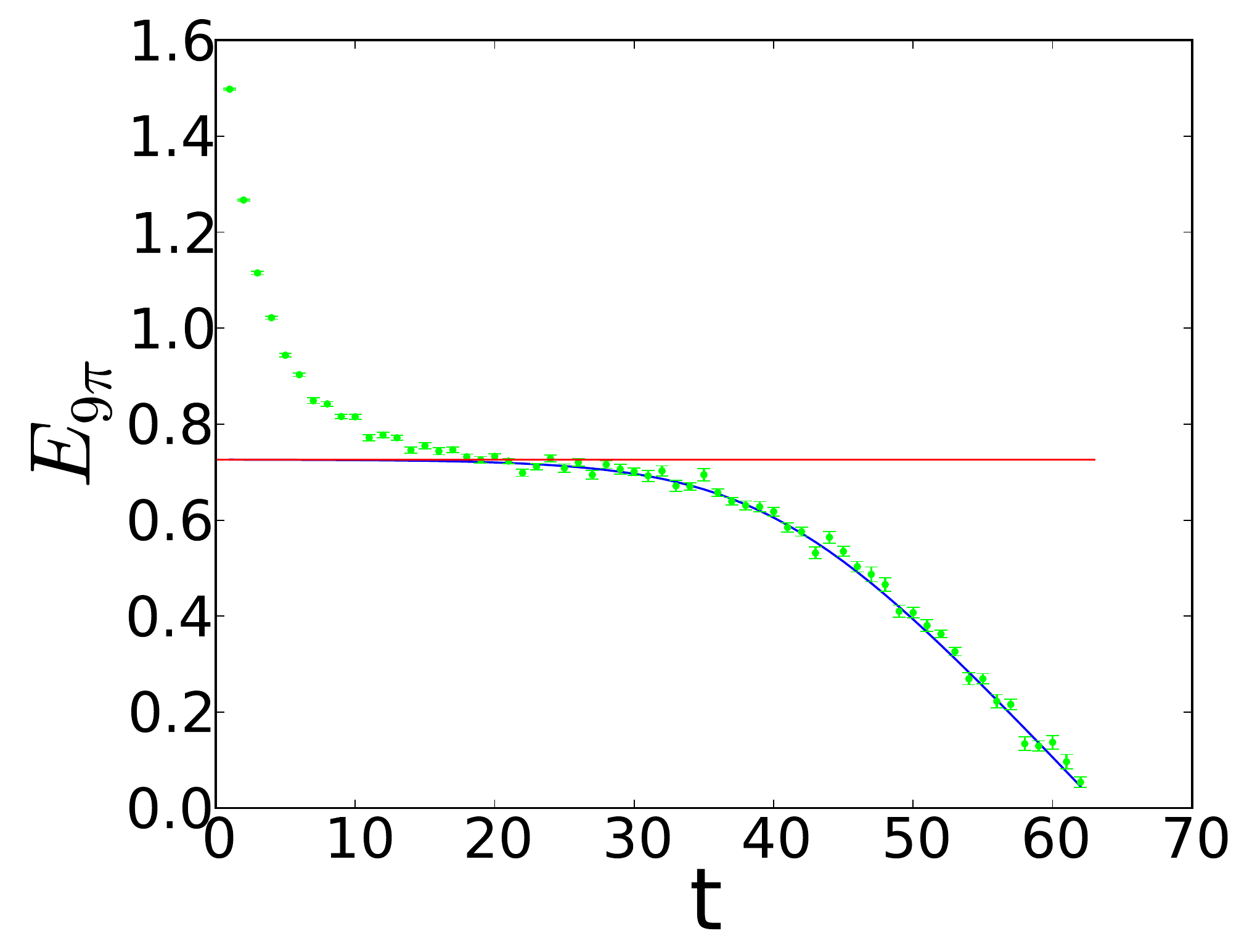}
    }} \mbox{\centerline{
      \includegraphics[width=5.4cm]{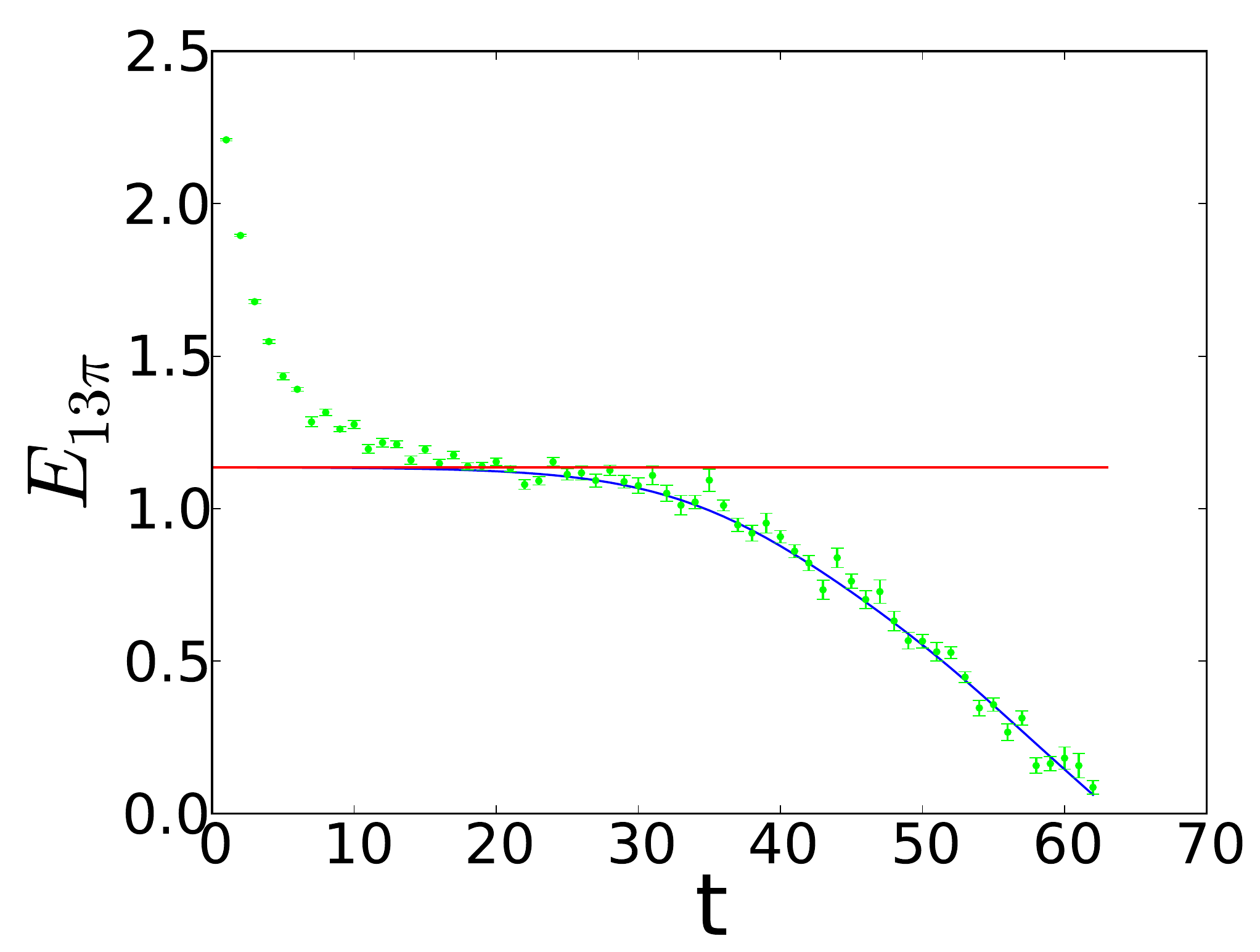}
      \includegraphics[width=5.4cm]{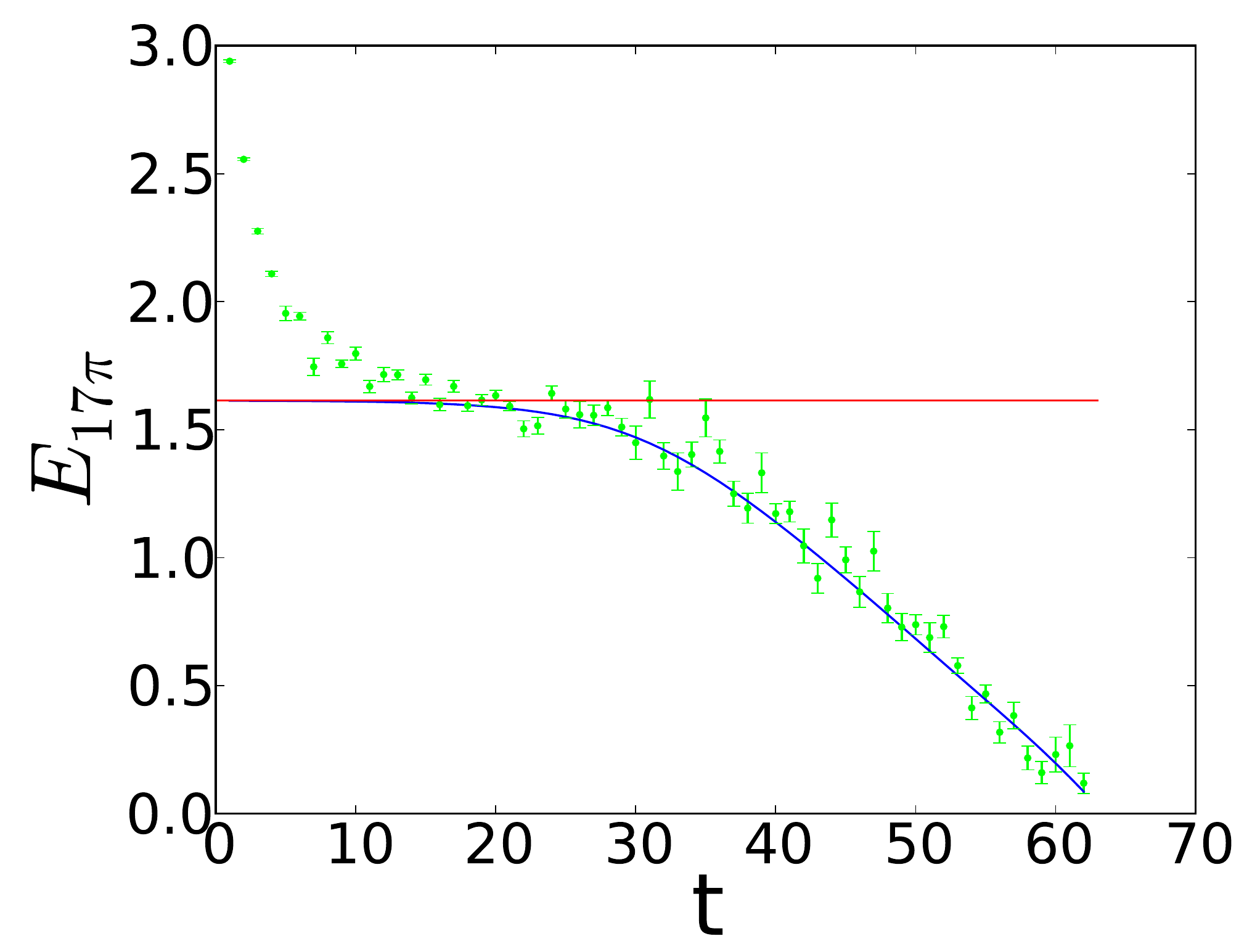}
      \includegraphics[width=5.4cm]{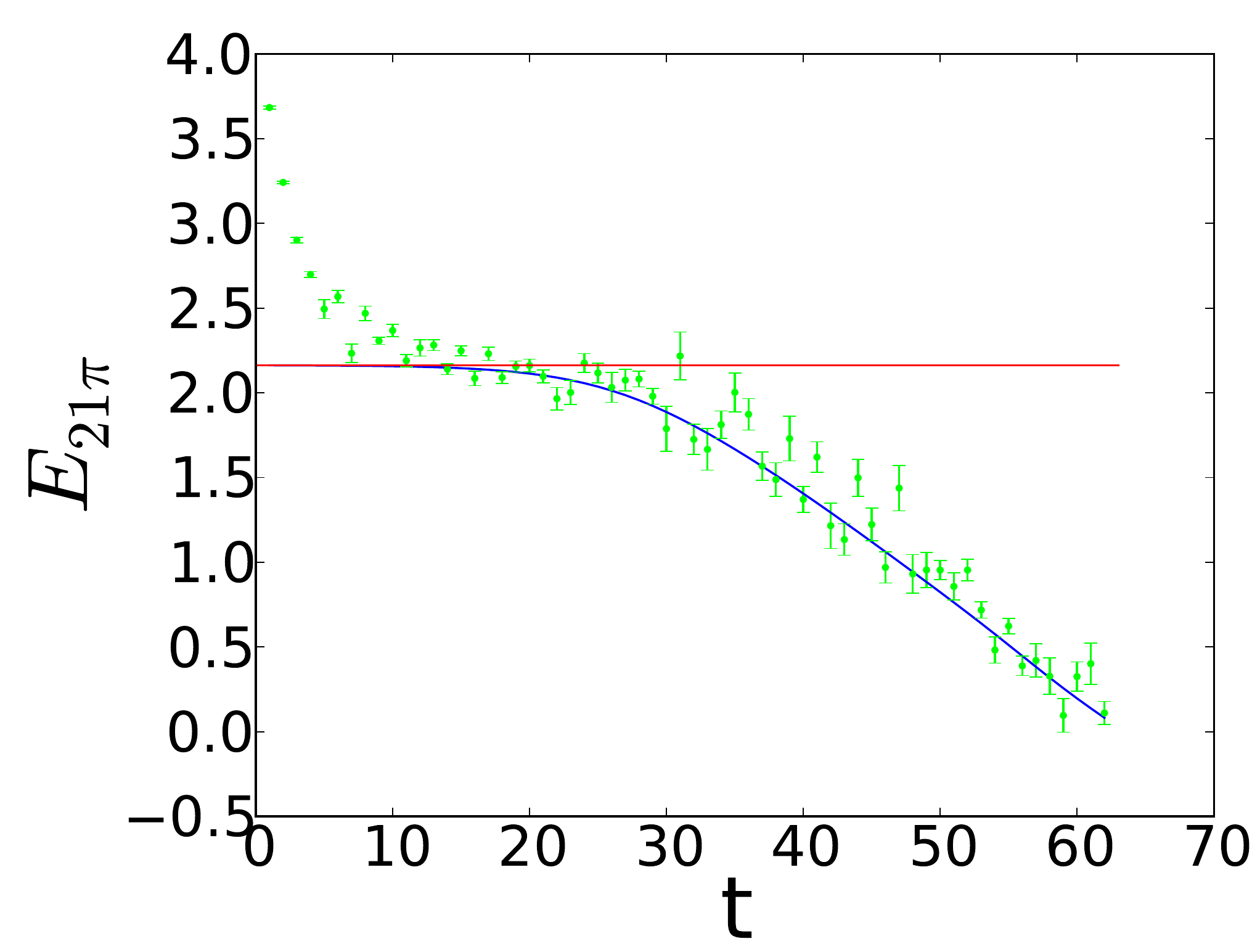}
    }}
  \caption{The green data is the effective mass calculated from the
    original data from ensemble B2, and blue line is reconstructed
    from the ground state energies extracted from the ensemble
    B4 as discussed in the main text.
    The red line is the fitted value of $E_{n\pi}$ extracted
    from the correlators of ensemble B4.}
  \label{fig:reconstruct_128_from_256}
\end{figure}

Previous studies of the energies and isospin chemical
potentials~\cite{Detmold_Brian:2011,Z_W_2011_proceeding} on ensemble
B2 showed that thermal states contribute significantly to
correlation functions and, even for $C_{12\pi}(t)$, the ground state
does not dominate in any region of Euclidean time.  The expected form
of correlation functions of an $n$-$\pi^+$ system with temporal extent
$T$ is~\cite{Detmold_Brian:2011}
\begin{eqnarray}
  \label{eqn:corr_expected}
  C_{n\pi}(t) &=& \sum_{m=0}^{\lfloor \frac{n}{2} \rfloor} 
     {n \choose m} A_m^n Z_m^n e^{-(E_{n-m}+E_{m})T/2}
  \cosh((E_{n-m}-E_{m}) (t-T/2))+\ldots, \ \ \ \ 
  \label{eq:c_all_thermal}
\end{eqnarray}
where $A_m^n = 1$ when $m = n/2$, otherwise $A_m^n=2$.
$E_{m}$ is the ground state energy of a $m$-$\pi^+$ system,
the $Z_m^n$ are the overlap factors for contribution with $m\ \pi$'s
propagating backward around the temporal boundary, and
the ellipsis denotes contributions from excited states.  The
ground state contribution comes from the $m=0$ term, and thermal states are
from the $m\ne 0$ terms in the sum, corresponding to contributions
where $m\ \pi^+$'s propagate backwards from the source to the sink
around the temporal boundary.  
For the $T=128$ B2 ensemble, effective
mass plots are shown in Fig.~\ref{fig:reconstruct_128_from_256} for
various $n$, and it is clear that correlation functions receive
significant contributions from thermal states. Their analysis requires
a fit including all thermal states, Eq.~({\ref{eq:c_all_thermal}}), in
order to extract the ground state energy.  Since the number of free
parameters in the fit grows with $n$, the systematic uncertainty of
$E_{n\pi}$ becomes large and we are unable to extract any accurate
information at large $n$.  In order to minimize contributions from
thermal states, a longer temporal extent is required.

\begin{figure}
  \includegraphics[width=5.4cm]{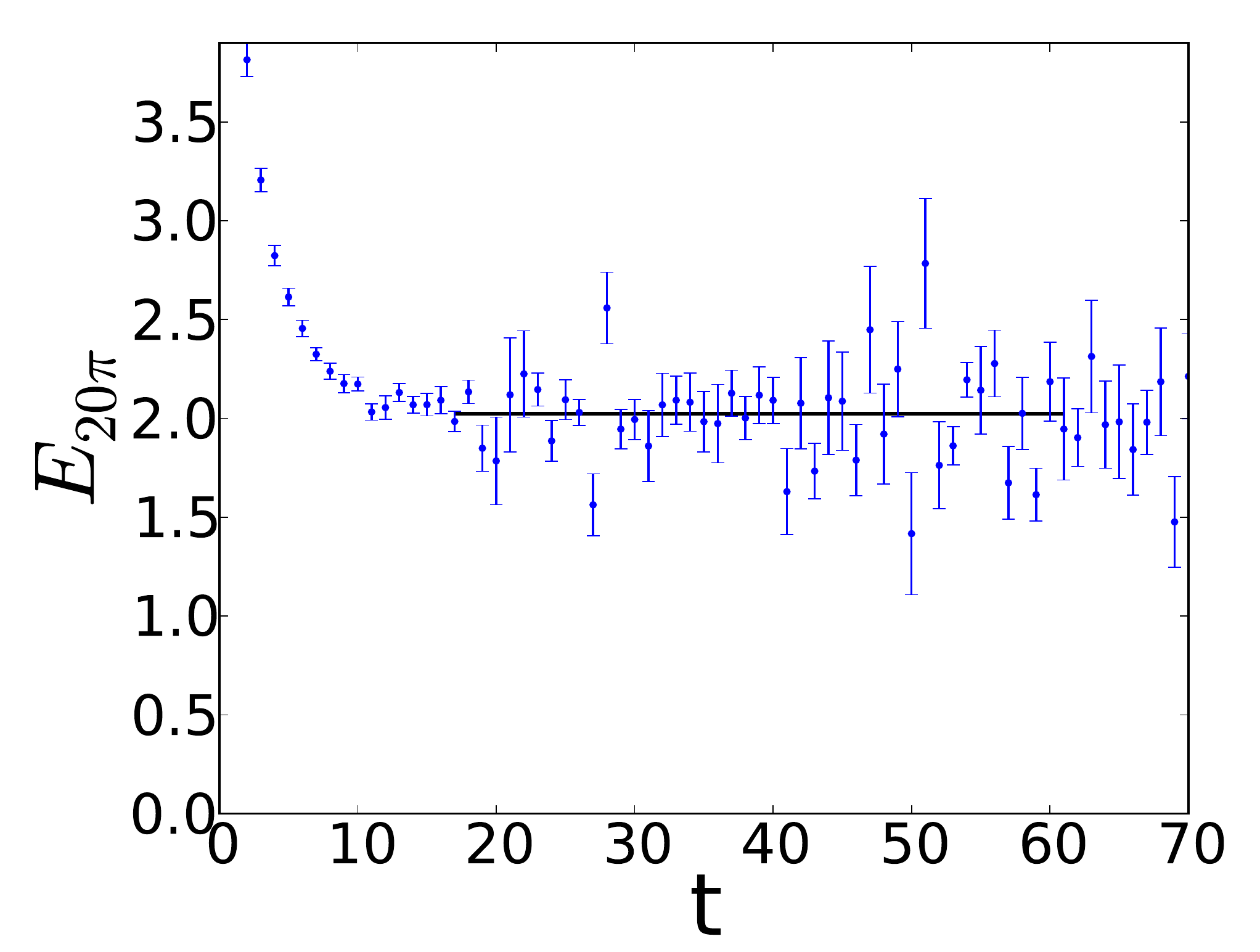}
  \includegraphics[width=5.4cm]{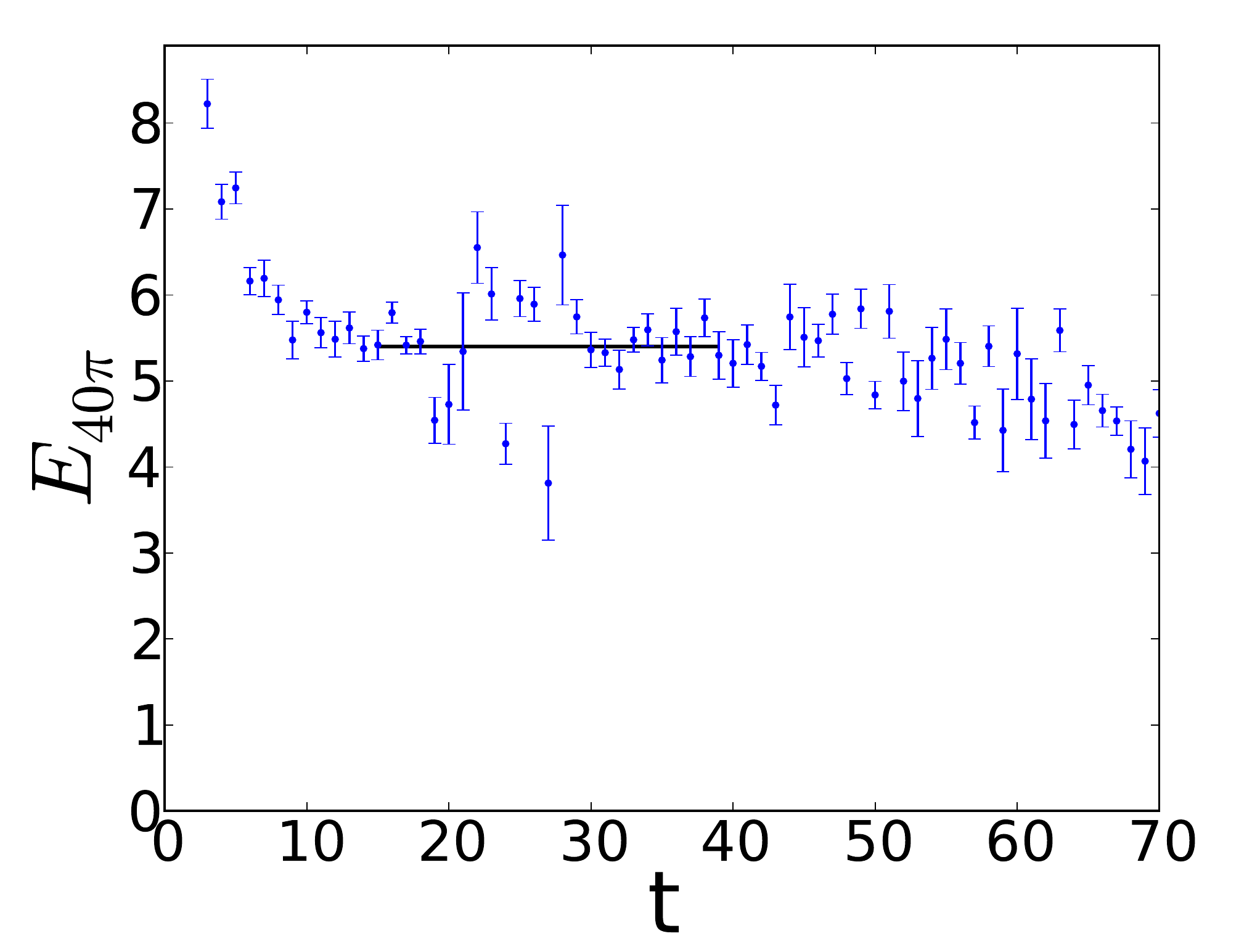}
  \includegraphics[width=5.4cm]{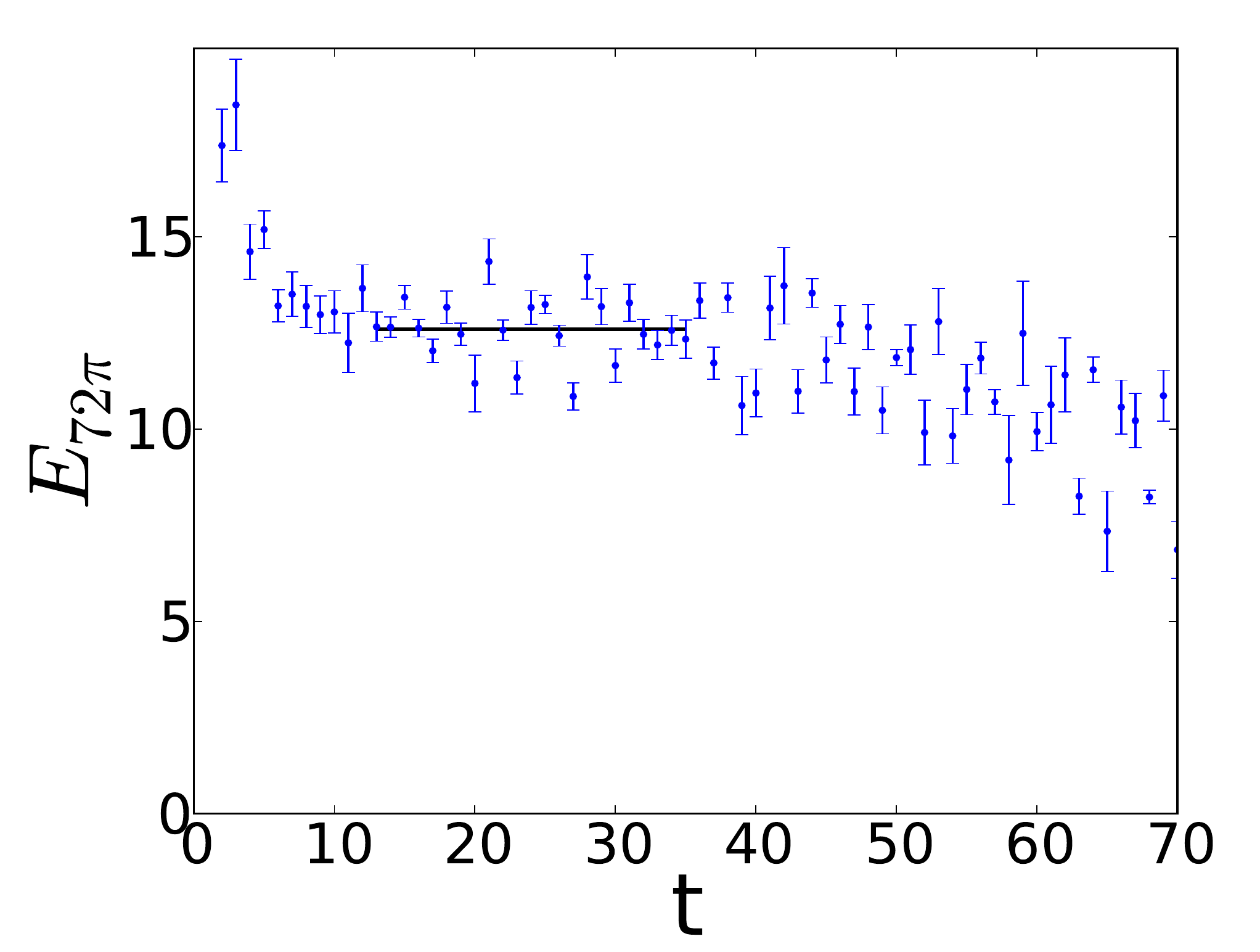}
  \caption{The effective mass of $C_{20\pi}(t)$ from the $2$-source
    calculation on the ensemble B4 is shown on the left along with
    the extracted ground state energy represented as a black band.
    Similarly, the effective mass of $C_{40\pi}(t)$ ($C_{72\pi}(t)$)
    from the $4$ ($6$) source calculation on the same ensemble and the
    corresponding extracted ground state energy is shown in the middle
    (on the right).  }
  \label {fig:6_source}
\end{figure}

Thermal effects are exponentially suppressed by the larger temporal
extent and the ensemble with $T=256$ has greatly reduced
contamination, and a simple single exponential fit at intermediate
times is sufficient to extract ground state energies, even for
$E_{72\pi}$, as shown in Fig.~\ref{fig:6_source}. Effective mass plots
of $C_{20\pi}$, $C_{40\pi}$ and $C_{72\pi}$ for this ensemble all show
a plateau region, and a single exponential fit, only including the term
in Eq.~(\ref{eqn:corr_expected}) with $m=0$, is enough to extract the
ground state energy $E_{n\pi}$.  However, for significantly larger
numbers of pions, a still larger temporal extents would again 
be necessary.

\subsection{Energies from $20^3 \times 256$ ensemble}
\label{sec:20_256}
\begin{figure}
  \includegraphics[width=6.4cm]{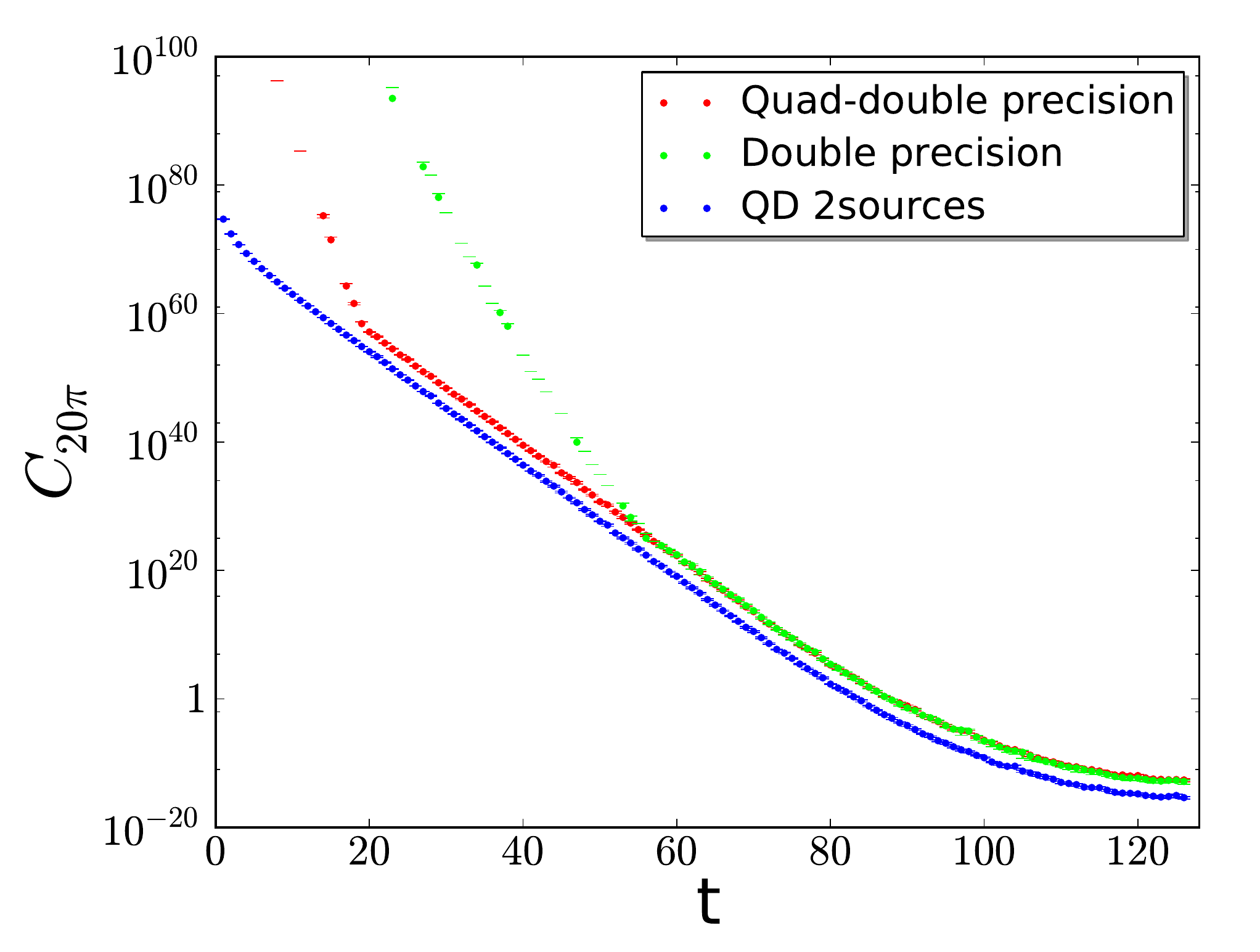}
  \includegraphics[width=6.4cm]{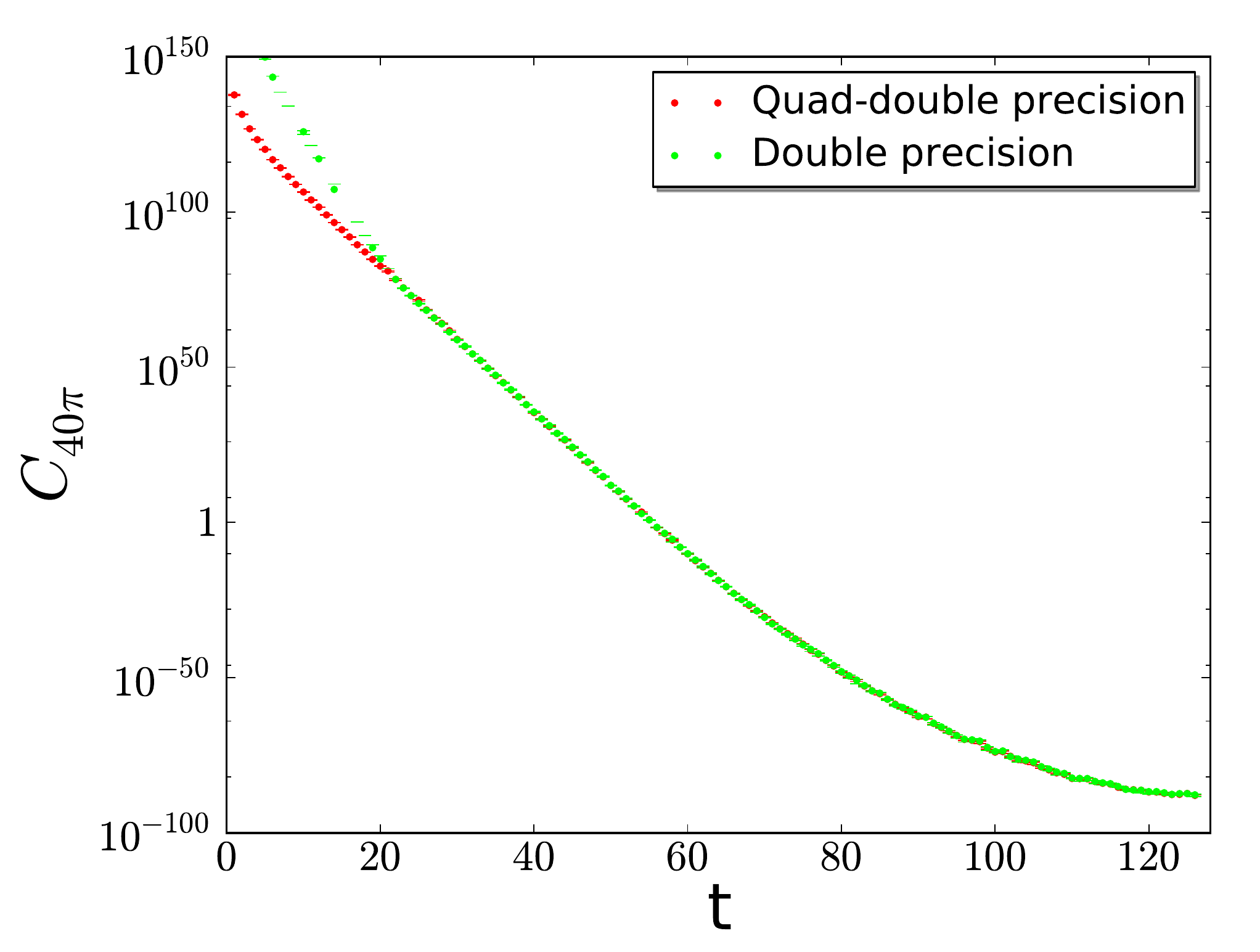}

  \includegraphics[width=6.4cm]{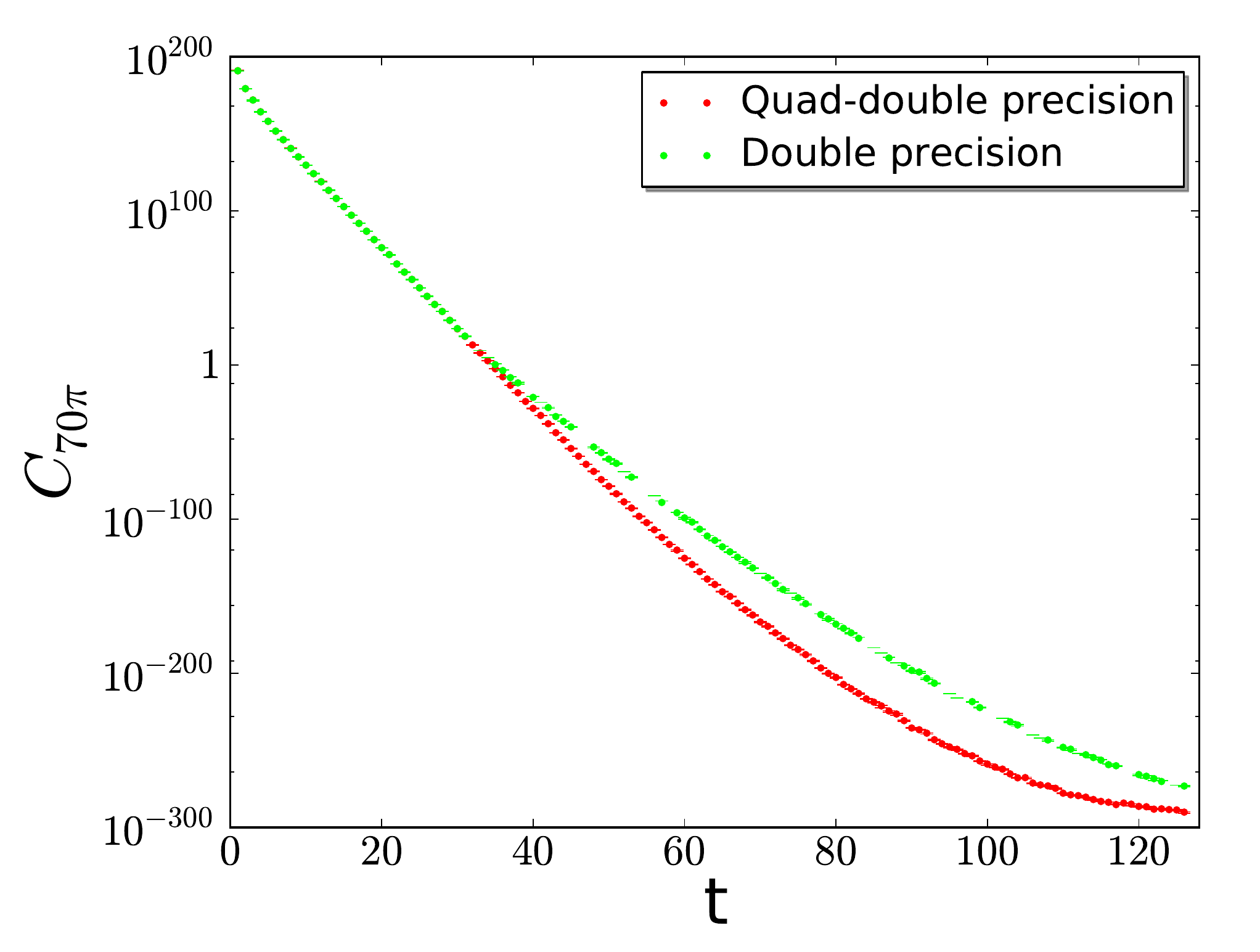}
  \caption{The correlation functions, $C_{20\pi}(t)$, $C_{40 \pi}(t)$
    and $C_{70\pi}(t)$, calculated from $6$-sources with quad-double
    precision and double precision are compared in the left, center,
    and right plots respectively.  The same calculations done with
    double precision shows even more severely breakdown, indicating
    that high precision is needed in order to study many pion
    systems.  Although $C_{20\pi}$ from $6$-sources with quad-double
    precision breaks down at earlier time slices, the rescaled $C_{20\pi}$ from
    $2$-source computations, which is shown also in the left plot, is
    free from precision issues and is used in extracting the
    $E_{20\pi}$.  }
  \label{fig:corre_compare_6source}
\end{figure}

Correlation functions, defined in Eq.~(\ref{eq:C_m_sources_mom}), for
systems with the quantum numbers of up to $72\ \pi^+$'s have been
computed on the B4 ensemble.  In this paper, only
systems having zero center of mass momentum are investigated.  For a
discussion of results for different total momenta, see
Ref~\cite{Z_W_2011_proceeding}.  Because of precision issues, we have
computed correlation functions from $2$, $4$, and $6$ sources, from
which $E_{1\pi\to 24 \pi}$, $E_{25\pi\to 48 \pi}$ and $E_{49\pi\to 72
  \pi}$ have been extracted respectively, where $E_{n\pi}$ is the
ground state energy of a \mbox{$n$-$\pi^+$} system at rest.
Fig.~\ref{fig:corre_compare_6source} shows $C_{20\pi}(t)$,
$C_{40\pi}(t)$ and $C_{70\pi}(t)$ from $6$-source contractions.  The
breakdown at earlier time slices of $C_{20\pi}(t)$ indicates that
computations with higher precision are required.  Computations with
arbitrary precisions are accessible with the ``arprec''
library~\cite{arprec_cite}, however at the same precision, they are
$\sim 5$ times more expensive than with the fixed quad-double precision
(implemented using the ``qd'' library \cite{qd}).  In our main
studies, we perform all contractions in quad-double precision, and
multiply the uncontracted propagators by a prefactor before performing
the contractions such that the particular $C_{n\pi}(t)$'s that we
focus on do not suffer from the limit of
the floating point dynamical range of Quad-double
precision (this prefactor is
removed at the end of the calculation).

As the correlation functions of systems containing many pions span a
large numerical range, $10^{250} \sim 10^{-250}$ for $C_{70 \pi}(t)$
for example, inverting the correlation matrix during a correlated fit
brings in significant instabilities, thus $E_{n\pi}$ for $n =
1,2,\ldots 72$ are extracted from uncorrelated fits in this study.
The fitting window is chosen between time slices where a clear plateau
region of the effective mass plot can be seen.  Statistical
uncertainties are constructed from fits to multiple bootstrap
resamplings of the ensemble (we use $N_s=88$ samples), and systematic
uncertainties are estimated by shifting the fitting window forward and
backward two time slices.

\begin{figure}
  \includegraphics[width=5.4cm]{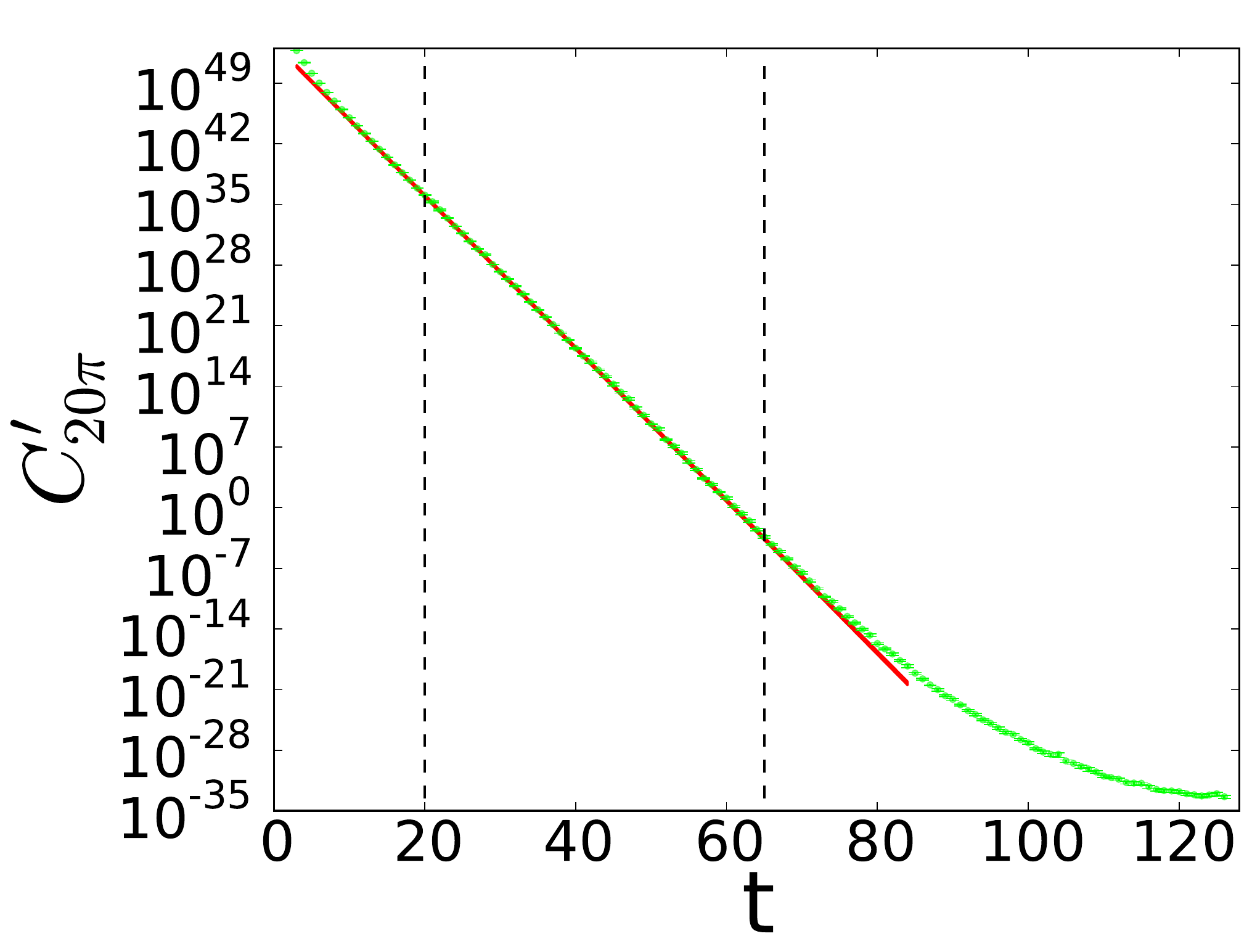}
  \includegraphics[width=5.4cm]{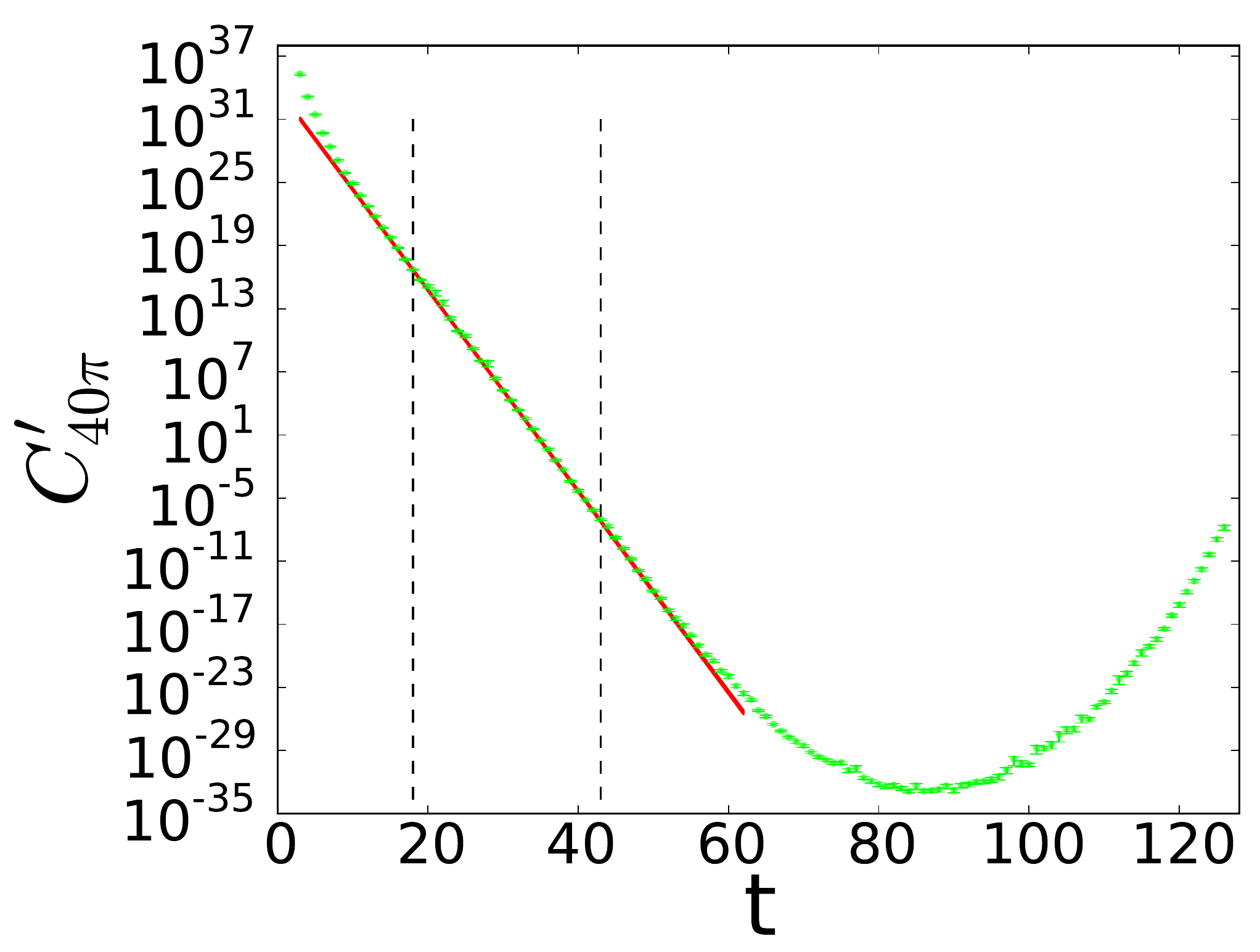}
  \includegraphics[width=5.4cm]{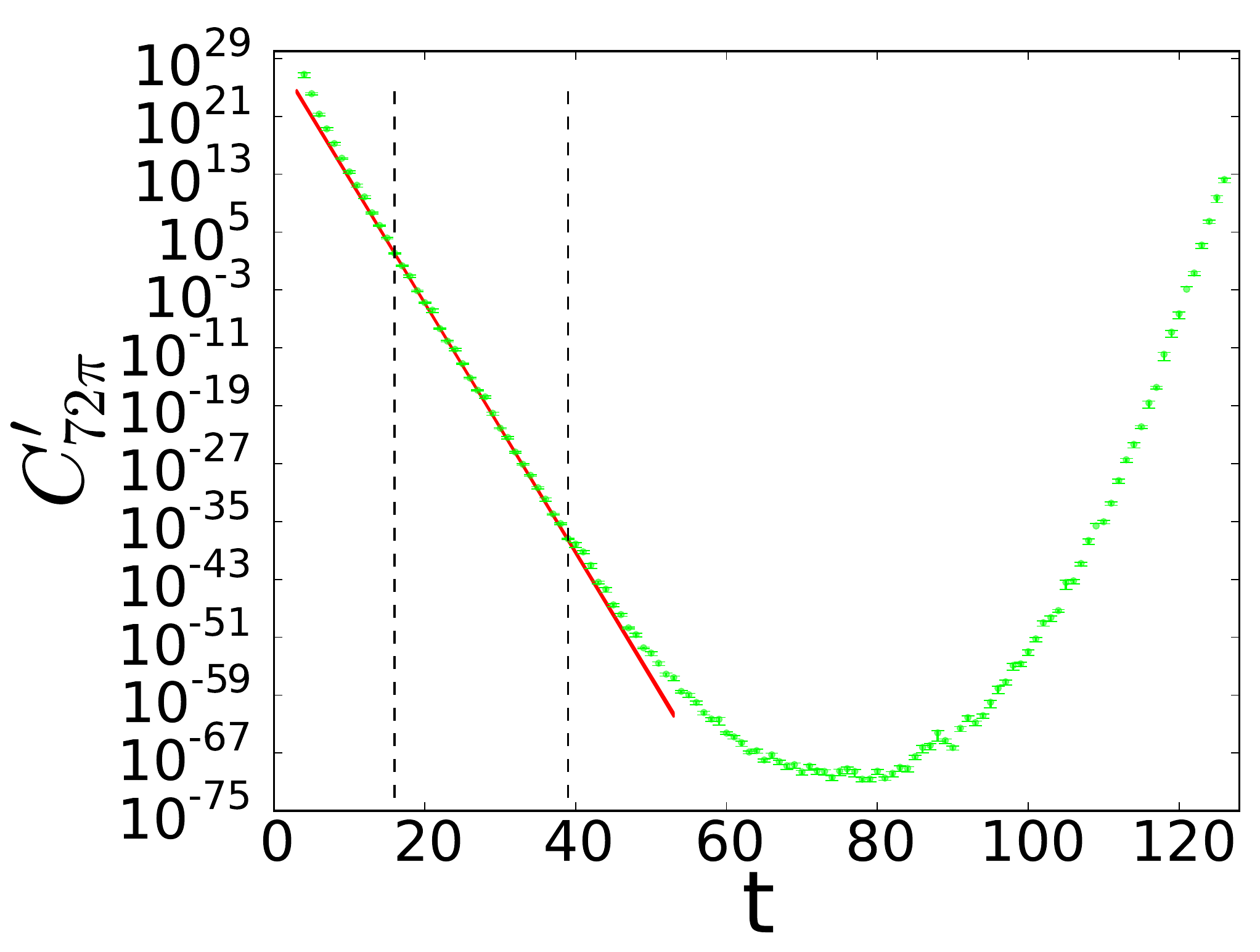}
  \caption{$C^{\prime}_{20\pi}(t)$ is shown on the left, where the
    green points are data, the red line is constructed from the
    fit, and two vertical dashed lines indicate the
    fitting window. Similar plots of preconditioned
    $C^{\prime}_{40\pi}(t)$ and $C^{\prime}_{72\pi}(t)$ are also
    shown.  }
  \label {fig:preconditioned_c}
\end{figure}

Since the ground state energy of a system containing many pions
becomes large, even fitting correlation functions with only one
exponential becomes problematic because of precision. Taking
the $25$-$\pi^+$ system for example, the ground state energy of this
system is $E_{25\pi} = 2.76$ in temporal lattice units, and the fit is
performed between $t/a_t=[15,58]\pm 2$. The correlation function varies
over $140$ orders of magnitude from $t=15$ to $t=58$. Such a large
change in magnitude requires care with precision and in order to
ameliorate this problem, instead of fitting correlation functions
directly, we fit the following preconditioned correlation functions:
\begin{eqnarray}
  \label{eqn:preconditioned_correlation_func}
  C_{n\pi}^\prime(t) = Z_n^\prime \exp(\delta E_n t) C_{n\pi}(t),
\end{eqnarray}
where $C_{n\pi}(t)$ is the original correlation function, and
$Z_n^\prime$, and $\delta E_n$ are fixed numbers, chosen so that
$C_{n\pi}^\prime (t)$ changes less dramatically inside the fitting
window.  Since the original correlation function behaves like a single
exponential inside the plateau region where the ground state
dominates, multiplying another exponential will not change this
feature. Furthermore, the extracted ground state energy should have no
dependence on $Z_n^\prime$ and $\delta E_n$, which is
numerically confirmed.  The preconditioned correlation
functions and the corresponding single exponential fits for $n=20, 40$
and $72$ are shown in Fig.~\ref{fig:preconditioned_c}.

\subsection{Antiperiodic $\pm$ Periodic propagator method ($A \pm P$ method)}

\begin{figure}
  \includegraphics[width=8.1cm]{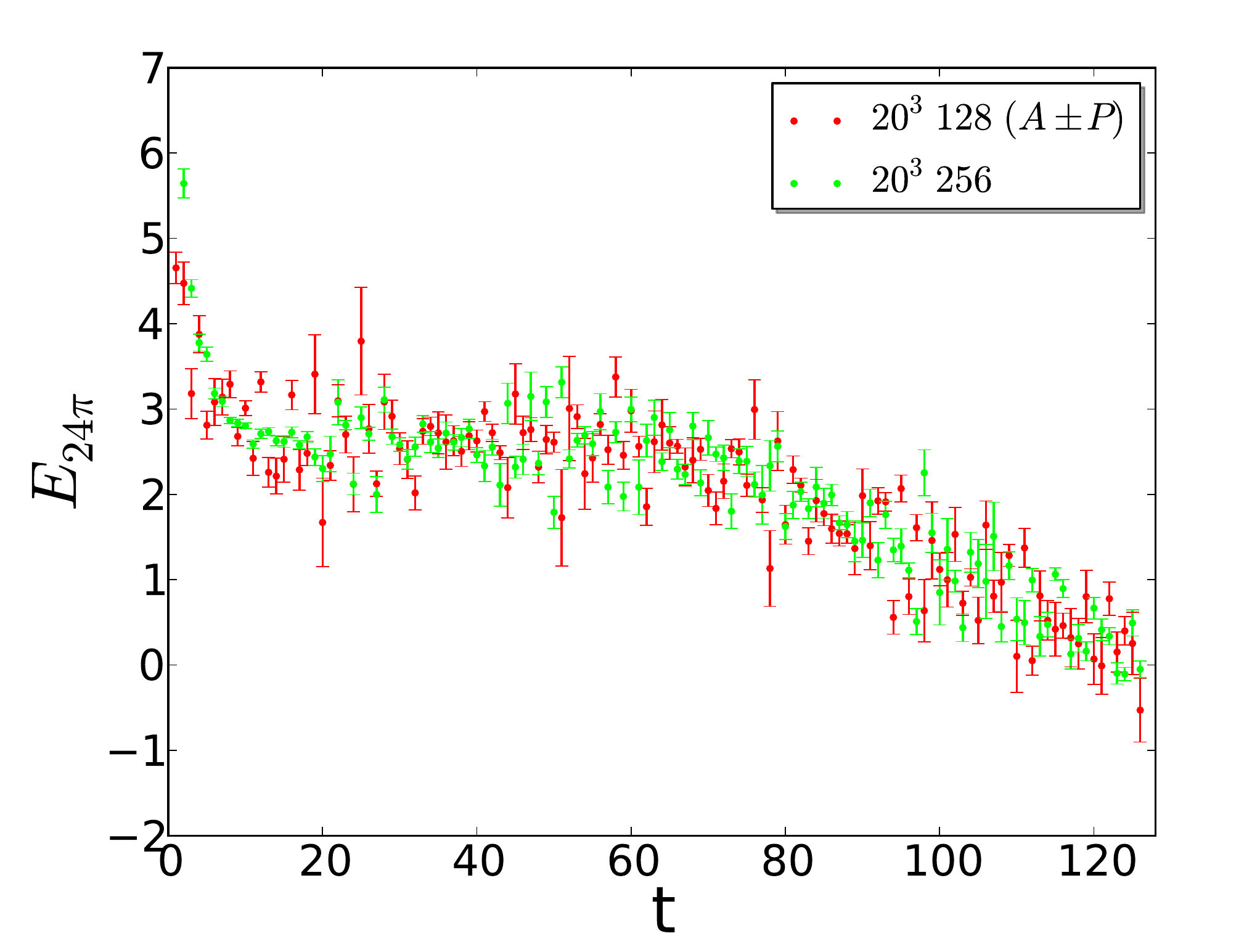}
  \includegraphics[width=8.1cm]{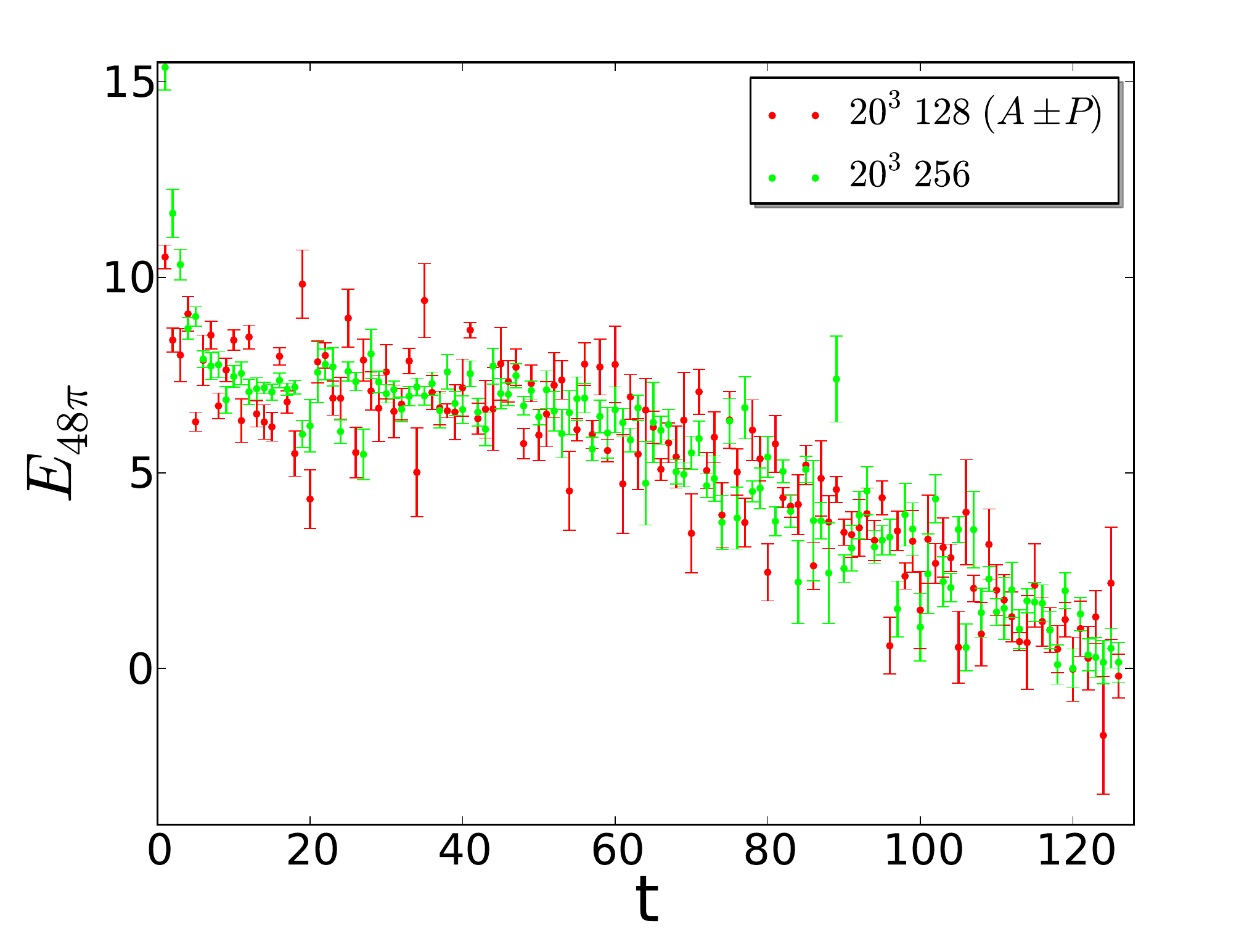}
  \caption{ Effective mass plots for $24 \pi^+$ and $48 \pi^+$
    correlators.  The green data are from ensemble B4 and the red
    data are from the $A\pm P$ method on ensemble B2.  Effective
    mass plots are consistent between these two calculations for all
    $n\ \pi^+$ systems.  }
  \label{fig:compare_20_256_128}
\end{figure}

\begin{figure}
  \includegraphics[width=15.cm]{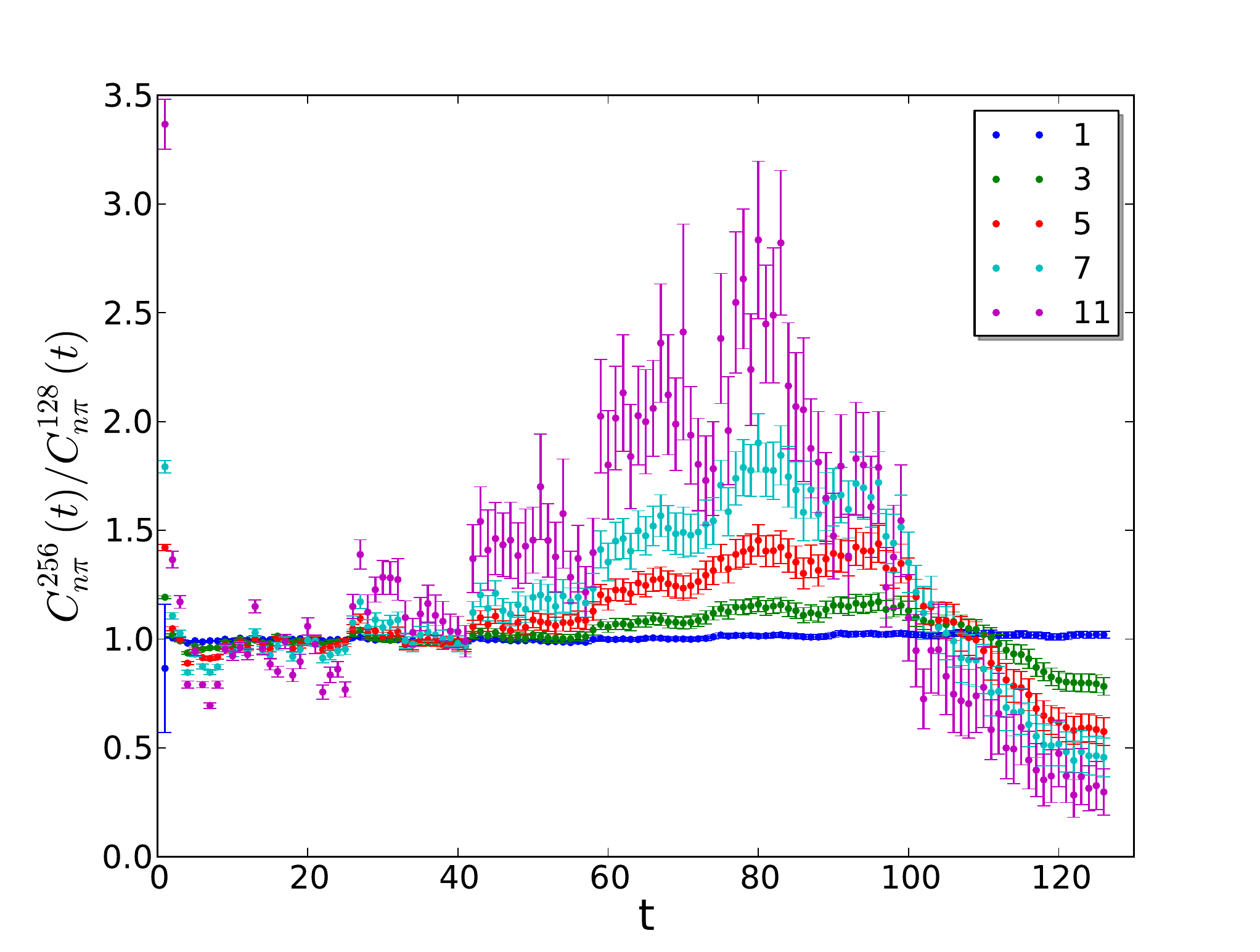}
   \caption{The ratio of 
          the correlation function of $n\ \pi^+$'s calculated by using the $A \pm P$ method
        on B2 ensemble, $C_{n\pi}^{128}(t)$, compared with that from B4 ensemble, $C_{n\pi}^{256}(t)$,
        for $n = 1, 3, 5, 7 , 11$, is shown.}
   \label{correlation_7_aplusp_vs_256}
\end{figure}

\begin{figure}
  \includegraphics[width=8.1cm]{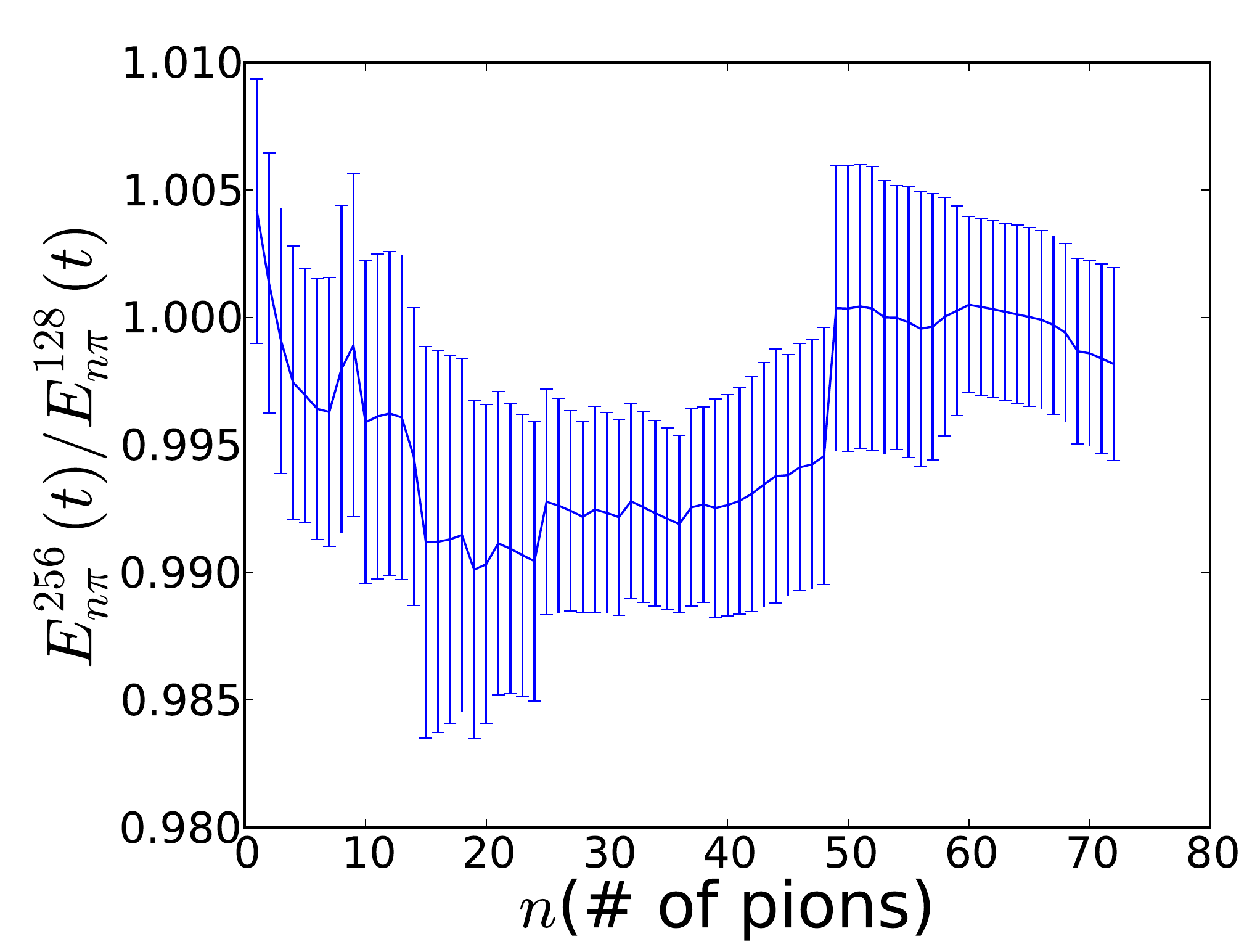}
  \includegraphics[width=8.1cm]{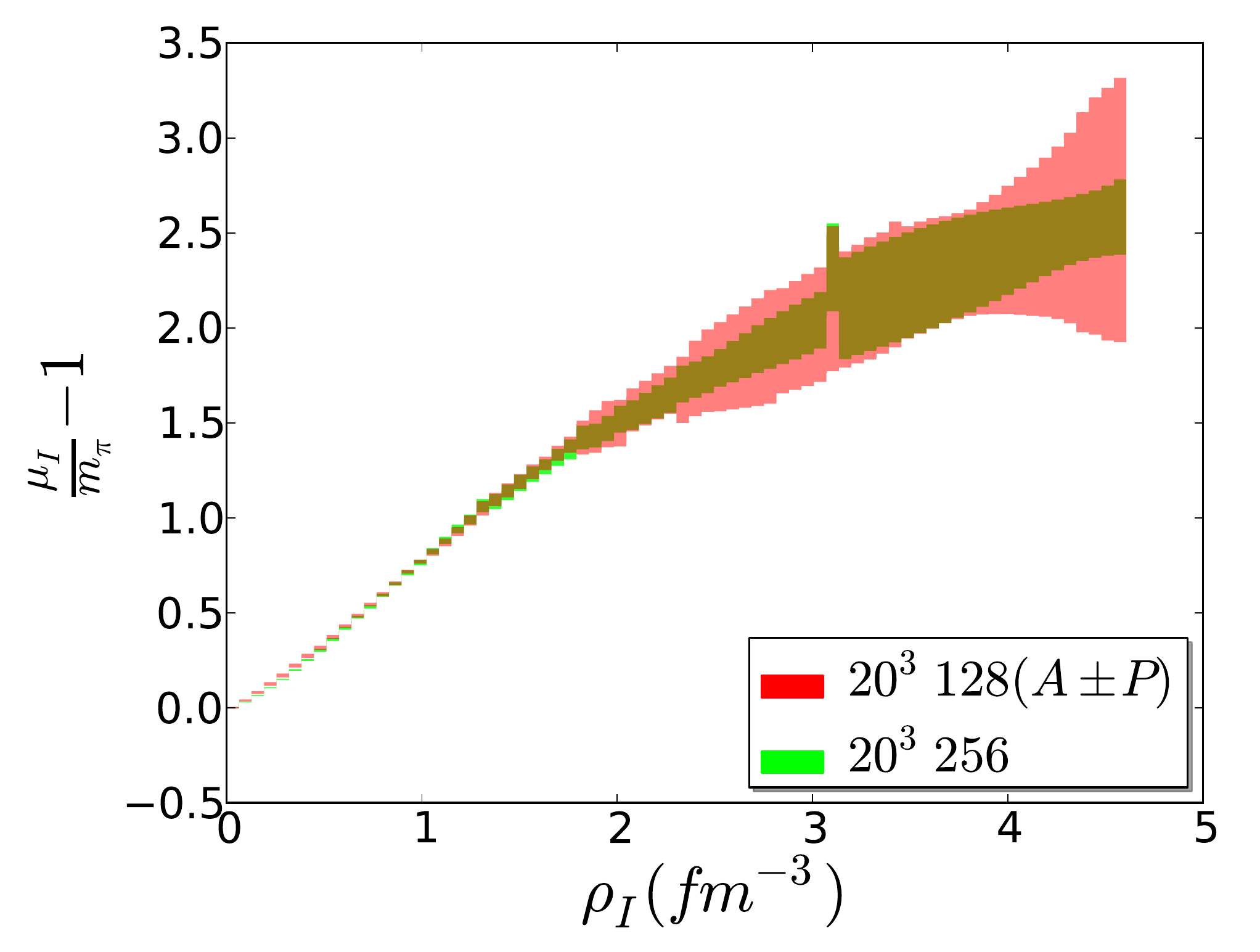}
  \caption{The ground state energies, $E_{n\pi}$, extracted from
    ensemble B2 ($E_{n\pi}^{128}$) with $A \pm P$ method are compared
    with those from ensemble B4 ($E_{n\pi}^{256}$) in the left plot,
    where the ratio of $E_{n\pi}^{256}/E_{n\pi}^{128}$ is plotted.
    The isospin chemical potentials, $\mu_I$, at different
    densities for the two ensembles are compared in the right plot.}
  \label{fig:20_128_vs_256_mu_E}
\end{figure}

By keeping all $Z_m^n$ factors the same as the ground state $Z_0^n$
extracted from the B4 ensemble, we have reconstructed the correlators
corresponding to the B2 ensemble by utilizing the ground state
energies extracted from the B4 ensemble\footnote{While we do not expect
$Z_m^n = Z_0^n$ for all $m$ because of the effects of pion interactions,
deviations are expected to be small (This is also supported by thermal
fits using Eq.~(\ref{eq:c_all_thermal}) for small $n$.).}.
 In Fig.~\ref{fig:reconstruct_128_from_256}, the reconstructed effective
masses are compared with those from the correlation functions computed
from the B2 ensemble, showing agreement within uncertainties.  The
contamination from the thermal states on the $T=128$ (B2) ensemble can
clearly be seen in the rate at which the plateau region (where the
ground state energy dominates) shrinks as $n$ increases.  For systems
with a large number of pions, excited states have not died out before
thermal states become important.

Since a temporal extent $T \ge 128$ is required to get a clean signal
for many-pion ground state energies, we have investigated the use of
the $A \pm P$ method (combining propagators that satisfy anti-periodic
and periodic boundary conditions in the temporal direction to cancel
certain modes~\cite{rbc:2001, rbc:2006, chris:jack2007}). 
On the $T=128$ B2 ensemble, we check the validity of
this method in comparison to the B4 ($T=256$) ensemble.  In order to
see the deviation of this method compared with those calculated
directly from the $T=256$ ensemble with anti-periodic boundary
conditions in the temporal direction, effective mass plots from the
two ensembles are compared in Fig.~\ref{fig:compare_20_256_128},
and the ratio of correlation functions from these two methods
are shown in Fig.~\ref{correlation_7_aplusp_vs_256}.
The $A \pm P$ method relays on the exact cancellation
of thermal contributions, and is seen to work very well
$1\ \pi^+$ system, see Fig.~\ref{correlation_7_aplusp_vs_256}. 
For systems with more than
$1\ \pi^+$, the $A \pm P$ method starts fail at later time slices,
however it still gives consistent results at earlier time slices, where ground 
state energies can be extracted.
Energies and isospin chemical potentials extracted from the $A \pm P$ method
are compared with those from ensemble B4 in
Fig.~\ref{fig:20_128_vs_256_mu_E}, which shows that
the disagreement of extracted ground energies below $1\%$,
and at our current precision, the $A \pm P$ 
method provides reliable results for the correlators we study.
This gives us confidence to use the $A\pm P$ method for ensembles B1
and B3, where we could otherwise not extract ground state energies
for large number of pions.

\subsection{Energies from $16^3 \times 128$ and $24^3 \times 128$
  ensembles}

\begin{figure}
  \includegraphics[width=5.4cm]{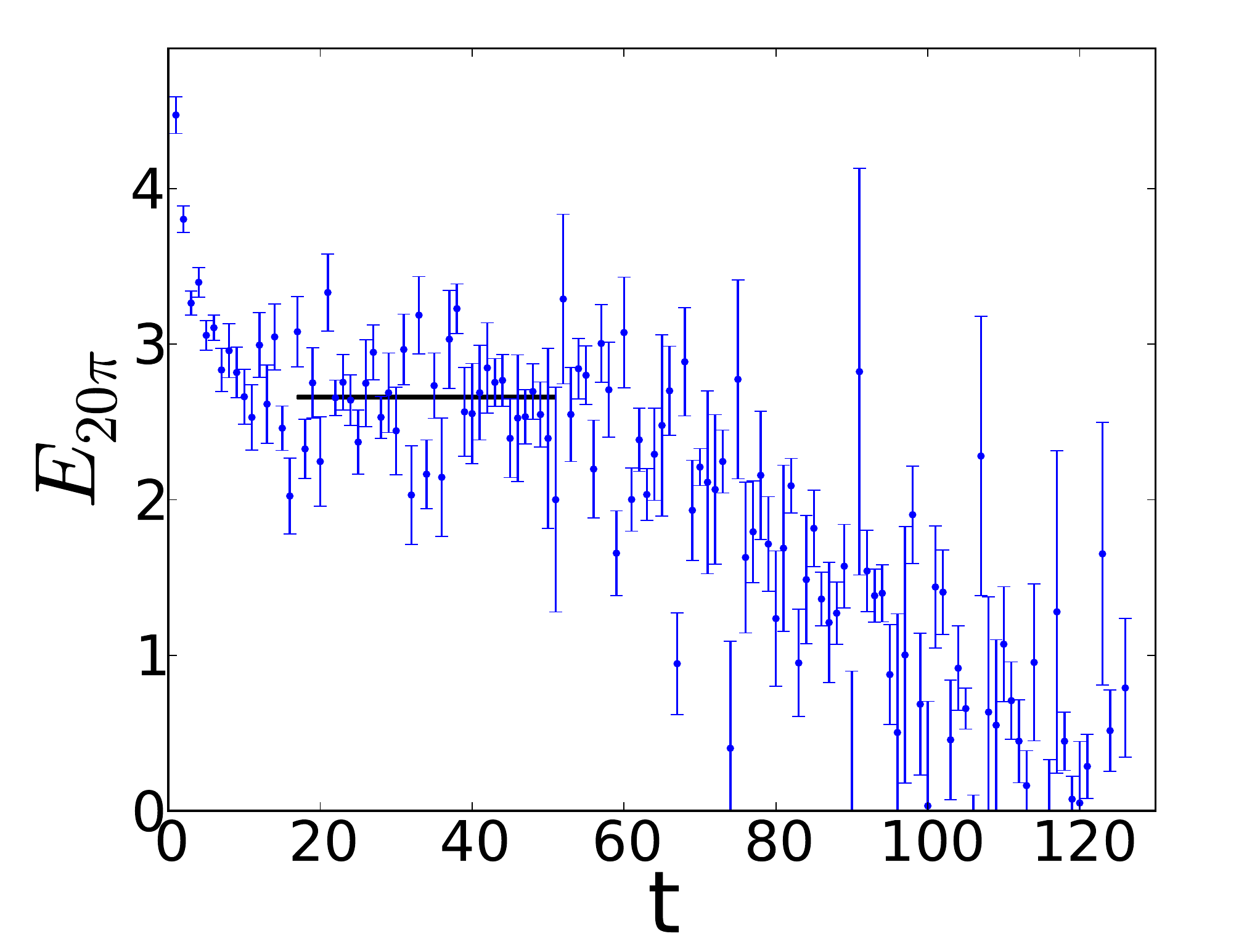}
  \includegraphics[width=5.4cm]{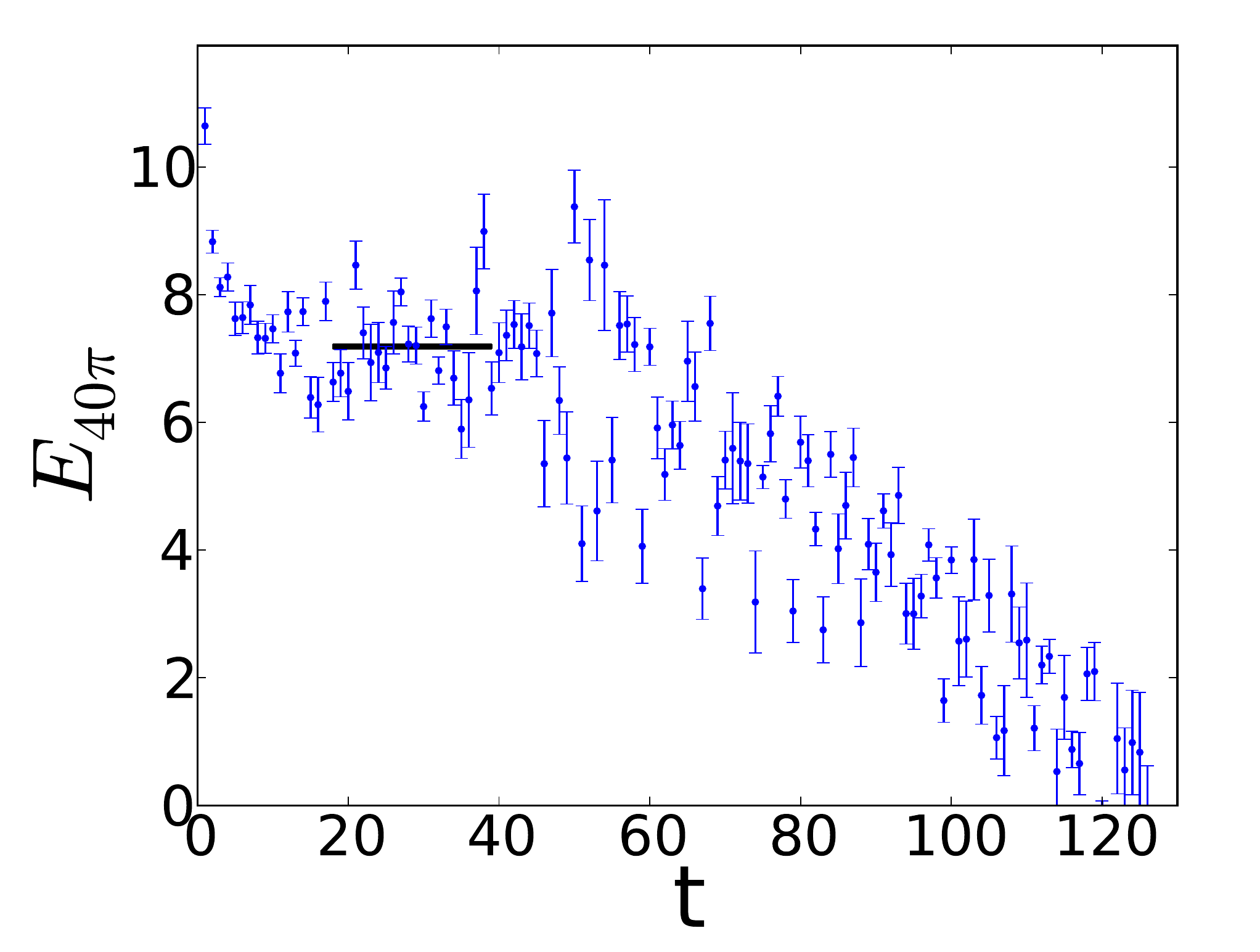}
  \includegraphics[width=5.4cm]{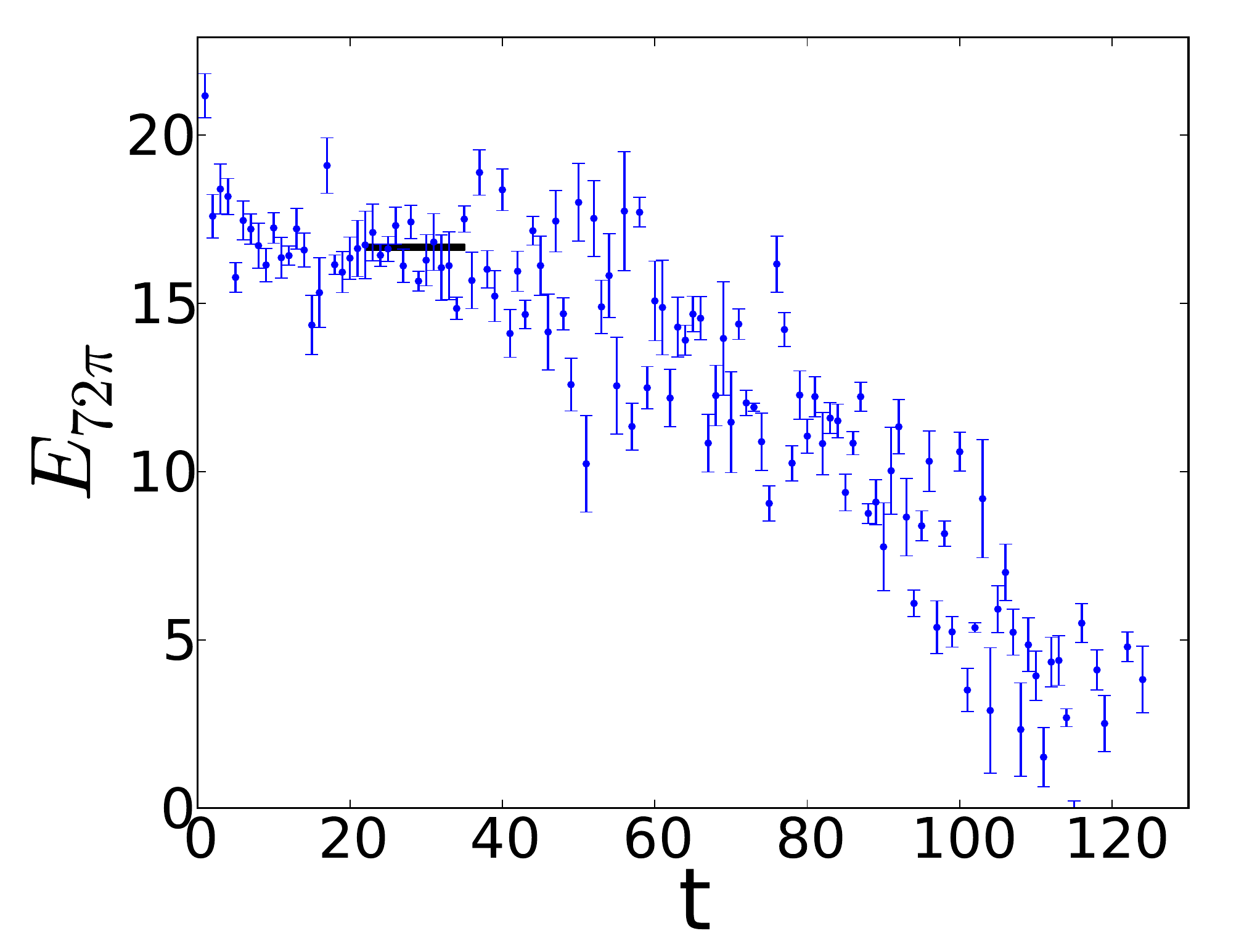}
  \caption{Effective mass plots with $A\pm P$ method on ensemble B1
    are shown here.  The effective mass of $C_{20\pi}(t)$ from the
    $2$-source calculation is shown on the left along with the ground
    state energy represented as a black band.  Similarly effective
    mass plots of $C_{40 \pi}$ from $4$-sources and $C_{72\pi}(t)$
    from $6$-sources calculations and the extracted ground state
    energies are shown in the middle and right respectively.  }
  \label {fig:6_source_16}
\end{figure}

\begin{figure}
  \includegraphics[width=5.4cm]{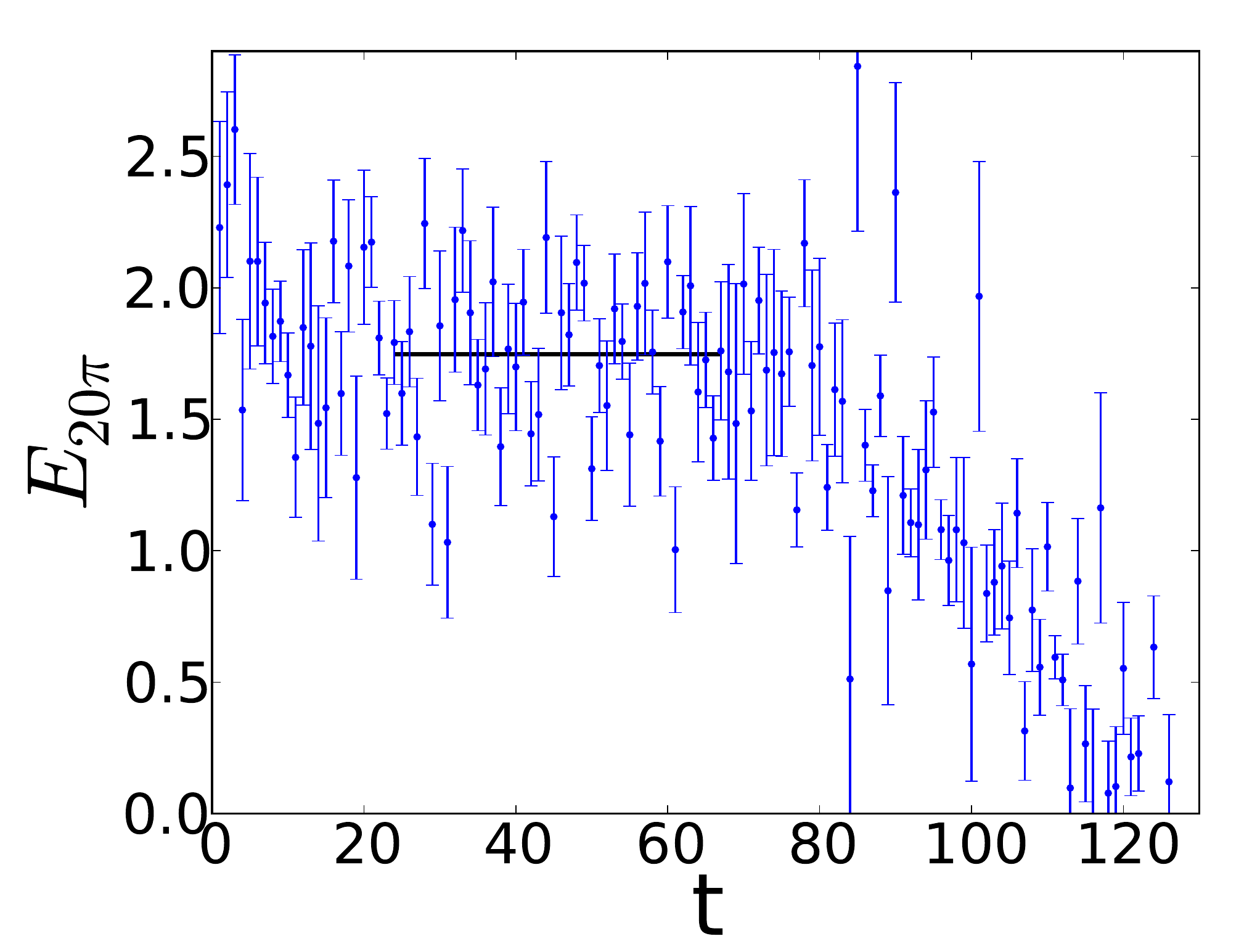}
  \includegraphics[width=5.4cm]{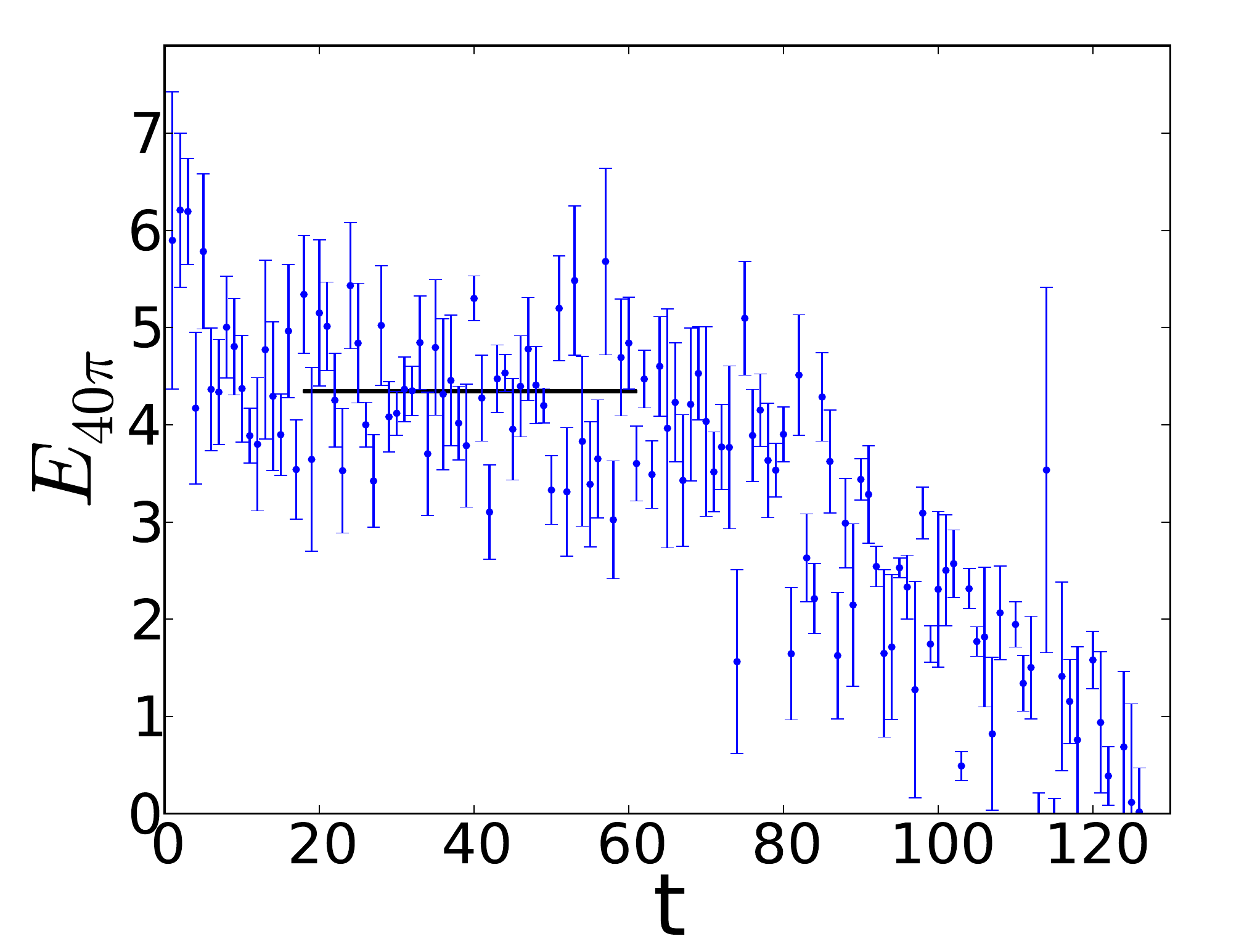}
  \includegraphics[width=5.4cm]{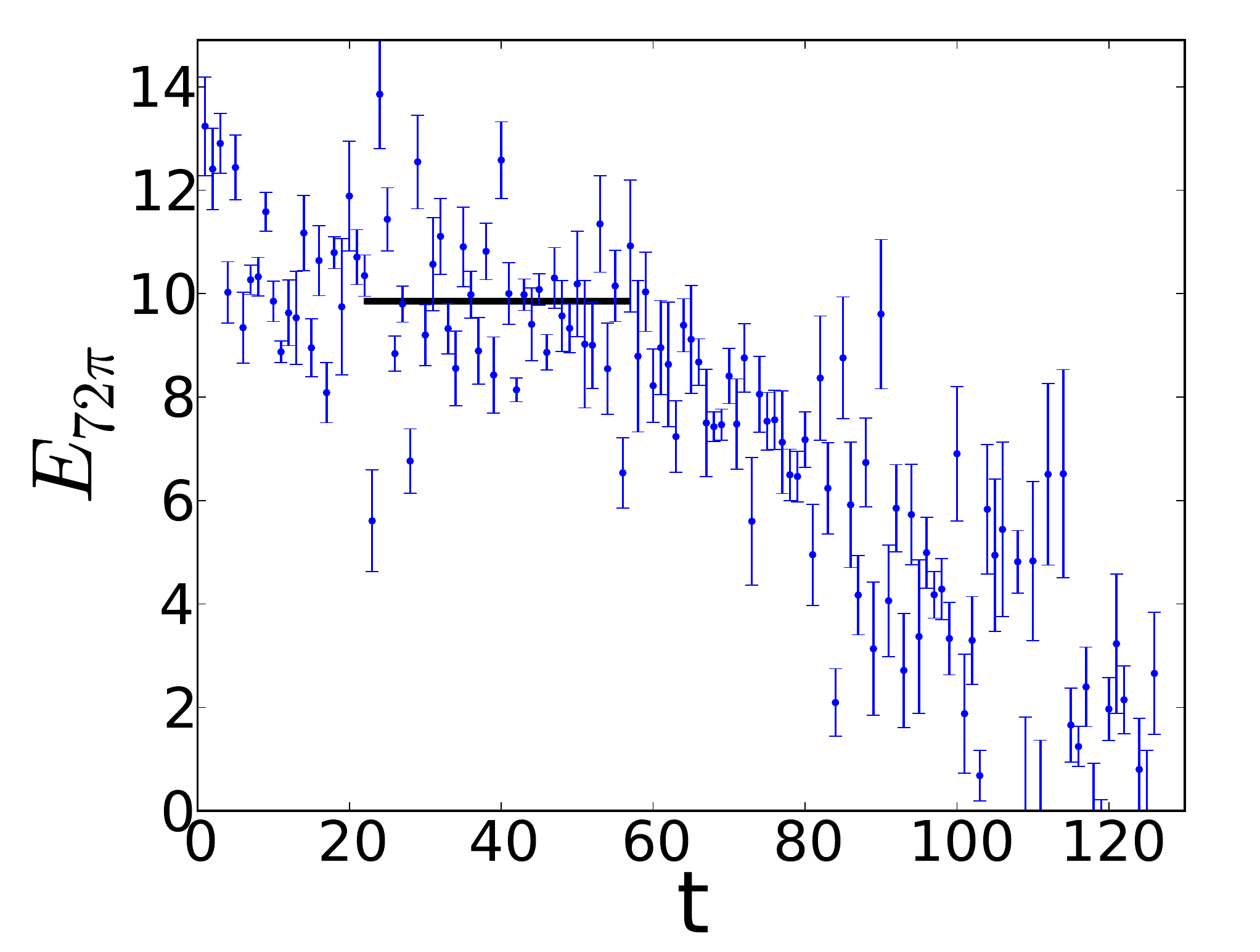}
  \caption{Effective mass plots with $A\pm P$ method on ensemble B3
    are shown.  }
  \label {fig:6_source_24}
\end{figure}

\begin{figure}
  \includegraphics[width=15.0cm]{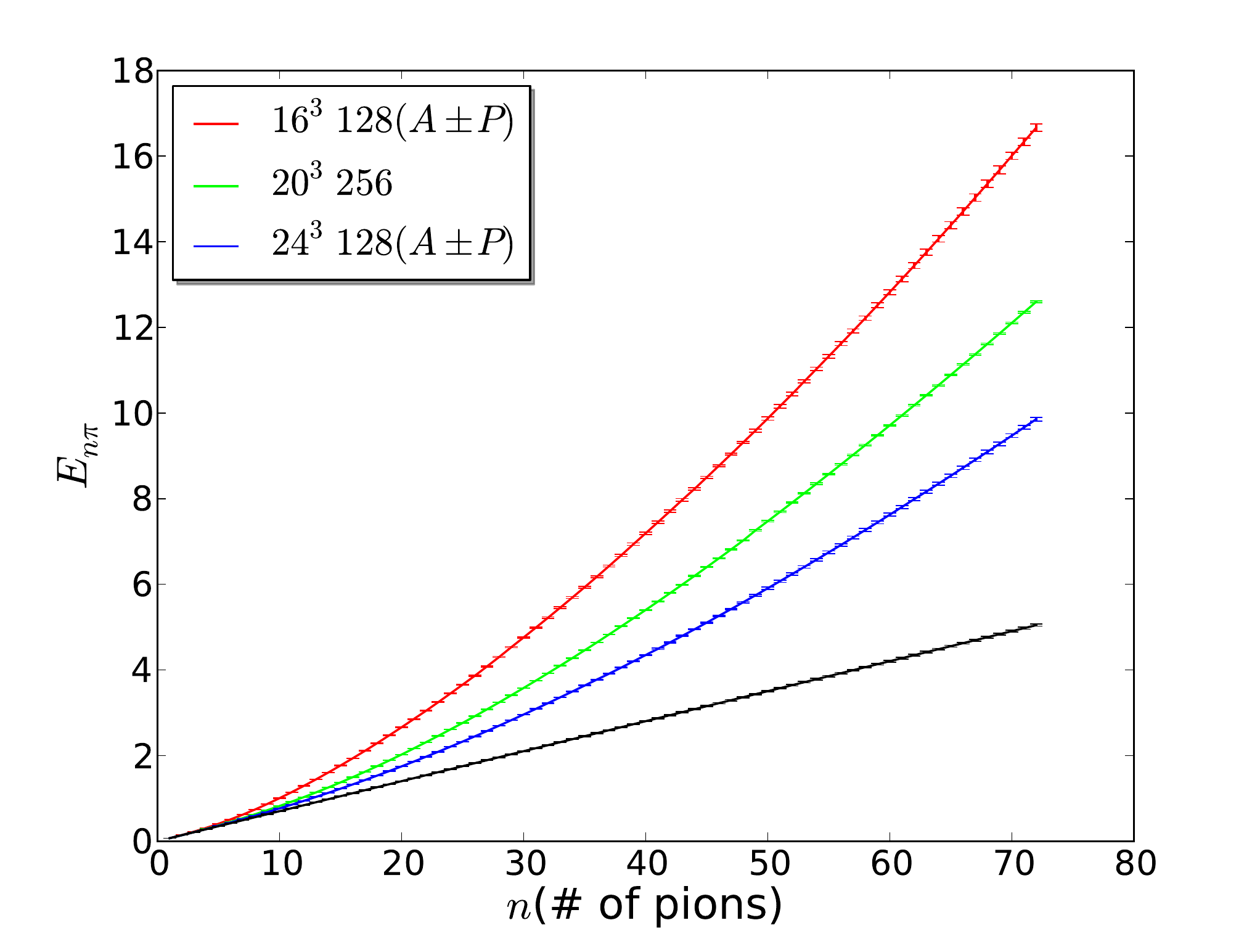}
  \caption{ The ground state energies of a system of
    $n$-$\pi^+$($E_{n\pi}$) extracted from ensembles B1 (red), B3
    (green) and B4 (blue) are shown.  The black line represents the
    total energy of $n$ non-interacting pions.  }
  \label{fig:energy_chemical_from_256}
\end{figure}

As the $A \pm P$ method has been validated on the B2 ensemble, systems
having up to $72\ \pi^+$'s has also been studied on ensembles B1 and
B3 using this method.  Effective mass plots with extracted ground
state energies from ensemble B1 are shown in
Fig.~\ref{fig:6_source_16} and those from ensemble B3 are shown in
Fig.~\ref{fig:6_source_24}.  All calculations are done with the ICm,
and ground state energies are extracted with the same statistical
method as those in the Section~\ref{sec:20_256}.  The extracted ground
state energies from all three volumes are shown in
Fig.~\ref{fig:energy_chemical_from_256}.

\section {Interaction parameters}
\label{scattering_parameters}
By considering the energy shifts of two particle states in a finite
volume, $\Delta E \equiv E_2 - 2E_1 = 2\sqrt{{\bf
    p}^2+m_{\pi}^2}-2E_1$ , L\"uscher derived a relationship between
the phase shift, $\delta(p)$, and the interacting momentum, $p = |{\bf
  p}|$, given by~\cite{Luscher:1986pf, Luscher:1990ux} (see also
\cite{Beane:2003da}),
\begin{eqnarray}
  p\cot\delta(p) \ =\ {1\over \pi L}\ {\bf
    S}\left(\,\left(\frac{p L}{2\pi}\right)^2\,\right)
  \ \ ,
  \label{eq:energies}
\end{eqnarray}
which is valid for momenta below the inelastic threshold.  The
regulated three-dimensional sum, ${\bf S}(x)$, is
\begin{eqnarray} {\bf S}\left(\, x \, \right)\ \equiv
  \stackrel{\Lambda\to\infty}{\lim}\left(\ \sum_{\bf j}^{
      |{\bf j}|<\Lambda}
    {1\over |{\bf j}|^2-x}\ -\  {4 \pi \Lambda}\right)
  \ \ \  ,
  \label{eq:Sdefined}
\end{eqnarray}
where the summation is over all triplets of integers ${\bf j}$ such
that $|{\bf j}| < \Lambda$.

By performing an expansion in small $1/L$, the energy
shift of \mbox{$n$ identical} bosons in a finite volume, $\Delta
E_n = E_n - n E_1$, has also been studied up to ${\cal O}(L^{-7})$ in
recent
work~\cite{Beane:2007qr,Tan:2007bg,Detmold:2008gh,Smi:Was_2009}.  The
resulting shift of energies due to both two-body and three-body
interactions is given by~\cite{Detmold:2008gh}:
\begin{eqnarray}
  \Delta E_n &=&
  \frac{4\pi\, \abar}{M\,L^3}\Choose{n}{2}\Bigg\{1
  -\left(\frac{\abar}{\pi\,L}\right){\cal I}
  +\left(\frac{\abar}{\pi\,L}\right)^2\left[{\cal I}^2+(2n-5){\cal J}\right]
  \nonumber 
  \\&&\hspace*{2cm}
  -
  \left(\frac{\abar}{\pi\,L}\right)^3\Big[{\cal I}^3 + (2 n-7)
  {\cal I}{\cal J} + \left(5 n^2-41 n+63\right){\cal K}\Big]
  \nonumber
  \\&&\hspace*{2cm}
  +
  \left(\frac{\abar}{\pi\,L}\right)^4\Big[
  {\cal I}^4 - 6 {\cal I}^2 {\cal J} + (4 + n - n^2){\cal J}^2 
  + 4 (27-15 n + n^2) {\cal I} \ {\cal K}
  \nonumber\\
  &&\hspace*{4cm}
  +(14 n^3-227 n^2+919 n-1043) {\cal L}\ 
  \Big]
  \Bigg\}
  \nonumber\\
  &&
  +\ \Choose{n}{3}\left[\ 
    {192 \ \abar^5\over M\pi^3 L^7} \left( {\cal T}_0\ +\ {\cal T}_1\ n \right)
    \ +\ 
    {6\pi \abar^3\over M^3 L^7}\ (n+3)\ {\cal I}\ 
  \right]
  \nonumber\\
  &&
  +\ \Choose{n}{3} \ {1\over L^6}\ \overline{\overline{\eta}}_3^L\ 
  \ \ + \ {\cal O}\left(L^{-8}\right)
  \ \ \ \ ,
  \label{eq:energyshift}
\end{eqnarray}
where $\Choose{m}{n} = m!/(n!(m-n)!)$, and the parameter $\abar$ is
the inverse phase shift at the binding momentum of the two body
system (below we will refer to this as the effective scattering length).
 This is related to the scattering length, $a$, and the
effective range, $r$, by
\begin{eqnarray}
  a
  & = & 
  \overline{a}\ -\ {2\pi\over L^3} \overline{a}^3 r \left(\ 1 \ -\
    \left( {\overline{a}\over\pi L}\right)\ {\cal I} \right)\ \ .
  \label{eq:aabar}
\end{eqnarray}
The geometric constants entering Eq.~(\ref{eq:energyshift}) are:
\begin{align}
  &{\cal I}\ =\ -8.9136329\,, &{\cal J}\ =\ 16.532316\,, \qquad\qquad
  {\cal K}\ = \ 8.4019240\,,
  \nonumber\\
  &{\cal L}\ = \ 6.9458079\,, &{\cal T}_0\ = -4116.2338\,,
  \qquad\qquad {\cal T}_1\ = \ 450.6392\,. &
  \label{eq:sums}
\end{align}

The three body parameter ${\overline {\overline \eta}}_3^L$ is
constructed from the volume dependent but renormalization group
invariant three body interaction parameter, ${\overline \eta}_3^L$,
the inverse phase shift, ${\overline a}$, and the effective range,
$r$, as
\begin{eqnarray}
  \overline{\overline{\eta}}_3^L & = & \overline{\eta}_3^L  \left(\ 1 \ -\
    6  \left({\overline{a}\over\pi L}\right)\ {\cal I} \right)
  \ +\ {72\pi \overline{a}^4 r\over M L} \ {\cal I}
  \ \ \ ,
  \label{eq:eta3barbar}
\end{eqnarray}
where
\begin{align}
  \overline{\eta}_3^L & = \eta_3(\mu)\ +\ {64\pi a^4\over
    M}\left(3\sqrt{3}-4\pi\right)\ \log\left(\mu L\right)\ -\ {96
    a^4\over\pi^2 M} {\cal S}_{\rm MS} \ \ \ .
  \label{eq:etathreebar}
\end{align}
and the renormalized scale dependent coupling $\eta_3(\mu)$ is responsible
for the three-body interactions.  The renormalization scheme dependent
quantity ${\cal S}$ defined in the Minimal Subtraction scheme is given
by ${\cal S}_{\rm MS} = -185.12506$.

\subsection{Two-body interactions from L\"uscher's method}
From the energy difference in the $2$-$\pi^+$ system, $\Delta E_{2\pi}
= E_{2\pi} - 2m_{\pi}$, the relative momentum of each $\pi^+$,
${\bp}$, in the center of mass frame (COM) can be calculated from the
dispersion relation.
We determine the effective scattering length\footnote{
As discussed above, ${\bar a}$ is the inverse phase shift at the binding
momentum of the two body system,
and the scattering length in
  Eq.~(\ref{eq:energies}) uses the Particle Physics sign convention,
  and it is negative for repulsive interactions.}, ${\overline a}$, by
calculating the interacting momenta $\{ {\bf p}_i\}$, on each bootstrap ensemble
 and applying Eq.~(\ref{eq:energies}),
and we average over all ensembles to get the mean value of ${\overline
  a}$, and the statistical uncertainty.  The systematic uncertainty is
determined by averaging the systematic uncertainty of ${\overline a}$
on each ensemble resulting from the systematic uncertainty of the
extracted energies from the choice of different fitting intervals. The
extracted effective scattering length for each volume is shown in
Table~\ref{table:scattering_length_luther}.  Our results are in
agreement with the extractions in
Ref.~\cite{Beane:Detmold:2011} from two-body systems studied on the
same ensembles.

\begin{table}[ht]
  \caption{The effective scattering length (${\overline a}$) from L\"uscher's method. 
    The first uncertainty is statistical uncertainty and 
    the second uncertainty is systematic.}
  \centering
  \begin{tabular}{c  c  c c}
    \hline\hline
    $V^3 \times T$ & $p^2/m_{\pi}^2$ & ${\overline a}({\rm fm})$ & $m_{\pi}{\overline a}$ \\
    \hline
    $16^3 \times 128$ & \hspace{1cm}$0.0668(45)(1) $ & \hspace{1cm}$-0.134(7)(5)$ & \hspace{1cm}$0.263(15)(9)$ \\
    \hline
    $20^3 \times 256$ & \hspace{1cm}$0.0301(9)(0)$ & \hspace{1cm}$-0.122(3)(1) $ & \hspace{1cm}$0.238(6)(1)$\\
    \hline
    $24^3 \times 128$ & \hspace{1cm}$0.0143(9)(1) $ & \hspace{1cm}$-0.106(6)(4) $  & \hspace{1cm}$0.203(12)(7)$\\
    \hline
    $32^3 \times 256$\footnote{Results for this ensemble are taken from Ref.~\cite{Beane:Detmold:2011}. } & \hspace{1cm}$0.00678(54)(81)$ &\hspace{1cm}$-0.114(9)(13)$  & \hspace{1cm} $0.223(17)(26)$ \\
    \hline
  \end{tabular}
  \label{table:scattering_length_luther}
\end{table}

\begin{table}[ht]
  \caption{The effective scattering length (${\overline a}$) and
    $m_{\pi}f_{\pi}^4{\overline {\overline \eta}}_3^L$ extracted from
    fits to different ranges of $n$. For a fixed $n_{\rm max}$, the
    $\chi^2/{\rm d.o.f.}$ is larger in smaller volumes, 
    indicating that Eq.~(\ref{eq:energyshift}) fails to describe
    systems of high densities.} 
  \centering
  \begin{tabular}{c|c|c|c|c|c|c}
    \hline\hline
    $\ $ & \multicolumn{3}{c}{$n = [3, 5] $}  & \multicolumn{3}{|c}{$n = [3, 6]$} \\
    \hline
    $V^3 \times T$ & $m_{\pi}{\overline a}$ & $m_{\pi}f_{\pi}^4{\overline {\overline \eta}}_3^L$ & $\chi^2/dof$
    & $m{\overline a}$ & $m_{\pi}f_{\pi}^4{\overline {\overline \eta}}_3^L$ & $\chi^2/dof$\\
    \hline
    $16^3 \times 128$ & $0.260(14)(2)$ & $0.70(10)(4)$ & $1.0$
    & $0.261(14)(1)$ & $0.67(9)(3)$ & $1.5$ \\
    \hline
    $20^3 \times 256$ & $0.234(6)(1)$ & $0.80(8)(3)$ & $0.25$
    & $0.235(6)(1)$ & $0.79(7)(1)$ & $0.5$\\
    \hline
    $24^3 \times 128$ & $0.209(11)(4)$ & $1.61(20)(20)$ & $0.26$
    & $0.209(11)(3)$ & $1.59(18)(12)$ & $0.25$\\
    \hline
    $\ $ & \multicolumn{3}{c}{$n = [3, 7] $}  & \multicolumn{3}{|c}{$n = [3, 8]$} \\
    \hline
    $V^3 \times T$ & $m_{\pi}{\overline a}$ & $m_{\pi}f_{\pi}^4{\overline {\overline \eta}}_3^L$ & $\chi^2/dof$
    & $m_{\pi}{\overline a}$ & $m_{\pi}f_{\pi}^4{\overline {\overline \eta}}_3^L$ & $\chi^2/dof$\\
    \hline
    $16^3 \times 128$ & $0.262(14)(1)$ & $0.64(9)(1)$ & $3.5$
    & $0.263(14)(1)$ & $0.62(8)(1)$ & $5.5$ \\
    \hline
    $20^3 \times 256$& $0.235(6)(5)$ & $0.79(7)(1)$ & $1.1$
    & $0.235(6)(1)$ & $0.76(7)(1)$ & $2.8$\\
    \hline
    $24^3 \times 128$ & $0.211(11)(2)$ & $1.56(17)(8)$ & $0.4$
    & $0.210(11)(2)$ & $1.50(16)(5)$ & $1.0$\\
    \hline
  \end{tabular}
  \label{table:eta3_diff_time}
\end{table}

\begin{figure}
  \includegraphics[width=7.0cm]{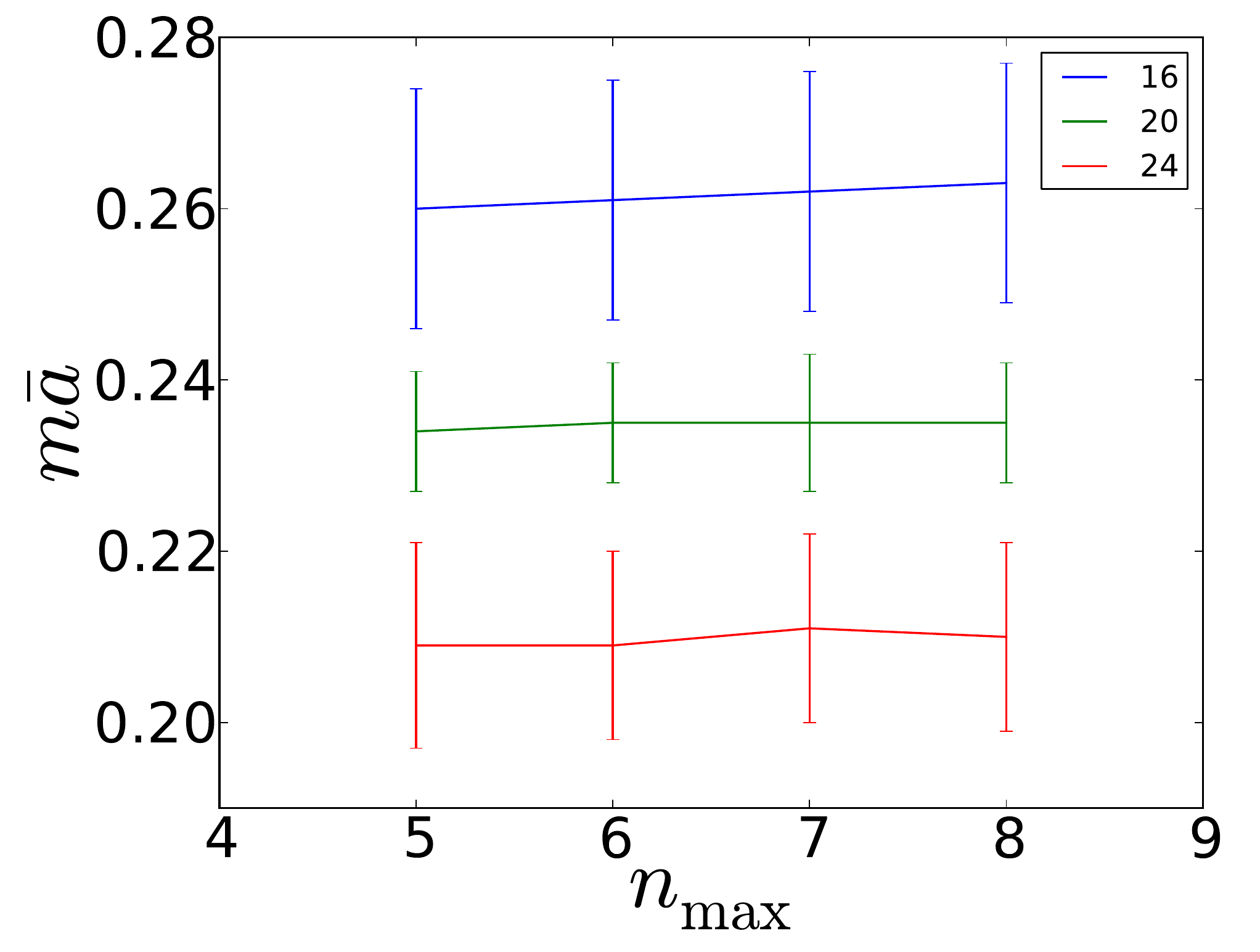}
  \includegraphics[width=7.0cm]{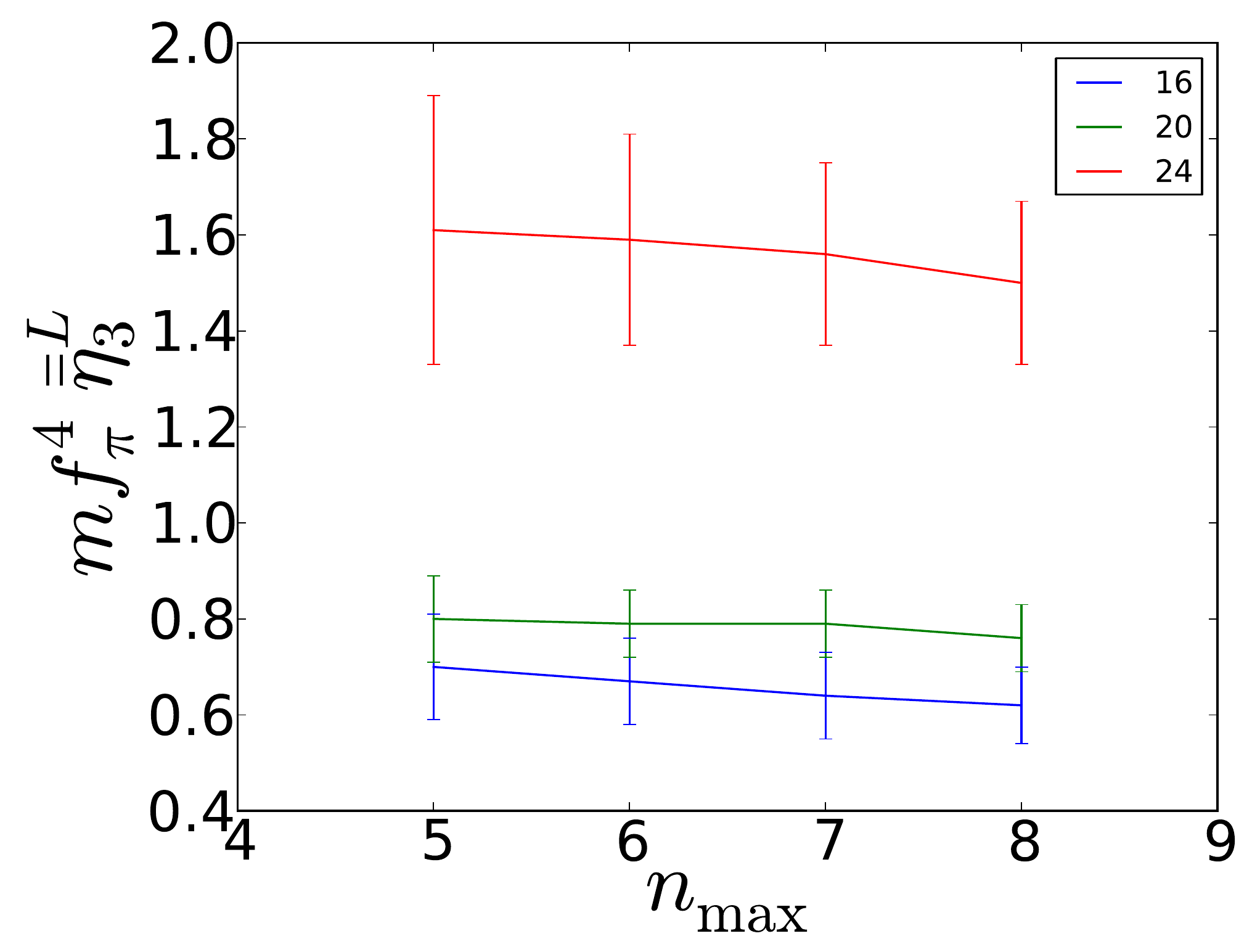}
  \caption{The $m{\overline a}$ and $mf_{\pi}^4{\overline{\overline
        \eta}}_3^L$ extracted from different fitting windows $[n_{\rm
      min}, n_{\rm max}]$ with $n_{\rm min} = 3$ fixed and varying
    $n_{\rm max}$.}
  \label{fig:eta3_diff_time}
\end{figure}

\subsection{Interaction parameters from small ${\overline a}/{L}$
  expansion}

\begin{figure}
  \includegraphics[width=5.4cm]{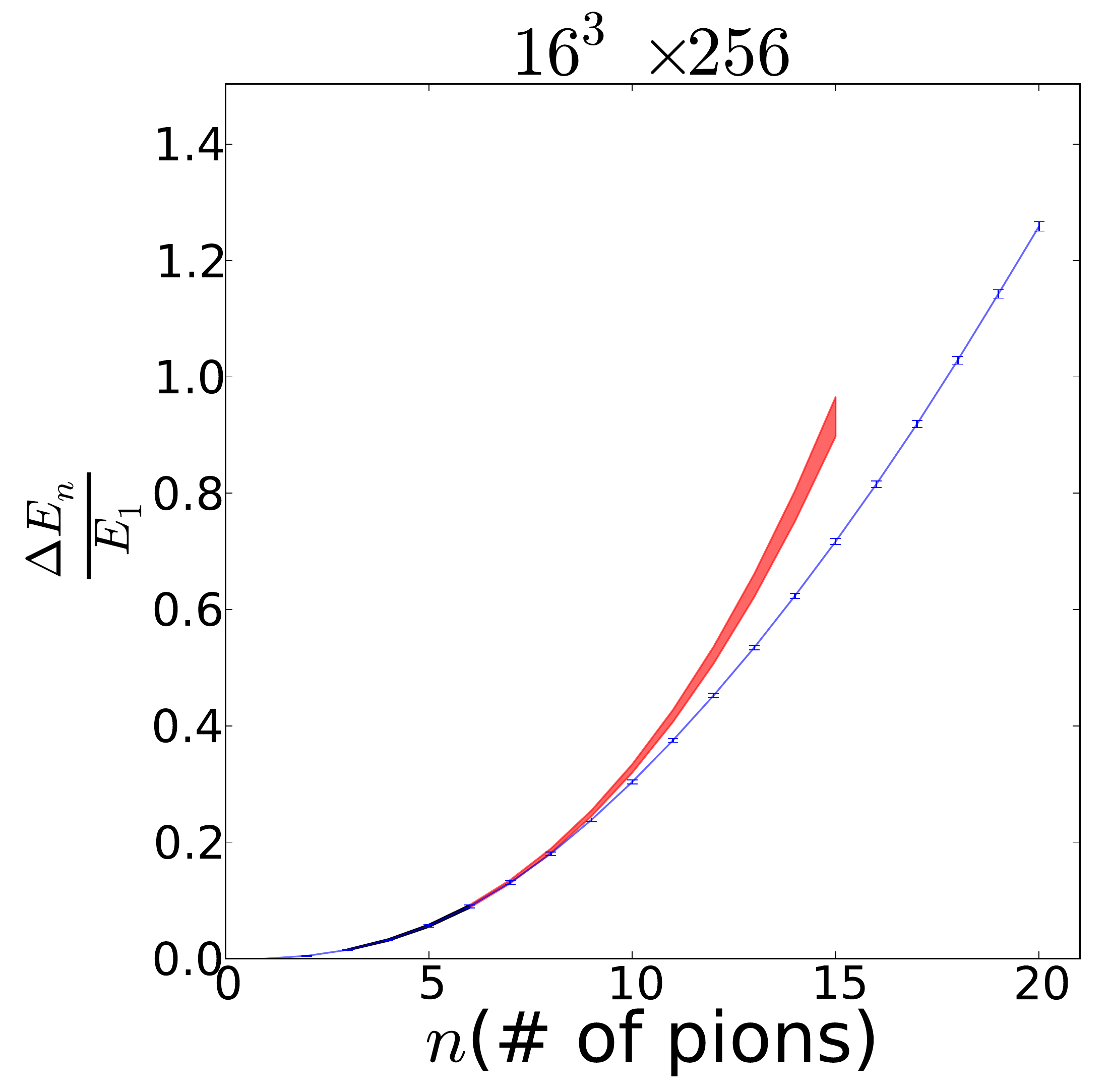}
  \includegraphics[width=5.4cm]{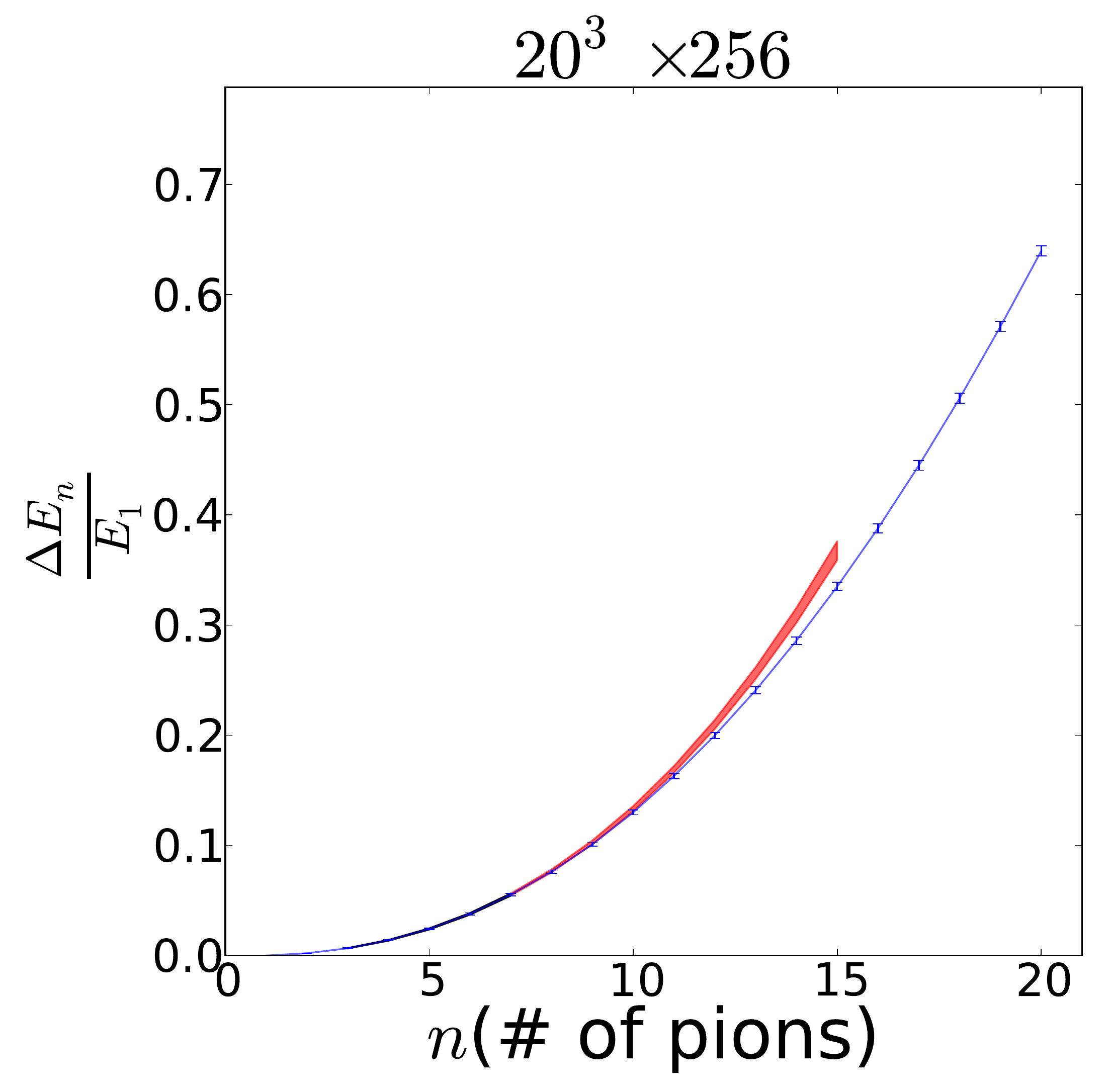}
  \includegraphics[width=5.4cm]{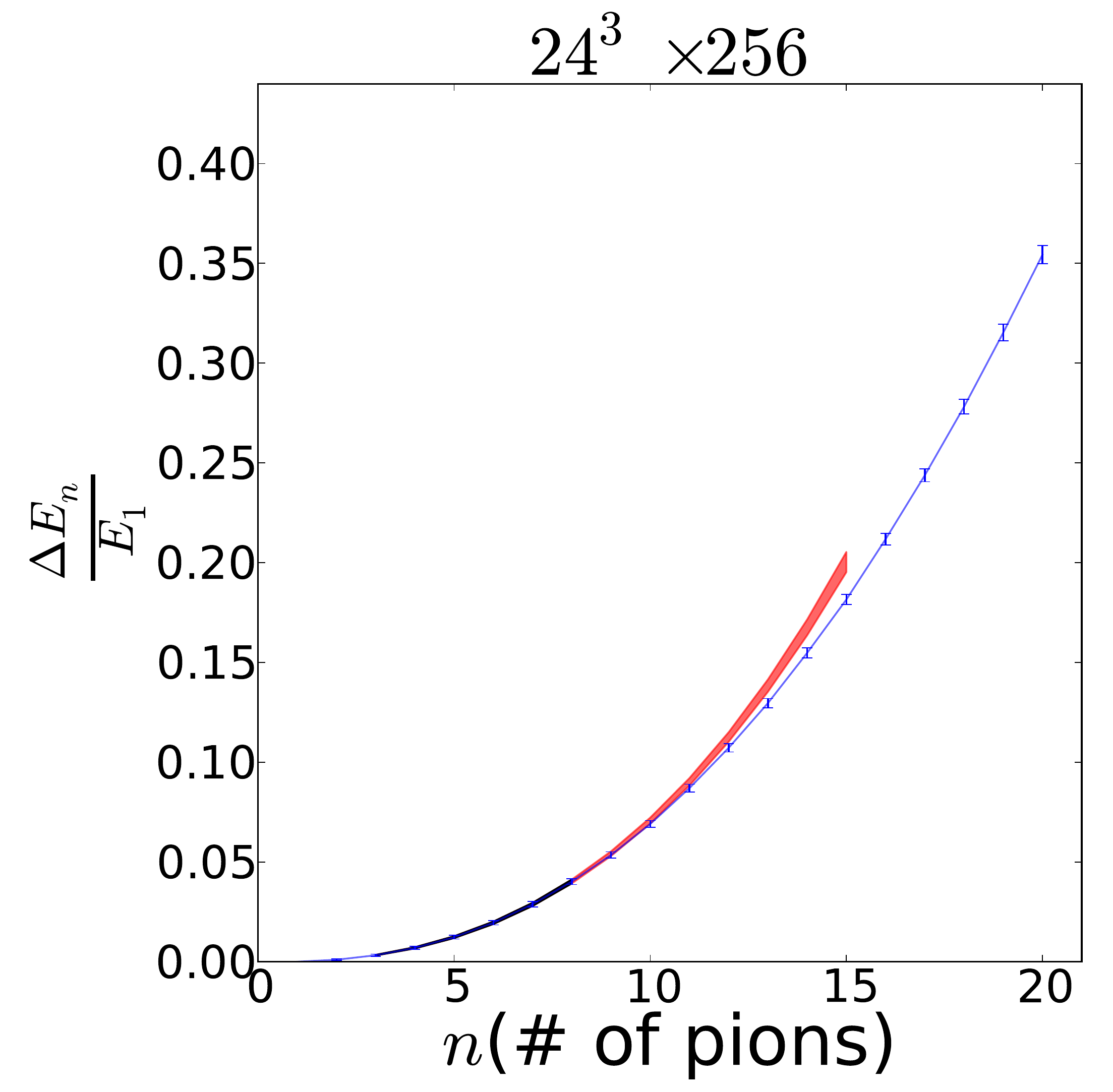}
  \caption{The energy differences, $\Delta E_n$, are plot as a
    function of the number of pions, $n$, where the blue points are
    the original data, the red bands are the fits, and the black bands
    are the regions where the fits are performed. From the left to
    right, $\Delta E_n$ from $16^3$, $20^3$, $24^3$ are shown.  }
  \label{fig:scattering_length_3_6}
\end{figure}

The dimensionless qualities $m_{\pi}{\overline a}$ and
$m_{\pi}f_{\pi}^4 {\overline{\overline \eta}}_3^L$ can be extracted by
fitting $\Delta E_n$ to the large volume expansion of
Eq.~(\ref{eq:energyshift}).  The fitting strategy is similar to that
used in L\"uscher's method by first fitting to each bootstrap ensemble
and then computing the distribution of fitted parameters in order to
get statistical and systematic uncertainties.  There are two ways to
extract $m{\overline a}$. One is by fitting only to $\Delta E_2$ using
Eq.~(\ref{eq:energyshift}) with the last two lines set to zero, and
the other way is by fitting multiple $\Delta E_n$'s, with $n \ge 3$,
and extracting $m_{\pi}f_{\pi}^4 {\overline{\overline \eta}}_3^L$ at
the same time as is shown in Table.~\ref{table:eta3_diff_time} and
Fig.~\ref{fig:eta3_diff_time}. The final ${\overline a}$ and
$mf_{\pi}^4 {\overline{\overline \eta}}_3^L$ extracted from the later
method are chosen from fits with $\chi^2 \sim 1$.  We are forced to
to use only few body systems as the quality of fit rapidly decreases
for large numbers of pions. This suggests that the weakly interacting
pion model of the system that Eq.~(\ref{eq:energyshift}) encodes is
becoming less valid.  Results for the two-body interaction extracted
in both ways agree within uncertainties with those extracted using
L\"uscher's method, and are shown in
Table~\ref{table:scattering_small_expansion}.  The original data for
the $\Delta E_n$'s and the results from the fits are shown in
Fig.~\ref{fig:scattering_length_3_6}.

\begin{table}[ht]
  \caption{The effective scattering length (${\overline a}$) from small ${\overline a}/L$ expansion.
    The symbol ``[2]'' indicates that only $\Delta E_2$ is used in the fitting,
    and ``[3,6]'' means that all $\Delta E_3$ to $\Delta E_6$ are used.}
  \centering
  \begin{tabular}{c  c  c c c}
    \hline\hline
    $V^3 \times T$ &$m_{\pi}{\overline a}[2]$ & $m_{\pi}{\overline a}[3,6]$ & $k\cot \delta /m_{\pi}$ & $m_{\pi}f_{\pi}^4{\overline{\overline \eta}}_3^L$[3,6] \\
    \hline
    $16^3 \times 128$ & \hspace{1cm}$0.259(14)(5) $ & \hspace{1cm}$0.260(14)(2)$ &\hspace{1cm} $-3.85(21)(3)$& \hspace{1cm}$0.70(10)(4)$ \\
    \hline
    $20^3 \times 256$ & \hspace{1cm}$0.234(6)(1)$ & \hspace{1cm}$0.235(6)(5) $ &\hspace{1cm} $-4.26(11)(10)$& \hspace{1cm}$0.79(7)(1)$\\
    \hline
    $24^3 \times 128$ & \hspace{1cm}$0.205(12)(5) $ & \hspace{1cm}$0.210(11)(2) $ &\hspace{1cm} $-4.78(25)(7)$ & \hspace{1cm}$1.50(16)(15)$\\
    \hline

  \end{tabular}
  \label{table:scattering_small_expansion}
\end{table}

\begin{figure}
  \includegraphics[width=12.0cm]{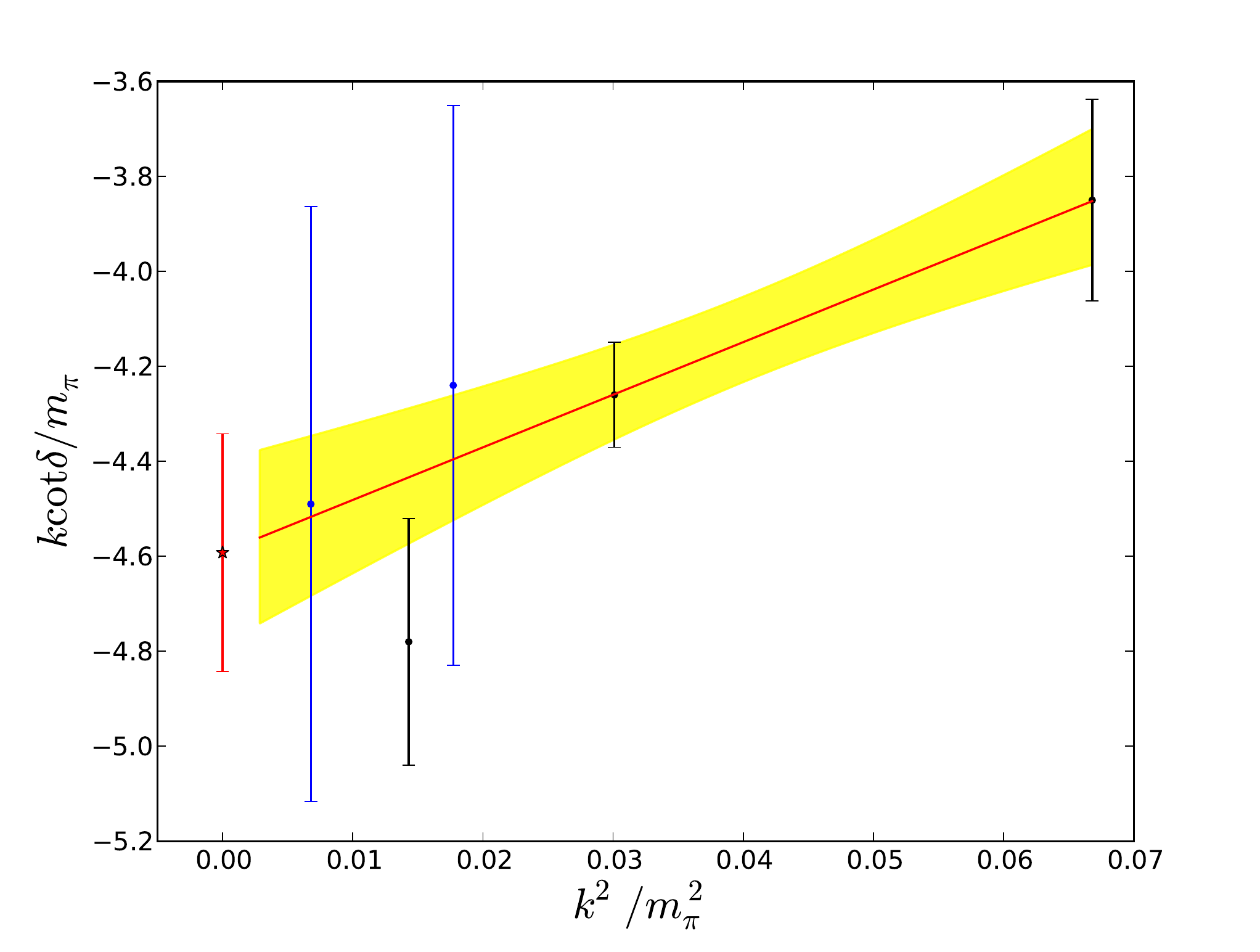}
  \caption{The scattering phase shifts from $16^3$, $20^3$, and $24^3$
     ensembles in this study, are shown as the black data points from right to left respectively. 
   The blue data points are the $24^3$ and $32^3$ ensemble
     results from Ref.~\cite{Beane:Detmold:2011} from right to left respectively.
     The $24^3$ data
    is excluded in the fit as discussed in the main text. 
      The shaded
    region is the uncertainty and the star is the infinite volume
    result.}
  \label{fig:scattering_fit}
\end{figure}

The effective scattering length, ${\overline a}$, extracted from the three
volumes depend on the volume. With multiple volumes,
Eq.~(\ref{eq:energyshift}) can be inverted to extract both the
scattering length, $a$, and the effective range, $r$.
In order to do this, we have also used $k\cot \delta/m_{\pi}$
determined on a matching $32^3 \times 256$ ensemble from
Ref.~\cite{Beane:Detmold:2011} with all lattice parameters the same,
except for the volume.  As the central value of the scattering length
from ensemble B3 deviates significantly from the trend of the other
ensembles (probably due to statistical fluctuations) and is lower
than the value extracted for this ensemble in
Ref.~\cite{Beane:Detmold:2011}, we exclude this point from our fit.
We are using the simplest form,
$k\cot\delta/m_\pi = -\frac{1}{m_\pi a} + \frac{m_\pi r}{2}(\frac{k^2}{m_\pi^2})$,
and neglecting higher order shape parameters as our interacting momenta
are small.
The infinite volume
results are $1/m_{\pi} a = -4.60(25)$ and $m_{\pi}r = 22.7(12.3)$,
which agree with the determinations of
Ref.~\cite{Beane:Detmold:2011}. The fit is shown in
Fig.~\ref{fig:scattering_fit}.

\begin{figure}
  \includegraphics[width=12cm]{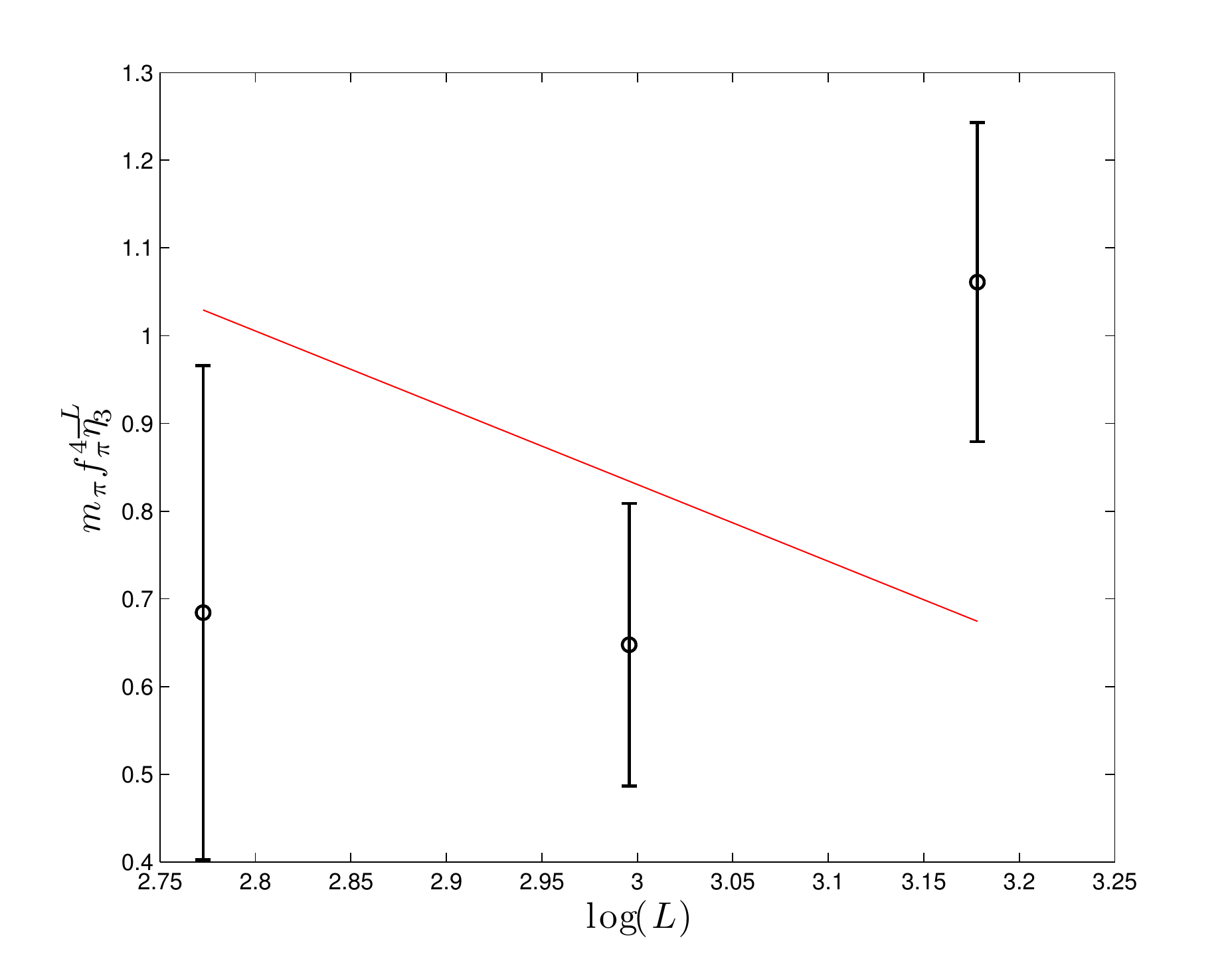}
  \caption{The extracted three-body interaction parameter, ${\overline
      \eta}_3^L(L)$, is plotted as a function of the spatial extent of
    the lattice, $L$, (black points).  The red line shows the
   expected dependence of ${\overline \eta}_3^L$ on $L$ from
    Eq.~(\ref{eq:simplify_eta3}) with $C = 4.3$, which clearly does
    not provide a good description of the data.}
  \label{fig:eta_3_bar}
\end{figure}

By utilizing the extracted effective range, $r$, and the effective scattering length,
${\overline a}(L)$, from the three different volumes, from
Eq.~(\ref{eq:eta3barbar}), the volume dependent parameter ${\overline
  \eta}_3^L$, responsible for the three-body interactions can be
determined for each volume. The extracted values of ${\overline
  \eta}_3^L$ are shown in Fig.~\ref{fig:eta_3_bar}.  The dependence of
${\overline \eta}_3$ on the volume can be rewritten from
Eq.~(\ref{eq:etathreebar}) into a simpler form
\begin{eqnarray} {\overline \eta}_3^L(L) = C + \frac{\alpha a^4
  }{M}\log (L), 
  \label{eq:simplify_eta3}
\end{eqnarray}
where $C$ contains contributions independent of $L$, and $\alpha = {64
  \pi}(3 {\sqrt 3} - 4\pi) = -1.48\times 10^{3}$.  We fit ${\overline
  \eta}_3^L$ to our data to determine $C$ and the best fit is shown in
Fig.~\ref{fig:eta_3_bar}. However the $\chi^2$ of the fit is poor and
it appears that Eq~(\ref{eq:simplify_eta3}) does not effectively
explain the volume dependence of our data. This might come from the
competing of higher order terms ${\cal O}(\frac{1}{L^3})$, but it also
may be a statistical effect.  The large value of ${\overline
  \eta}_3^L$ for $L=24$ is correlated with a down shift of the
scattering length ${\overline a}$. In Ref.~\cite{Beane:Detmold:2011},
a value of $m{\overline a} = 0.236(18)(27)$ was found for $L=24$,
which agrees with the value $m{\overline a} = 0.210(16)(5)$ found
above, but with a large central value,
 perhaps indicating a statistical fluctuation.

\section{QCD phase diagram at non-zero $\mu_I$}
\label{sec:qcd_phase_diagram}


\begin{figure}
  \includegraphics[width=15cm]{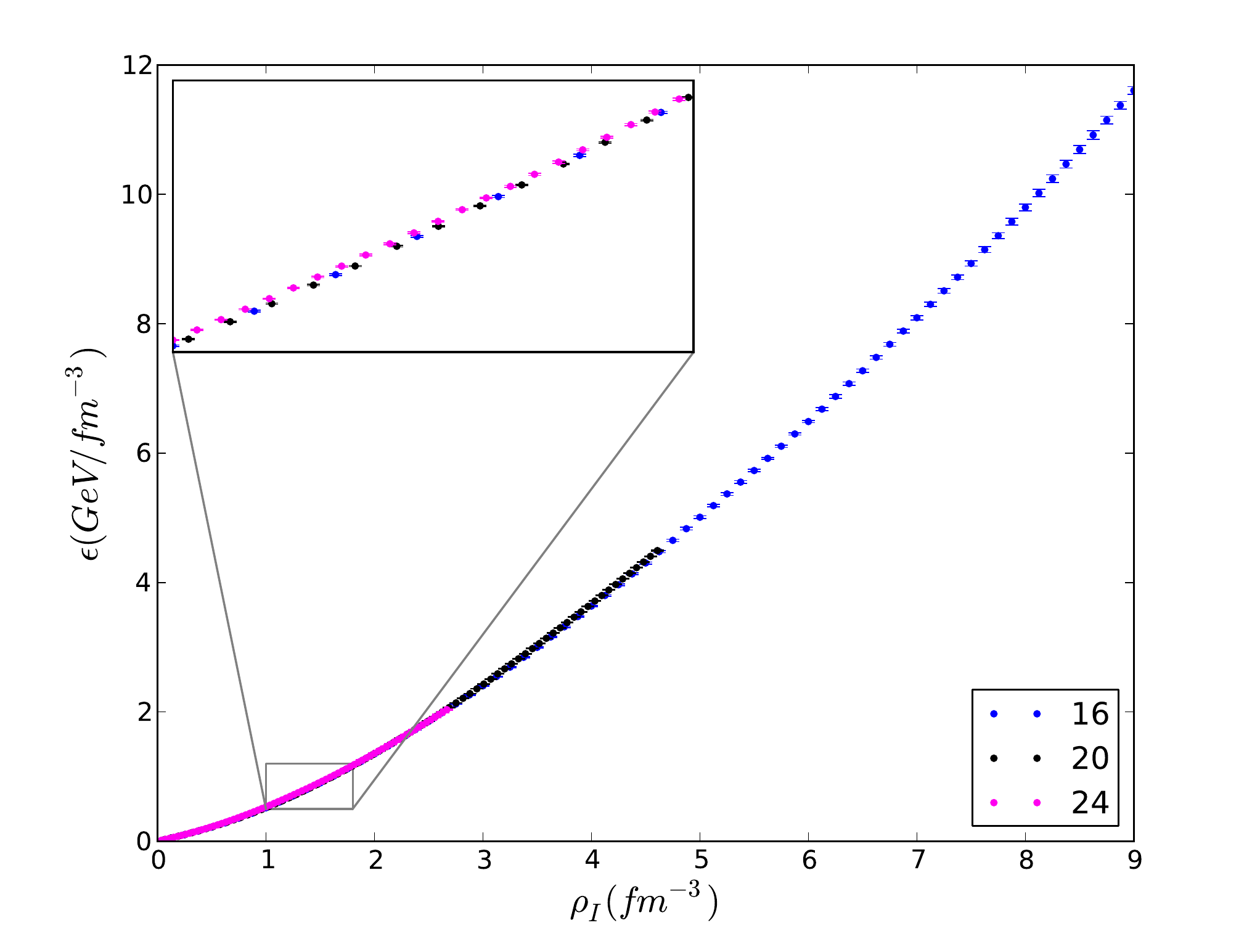}
  \caption{Energy densities ($\epsilon$) calculated on $3$ different
    volumes are shown as a function of isospin density.  The blue
    points are from the $16^3$ ensemble, the black ones are from the
    $20^3$ ensemble and the pink one are from the $24^3$ ensemble.
    The inset show the slight difference in energy density on three
    ensembles.  }
  \label{fig:energy_density_16_20_24}
\end{figure}

In Fig.~\ref{fig:energy_density_16_20_24}, we show the energy density,
$\epsilon = \frac{E}{V}$, determined from the ground state energies,
$E_{n\pi}$ that have been computed on each of the three volumes.  For
a fixed $n$, the pions are forced to be closer to each other in a
smaller volume, and the repulsive interactions between them become
stronger. This drives up the energy of the whole system.
The energy densities are weakly dependent on the volume, however
there are slightly differences as shown in
the inset of Fig.~\ref{fig:energy_density_16_20_24}.

\begin{figure}
  \includegraphics[width=15.0cm]{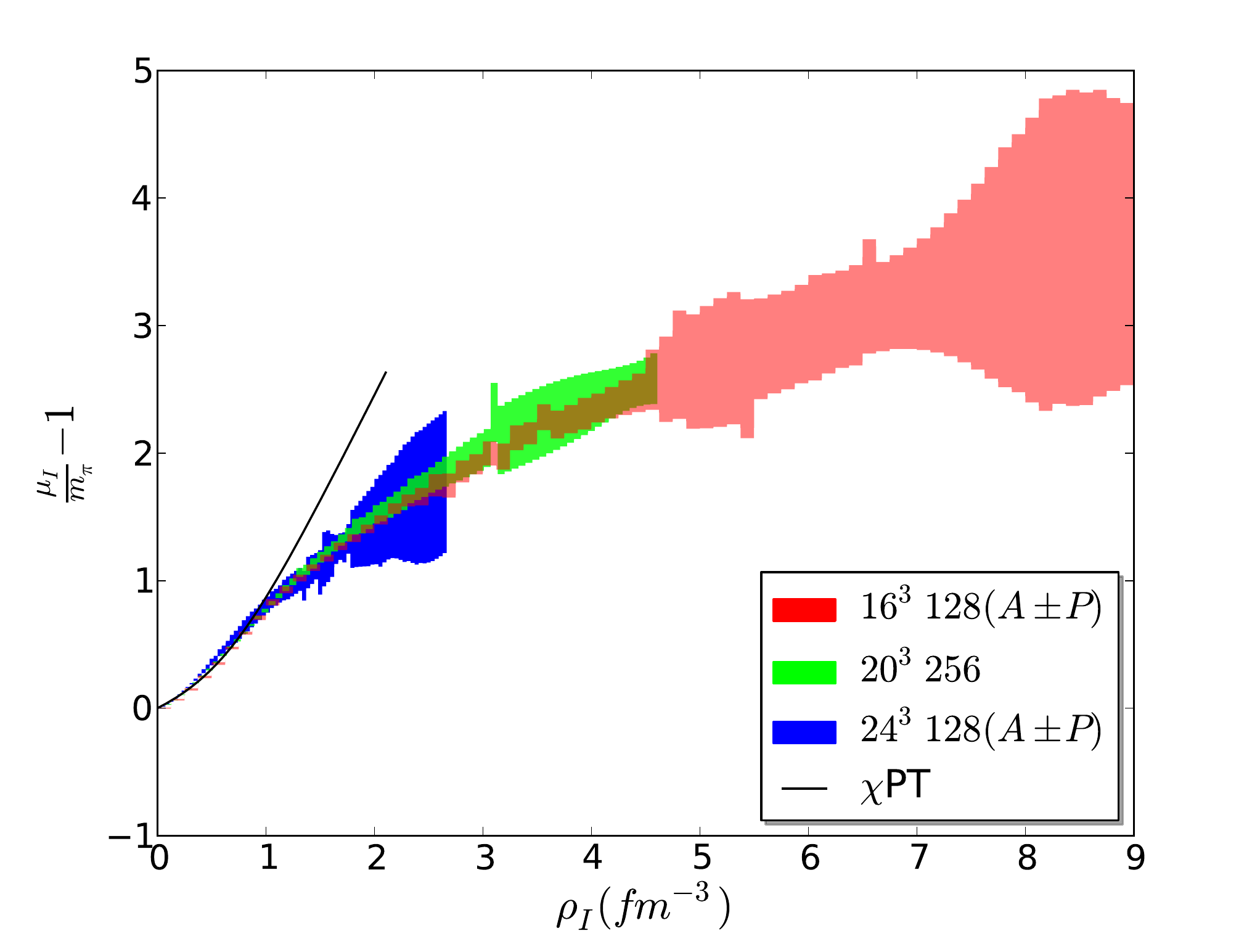}
  \caption{The isospin chemical potential, $\mu_I$, is plotted as a
    function of the isospin density, $\rho_I$, from three lattice
    ensembles, B1 (red), B3 (blue) and B4 (green).  The solid
    black line is from expectations of $\chi$PT \cite{Son:Stephanov} }
  \label{fig:chemical_from_256}
\end{figure}

From the extracted ground state energies, the isospin chemical
potential can also be determined by a backward finite difference,
${\mu}_I\left( n \right ) = {{dE}\over{dn}} \sim E_n-E_{n-1}$.  We
calculate $\mu_I(n)$ on each bootstrap ensemble, which accommodates
correlations between $E_{n\pi}$'s extracted on the same ensemble, and
the systematic uncertainty of the ${\mu}_I(n)$ from each ensemble is
evaluated by adding systematic uncertainty from varying the fit ranges
used to determine $E_{n}$ and $E_{n-1}$ in quadrature.  The final
systematic uncertainty on $\mu_I(n)$ is from averaging the systematic
uncertainties of all the bootstrap ensembles, and the statistical
uncertainty is the standard deviation of the values of $\mu_I(n)$ on
the individual bootstrap ensembles.

In Fig.~\ref{fig:chemical_from_256}, the dependence of
$\mu_I/m_{\pi}-1$ on the isospin density $\rho_I$ is shown for
the three volumes.  The isospin chemical potential exhibits similar
behaviour in all three volumes, where they overlap.
At small $\rho_I$, $\mu_I$ increases
at an accelerating rate, in agreement with the prediction from chiral
perturbation theory ($\chi$PT)~\cite{Son:Stephanov}, however at around
$\rho_I \approx 0.5\ {\rm fm}^{-3}$ the behaviour of $\mu_I$ starts to
change, and the accelerating rate gradually decreases, and at even
higher isospin density the $\mu_I$ starts to flatten off.  This change
of behaviour of $\mu_I$ indicates that the physical state of the
system may be altering.

\begin{figure}
  \includegraphics[width=12.0cm]{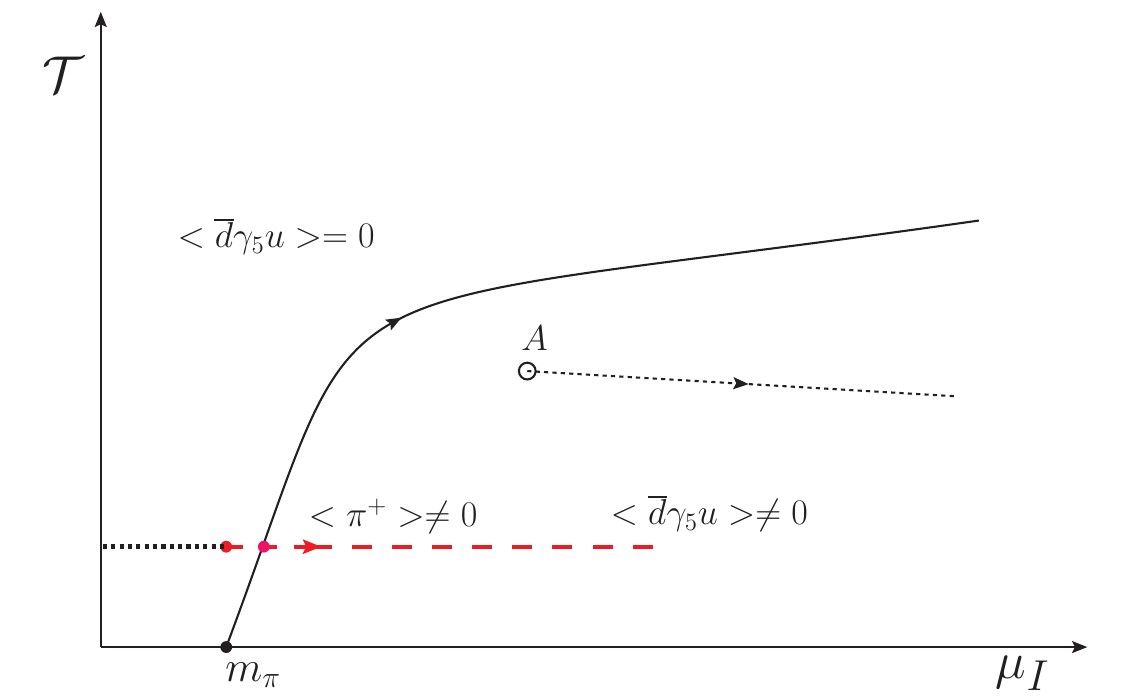}
  \caption{Expected QCD phase diagram following
    Ref.~\protect{\cite{Son:Stephanov}}. Our calculations at a fixed
    temperature, ${\cal T}\sim 20\ {\rm MeV}$ probe the phase structure
    along the red dashed line from $\mu_I = m_{\pi}$ to $\mu_I = 4.5\
    m_{\pi}$. }
  \label{fig:phase_this_study}
\end{figure}

The expected phase structure of QCD at non-zero isospin chemical
potential has been discussed in Ref.~\cite{Son:Stephanov}. At zero
temperature, when $\mu_I < m_{\pi}$, there is not enough energy to
excite a pion out of the vacuum.  As soon as $\mu_I$ reaches
$m_{\pi}$, pions can be produced and the system enters a phase
with a pion condensate (BEC) via a second order phase transition. At
asymptotically large values of $\mu_I$, the attractive nature of one
gluon exchange guarantees the existence of a colour-superconducting
BCS-like state in which quark--anti-quark Cooper pairs are formed. At
an intermediate value of $\mu_I$ a BEC-BCS crossover is
conjectured~\cite{Son:Stephanov}.

In this paper, our calculations are performed at a small but nonzero
temperature, ${\cal T}\sim 20$ MeV. With the canonical method used in
the current calculation, the lowest isospin chemical potential that we
probe is $\mu_I = m_{\pi}$ by definition as we directly add $\pi^+$'s
into the system.  In the smallest volume, for $n=72$ $\pi^+$'s (the
largest value we consider), an isospin density of $\rho_I \sim 9\ {\rm
  fm}^{-3}$ is achieved, and the phase diagram is explored from $\mu_I
= m_{\pi}$ up to $\mu_I \approx 4.5\ m_{\pi}$ in this paper as shown
by the red dashed line in Fig.~\ref{fig:phase_this_study}.

\begin{figure}
  \includegraphics[width=12cm]{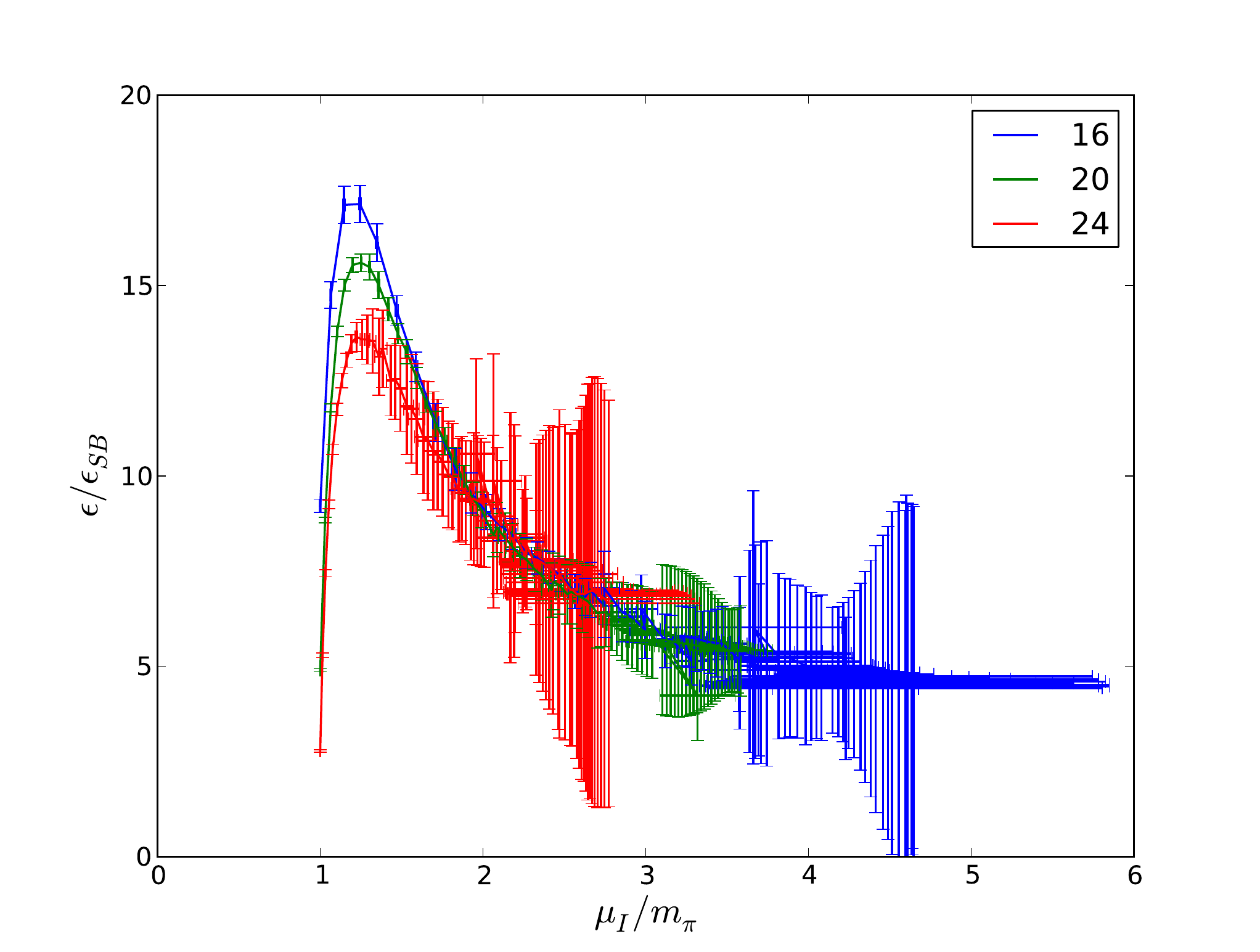}
  \caption{The $\epsilon/\epsilon_{SB}$ is plotted as a function of
    $\mu_I/m_\pi$.}
  \label{fig:epsilon_mu4_mu}
\end{figure}

In order to investigate the possible phase transition suggested by the
behaviour of the isospin chemical potential in more detail, we have
also compared the extracted energy density with the energy density of
a cold degenerate system described by a model of weakly interacting
quarks filling their Fermi sphere up to a maximum momentum $k_F
\approx E_F = \mu_I$~\cite{simon:seyong}.  This Stefan-Boltzmann
energy density is given by
\begin{eqnarray}
  \epsilon_{SB} = \frac{N_f N_c}{4 \pi^2}\mu_I^4
\end{eqnarray}
where $N_f = 3$ and $N_c = 3$.  The ratio of $\epsilon /
\epsilon_{SB}$ is plotted in Fig.~\ref{fig:epsilon_mu4_mu}, and
exhibits similar behaviours in all three volumes.  The ratio increases
from $\mu_I = m_{\pi}$ to a peak around $\mu_I \approx 1.3\ m_{\pi}$,
and then drops and eventually begins to plateau at around $\mu_I
\approx 3\ m_{\pi}$.  Peak positions, $\mu^I_{peak}$, for each volume
identified from Fig.~\ref{fig:epsilon_mu4_mu} are $\mu^I_{peak} = \{
1.20(5), 1.25(5), 1.27(5) \}\ m_{\pi}$ for $L = \{ 16, 20, 24\}$
respectively. With an extrapolation linear in $1/L^3$, the peak
position in infinite volume is $\mu^I_{peak} = 1.30(7)\ m_{\pi}$.
The system for $\mu_I < 1.3\ m_{\pi}$, can be identified as a pion
gas.  When $\mu_I\sim\mu^I_{peak}$, pions start to condense and the
system resides in the BEC state.  The plateau beginning to form beyond
$\mu_I \approx 3\ m_{\pi}$, may indicate a crossover from the BEC to BCS
state, however higher precision and larger $\mu_I$ is required to
make a definite statement. Discretization effects also remain to be 
investigated.

Two flavour QCD with finite $\mu_I$ at large ${\cal T}$ has been
investigated in Ref.~\cite{Kogut:2004zg}, where a finite temperature
deconfinement phase transition was identified at $\mu_I < m_{\pi}$,
however for $\mu_I > m_{\pi}$ no results were presented.  In
Ref.~\cite{deForcrand:Wenger2007}, the phase diagram of $N_f = 4+4$
QCD was investigated at different temperatures and values $\mu_I$ using
the grand canonical approach, and a phase transition from a pion gas
to a BEC state has also been suggested at $\mu_I$ slightly higher than
$m_{\pi}$, in agreement with the results found here.  Two color QCD
has been studied in Ref.~\cite{simon:seyong}, where the authors
identified the transition from vacuum to BEC state and the BEC/BCS
transition.  Somewhat interestingly, 
the ratio of the energy density and its Stefan-Boltzmann
limit has also been studied (inset of Fig.~1 in
Ref.~\cite{simon:seyong}), showing qualitatively similar behaviour to
that found in the current study.

\section{Conclusion}

In this work, we have studied lattice QCD at non-zero isospin chemical
potential using a canonical approach in which we have investigated
systems with the quantum numbers of up to 72 $\pi^+$'s in three
lattice volumes, $L^3 \sim$ (2.0, 2.5 and 3.0 ${\rm fm})^3$ at a pion
mass of $m_\pi\sim390$ MeV at a single lattice spacing. In order to
perform this study, we have developed several new methods for
performing the requisite Wick contractions of quark field operators.
These methods are an enormous computational improvement over previous
approaches and their accuracy and performance have been carefully
investigated.

In our analysis, we have determined the ground state energies of
multi-pion systems in three different volumes and have used this to
extract the isospin chemical potential of the states that are
produced. In the smallest volumes, systems with isospin chemical potentials of
up to $\mu_I\sim 1600$ MeV are created. By considering the energy
density as a function of the isospin chemical potential, we provide
strong evidence for the transition of the system from a weakly
interacting pion gas to a Bose-Einstein condensed (BEC) phase at
$\mu\sim m_\pi$ as expected from $\chi$PT. At higher values of the
chemical potential the system is expected to transition to a BCS
superconducting state and we have sought numerical evidence for this
but do not have conclusive results. It is interesting to note that the
behaviour of the energy density as a function of the isospin chemical
potential is very similar to that recently found in two-colour QCD
with a baryon chemical potential by Hands et al.~\cite{simon:seyong}.

By focusing on few pion systems, we have extracted the two and three
pion interactions, determining the scattering length, effective range
and the renormalisation group invariant effective three-body
interaction. The scattering parameters were found to be in good
agreement with other recent determinations and we have attempted to
investigate the intrinsic volume dependence of the renormalisation
group invariant three-pion interaction. We have also found that as the
density increases and the system transitions to a BEC, it can no
longer be well described in terms of weak few-body interactions.

\section{Acknowledgment}
We thank Josh Erlich and Martin Savage for valuable discussions.  
We appreciate the support of the DOE NERSC facility, NSF XSEDE resources and 
in particular TG-PHY080039N, as well as the Sporades cluster at
the College of William \& Mary.
The work of WD, KO and ZS was supported in part by JSA, LLC under DOE
contract No. DE-AC05-06OR-23177 and by the Jeffress Memorial Trust,
J-968.  WD and KO were supported in part by DOE grant
DE-FG02-04ER41302.  WD was also supported by DOE OJI Award
DE-SC000-1784.


\begin{thebibliography}{99}
    \bibitem{Beane:2009gs} S.~R.~Beane {\it et al.} [NPLQCD],
  Phys.\ Rev.\ D {\bf 80}, 074501 (2009) [arXiv:0905.0466].


    \bibitem{Yamazaki:2009ua} T.~Yamazaki, Y.~Kuramashi, and A.~Ukawa,
  Phys.\ Rev.\ D {\bf 81}, 111504 (2010) [arXiv:0912.1383].

    \bibitem{Beane:2007es} S.~R.~Beane, W.~Detmold, T.~C.~Luu,
  K.~Orginos, M.~J.~Savage and A.~Torok [NPLQCD],
  Phys.\ Rev.\ Lett.\ {\bf 100}, 082004 (2008) [arXiv:0710.1827].

    \bibitem{Detmold:2008fn} W.~Detmold, M.~J.~Savage, A.~Torok,
  S.~R.~Beane, T.~C.~Luu, K.~Orginos and A.~Parre\~no [NPLQCD],
  Phys.\ Rev.\ D {\bf 78}, 014507 (2008) [arXiv:0803.2728].

    \bibitem{Detmold:2008yn} W.~Detmold, K.~Orginos, M.~J.~Savage and
  A.~Walker-Loud [NPLQCD],
  Phys.\ Rev.\ D {\bf 78}, 054514 (2008) [arXiv:0807.1856].

    \bibitem{Detmold_Brian:2011} W.~Detmold,\ B.~Smigielski
  Phys.\ Rev.\ D {\bf 84}, 014508 (2011) [arXiv:1103.4362]


    \bibitem{Kaplan:1986} D.~B.~Kaplan,\ A.~E.~Nelson
  Phys.\ Lett.\ {\bf B175},57 (1986)


    \bibitem{Son:Stephanov} D.~T.~Son and M.~A.~Stephanov,
  Phys.\ Rev.\ Lett.\ {\bf 86}, 592 (2001) [arxiv:hep-lat/0005225].
 

    \bibitem{Kogut:2004zg} J.~B.~Kogut and D.~K.~Sinclair,
  Phys.\ Rev.\ D {\bf 70}, 094501 (2004) [arXiv:hep-lat/0407027].

    \bibitem{Sinclair:2006zm} D.~K.~Sinclair and J.~B.~Kogut,
  PoS {\bf LAT2006}, 147 (2006) [arXiv:hep-lat/0609041].

    \bibitem{deForcrand:Wenger2007} Ph.~de~Forcrand,\
  M.~A.~Stephanov,\ U.~Wenger
  PoS {\bf LAT2007}, 237 (2007) [arXiv:hep-lat/0711.0023]

    \bibitem{Detmold:Savage} W.~Detmold and M.~J.~Savage,
  Phys.\ Rev.\ D{\bf 82}, 014511 (2010) [ arxiv:1001.2768].

    \bibitem{Z_W_2011_proceeding} Z.~Shi,\ W.~Detmold,
  PoS\ {\bf LAT 2011}, 328 (2011) [arXiv:1111.1656]

    \bibitem{heuy:david2008} H.~Lin,\ S.~D.~Cohen,\ J.~Dudek,\
  R.~G.~Edwards,\ B.~Jo\'o,\ D.~G.~Richards,\ J.~Bulava,\ J.~Foley,\
  C.~ Morningstar,\ E.~Engelson,\ S.~Wallace,\ K.~J.~Juge,\
  N.~Mathur,\ M.~J.~Peardon,\ S.~M.~Ryan,
  Phys.\ Rev.\ D {\bf 79}, 034502 (2009) [arXiv:0810.3588]

    \bibitem{V_M_inverse:cite} N.~Macon,\ A.~Spitzbart,
  The American Mathematical Monthly,\ Vol.\ {\bf 65}, No. 2: 95–100
  (1958)

    \bibitem{arprec_cite} D.~H.~Bailey,\ Y.~Hida,\ X.~S.~Li,\
  B.~Thompson, ``ARPREC: An arbitrary precision computation package,''
  September 2002.  Available at
  http://crd.lbl.gov/$\sim$dhbailey/dhbpapers/arprec.pdf.


    \bibitem{qd} Y. Hida, X. S. Li, and D. H. Bailey. 
  ``Quad-double arithmetic: Algo- rithms, implementation, and
  application.'' 
  Technical Report LBNL-46996, (2000). 

    \bibitem{shei_wohl_83} B.~Sheikholeslami,\ R.~Wohlert,
  Nucl.\ Phys.\ B{\bf 259}, 572 (1985)

    \bibitem{Okamoto:2001jb} M.~Okamoto {\it et al.}  [CP-PACS
  Collaboration],
  Phys.\ Rev.\ D {\bf 65}, 094508 (2002) [arXiv:hep-lat/0112020].

    \bibitem{Chen:2000ej} P.~Chen,
  Phys.\ Rev.\ D {\bf 64}, 034509 (2001) [arXiv:hep-lat/0006019].

    \bibitem{Sheik_Woh:85} B.~Sheikholeslami,\ R.~Wohlert,
  Nucl.\ Phys.\ B{\bf 259}, 572 (1985)

    \bibitem{colin:mike2004} C.~Morningstar,\ M.~Peardon,
  Phys.\ Rev.\ D{\bf 69}, 054501 (2004) [arXiv:hep-lat/0311018]


    \bibitem{rbc:2001} T.~Blum,\ P.~Chen,\ N.~Christ,\ C.~Cristian,\
  C.~Dawson,\ G.~Fleming, R.~Mawhinney,\ S.~Ohta,\ G.~Siegert,\
  A.~Soni,\ P.~Vranas,\ M.~Wingate,\ L.~Wu,\ Y.~Zhestkov,
  Phys.\ Rev.\ D {\bf 68}, 114506 (2003) [arXiv:hep-lat/0110075]

    \bibitem{rbc:2006} Y.~Aoki,\ T.~Blum,\ N.~H.~Christ,\ C.~Dawson,\
  T.~Izubuchi,\ R.~D.~Mawhinney, J.~Noaki,\ S.~Ohta,\ K.~Orginos,\
  A.~Soni,\ N.~Yamada,
  Phys.\ Rev.\ D {\bf 73}, 094507 (2006) [arXiv:hep-lat/0508011]

    \bibitem{chris:jack2007} C.~Aubin,\ J.~Laiho,\ R.~S.~Van de Water,
  Phys.\ Rev.\ D {\bf 81}, 014507 (2009) [arXiv:0905.3947]

    \bibitem{Luscher:1986pf} M.~L\"uscher,
  Commun.\ Math.\ Phys.\ {\bf 105}, 153 (1986).

    \bibitem{Luscher:1990ux} M.~L\"uscher,
  Nucl.\ Phys.\ B {\bf 354}, 531 (1991).


    \bibitem{Beane:2003da} S.~R.~Beane, P.~F.~Bedaque, A.~Parre\~no
  and M.~J.~Savage,
  Phys.\ Lett.\ B {\bf 585}, 106 (2004) [arXiv:hep-lat/0312004].

    \bibitem{Beane:2007qr} S.~R.~Beane, W.~Detmold and M.~J.~Savage,
  Phys.\ Rev.\ D {\bf 76}, 074507 (2007) [arXiv:0707.1670].

    \bibitem{Detmold:2008gh} W.~Detmold and M.~J.~Savage,
  Phys.\ Rev.\ D {\bf 77}, 057502 (2008) [arXiv:0801.0763].

    \bibitem{Tan:2007bg} S.~Tan,
  Phys.\ Rev.\ A {\bf 78}, 013636 (2007) [arXiv:0709.2530]

%
    \bibitem{Smi:Was_2009} B.~Smigielski and J.~Wasem
  Phys.\ Rev.\ D {\bf 79}, 054506 (2009) [arXiv:0811.4392]

    \bibitem{Beane:Detmold:2011} S.~R.~Beane, E.~Chang, W.~Detmold,
  H.~W.~Lin, T.~C.~Luu, K.~Orginos, A.~Parre\~no, M.~J.~Savage,
  A.~Torok, A.~Walker-Loud,
  Phys.\ Rev.\ D {\bf 85}, 034505 (2012)
  [arXiv:1107.5023]


    \bibitem{simon:seyong} S.~Hands,\ S.~Kim,\ J.~Skullerud,
  Phys.\ Rev.\ D{\bf 81}, 091502 (2010) [arXiv:1001.1682]





\end{thebibliography}
\end{document}